\documentclass[10pt,conference]{IEEEtran}
\IEEEoverridecommandlockouts
\usepackage{cite}
\usepackage{amsmath,amssymb,amsfonts}
\usepackage{algorithmic}
\usepackage{algorithm}
\usepackage{graphicx}
\usepackage{textcomp}
\usepackage[table]{xcolor}
\usepackage{multirow}
\usepackage{array}
\usepackage{tikz}
\usepackage{tcolorbox}
\usepackage{booktabs}
\usepackage{adjustbox}
\usepackage{csquotes}
\usepackage{standalone}
\usepackage{subcaption}
\usepackage[hidelinks]{hyperref}
\usepackage{pgfplots}
\usepackage{balance}
\usepackage[flushleft]{threeparttable}
\def\BibTeX{{\rm B\kern-.05em{\sc i\kern-.025em b}\kern-.08em
    T\kern-.1667em\lower.7ex\hbox{E}\kern-.125emX}}

\newtcolorbox{rbox}{colback=red!10,boxrule=0.2pt,colframe=black,top=1pt,bottom=1pt,left=1pt,right=1pt}

\newcommand{\revisioncolor}{black} % Set the default color to blue
\newcommand{\revision}[1]{{\color{\revisioncolor}#1}}

\pgfplotsset{compat=1.11,
    /pgfplots/ybar legend/.style={
    /pgfplots/legend image code/.code={%
       \draw[##1,/tikz/.cd,yshift=-0.25em]
        (0cm,0cm) rectangle (3pt,0.8em);},
   },
}

\usepgfplotslibrary{fillbetween}

\usepackage{listings}
\usepackage{url}

\definecolor{mGreen}{rgb}{0,0.6,0}
\definecolor{mGray}{rgb}{0.5,0.5,0.5}
\definecolor{mPurple}{rgb}{0.58,0,0.82}
\definecolor{backgroundColour}{rgb}{0.95,0.95,0.92}

\lstdefinestyle{CStyle}{
    backgroundcolor=\color{backgroundColour},   
    commentstyle=\color{mGreen},
    keywordstyle=\color{magenta},
    numberstyle=\tiny\color{mGray},
    stringstyle=\color{mPurple},
    basicstyle=\footnotesize,
    breakatwhitespace=false,         
    breaklines=true,                 
    captionpos=b,                    
    keepspaces=true,                 
    numbers=left,                    
    numbersep=5pt,                  
    showspaces=false,                
    showstringspaces=false,
    showtabs=false,                  
    tabsize=2,
    language=C,
    frame=none,
    aboveskip=0pt,
    belowskip=0pt
}

\begin{document}

\title{Faster Configuration Performance Bug Testing with Neural Dual-level Prioritization}

\author{
\IEEEauthorblockN{Youpeng Ma$^{1}$, Tao Chen$^{2\ast}$, Ke Li$^{3}$}

\IEEEauthorblockA{$^1$ School of Computer Science and Engineering, University of Electronic Science and Technology of China, China}
\IEEEauthorblockA{$^2$ IDEAS Lab, School of Computer Science, University of Birmingham, United Kingdom}
\IEEEauthorblockA{$^3$ Department of Computer Science, University of Exeter, United Kingdom}

\IEEEauthorblockA{myp@std.uestc.edu.cn, t.chen@bham.ac.uk, k.li@exeter.ac.uk}

\thanks{$^{\ast}$Tao Chen is the corresponding author. Youpeng Ma is also supervised in the IDEAS Lab.}
%\thanks{$^{\ast}$Corresponding author.}
}

\newcommand{\approach}{\texttt{NDP}}

\maketitle

\begin{abstract}
As software systems become more complex and configurable, more performance problems tend to arise from the configuration designs. This has caused some configuration options to unexpectedly degrade performance which deviates from their original expectations designed by the developers. Such discrepancies, namely configuration performance bugs (CPBugs), are devastating and can be deeply hidden in the source code. Yet, efficiently testing CPBugs is difficult, not only due to the test oracle is hard to set, but also because the configuration measurement is expensive and there are simply too many possible configurations to test. As such, existing testing tools suffer from lengthy runtime or have been ineffective in detecting CPBugs when the budget is limited, compounded by inaccurate test oracle.

In this paper, we seek to achieve significantly faster CPBug testing by neurally prioritizing the testing at both the configuration option and value range levels with automated oracle estimation. Our proposed tool, dubbed \approach, is a general framework that works with different heuristic generators. The idea is to leverage two neural language models: one to estimate the CPBug types that serve as the oracle while, more vitally, the other to infer the probabilities of an option being CPBug-related, based on which the options and the value ranges to be searched can be prioritized. Experiments on several widely-used systems of different versions reveal that \approach~can, in general, better predict CPBug type in 87\% cases and find more CPBugs with up to 88.88$\times$ testing efficiency speedup over the state-of-the-art tools.

%We consider the mismatch between expected and actual performance tendencies as an oracle to expose configuration-related performance bugs (CPBugs), which often reflect errors at the code level. To achieve cost-effective software iterations, testers require accurate and efficient testing methods, yet existing approaches still have room for improvement in exposing CPBugs.

%In this paper, we conduct an in-depth study of known real-world CPBugs and propose a novel, faster testing framework for configuration-related performance bugs, called Neural Dual-level Prioritizations (NDLP). By employing RoBERTa to repeatedly classify the relevant attributes of configuration options, combined with genetic algorithms, NDLP guides the processes of test sequencing, sampling, and result analysis. Additionally, for options with larger search spaces, we introduce several search methods to accelerate the exposure of CPBugs. We evaluate NDLP on five large-scale open-source software projects, demonstrating that NDLP can achieve a performance improvement of up to XXX.
\end{abstract}


\section{Introduction}

Modern software systems typically have a high degree of configurability wherein the configuration options directly (e.g., certain optimization) or indirectly (e.g., resource allocation) affect software performance, such as throughput and latency~\cite{10.1145/3702986,DBLP:conf/kbse/HeJLXYYWL20,DBLP:journals/tse/ChenGongChen25,DBLP:journals/tse/GongChen25,DBLP:conf/sigsoft/Gong023,DBLP:journals/tosem/ChenL23a,DBLP:journals/pacmse/Gong024}. As software configurability continues to improve, they are also more likely to be buggy. We refer to these performance bugs caused by configuration errors as Configuration Performance Bugs (CPBugs). It is worth noting that CPBugs differ from the typical misconfigurations that concern user-induced configuration errors~\cite{DBLP:conf/issta/ChengZMX21}; instead, they are the errors of the configuration design in the source code that are unintentionally introduced by the developers of the systems~\cite{DBLP:conf/kbse/HeJLXYYWL20}. A typical example of CPBugs has been illustrated in Table~\ref{tb:keyexp}. Here, we can see that the option \verb|read_buffer_size| in \textsc{MySQL} is mainly used to change the size of the read buffer allocated for each sequential table scan request. In the developers' expectation, increasing this option value should allow \textsc{MySQL} to cache more results, hence a larger buffer should improve performance. However, the performance actually drops when increasing the value beyond 256K. This is because, in the code logic, \verb|MyISAM| initializes an \verb|IO_CACHE| for writing and uses \verb|read_buffer_size| for bulk inserts. \verb|my_malloc| is called with the \verb|MY_ZEROFILL| flag, which causes \verb|memset| to be called on the size of \verb|read_buffer_size|, hence mistakenly restricting the permitted memory quote for processing SQL commands and resulting in a large performance decrease.

% \begin{figure}[htbp]
% \centerline{\includegraphics[width=\linewidth]{fig/图片 1.png}}
% \caption{A real-world CPBug.\textit{The incorrect use of a flag in the program results in a call to memset based on the size of read\_buffer\_size.}}
% \label{fig}
% \end{figure}

CPBugs can lead to devastating outcomes~\cite{DBLP:conf/icse/VelezJSAK22,DBLP:conf/icse/WangJLZYXPL23,DBLP:conf/kbse/HeJLXYYWL20}. For example, there have been several large-scale flight delays, which were mainly caused by problematic configuration designs of the systems~\cite{nygard2018release}. Systems from Google and Meta~\cite{DBLP:series/synthesis/2018Barroso,DBLP:conf/sosp/TangKVCWNDK15} have also suffered performance degradation or outage due to configuration issues, leading to a huge loss of revenue.

% For example, large-scale flight delays occurred due to anticipated system performance degradation or interruptions, which were primarily caused by issues in configuration design.
% For example, in 2023, there were several large-scale flight delays due to expected system performance degradation or outages,

 \begin{table}[t!]
\centering

\caption{A real-world example of CPBugs for \textsc{MySQL}.}

\label{tb:keyexp}
\adjustbox{max width=\columnwidth}{
\begin{tabular}{|p{8.5cm}|}
 \toprule

\textbf{ID:} MySQL-44723; \textbf{CPBug-related Option:} \verb|read_buffer_size|\\
%\midrule

\midrule
\textbf{Expected Performance:} Each thread that does a sequential scan for a \texttt{MyISAM} table allocates a buffer of this size (in bytes) for each table it scans. If you do many sequential scans, you might want to increase this value, which defaults to 131072.......\\

\midrule
\textbf{Actual Performance:} The performance decreases if the option \verb|read_buffer_size| is set to be larger than 256K......\\
%while an \texttt{INSERT} statement goes beyond 256K in size

\bottomrule
\end{tabular}
}
\vspace{-0.3cm}
\end{table}

%Performance issues are a major obstacle in today's software development, especially in the current DevOps era of continuous deployment where the time available for developers and testers is extremely limited. These situations not only affect the operation and maintenance of software systems but can also lead to performance degradation or even service interruptions, resulting in significant economic losses. Such occurrences are not uncommon in the real world. A typical example [26] is the Centralized Flow Management System used for airport management. This system is designed to optimize and coordinate air traffic flow, ensuring flights operate safely and efficiently. Due to a configuration in the system's JDBC specification that allows \verb|java.sql.Statement.close()| to throw an \verb|SQLException|, if an exception is thrown when closing the statement, the database connection is not closed, leading to resource leaks. After repeated calls, the system's performance gradually degrades until the resource pool is exhausted, causing a system outage. This resulted in thousands of passengers being stranded at the airport and cost the company hundreds of thousands of dollars. Such incidents are not isolated; in 2023 alone, several large-scale flight delays due to software system performance degradation or outages were reported. Production service systems of companies like Google (approximately 27\%) and Facebook (approximately 16\%) [27,28] have also experienced performance degradation or service interruptions due to configuration issues.

\begin{table*}[t!]
\centering

\caption{Categorization of CPBugs types from He et al.~\cite{DBLP:conf/kbse/HeJLXYYWL20}. Anti-Performance implies better performance at the cost of lower security, consistency, integrity, and etc; Pro-Performance means otherwise. Their value change can be in any direction.}

%Source and target denotes the value of an option in the source and target configuration, respectively

\label{tb:cpbugs-type}
\adjustbox{max width=\textwidth}{
\begin{tabular}{ll|ll|ll}
 \toprule

\textbf{CPBug Type} & \textbf{Option Purpose} & \textbf{Source Option Value}  & \textbf{Target Option Value} & \textbf{Expected Performance} & \textbf{Actual Performance}  \\

\midrule

Type-1&Optimization on-off&OFF&ON&Rise&Drop\\
Type-2&Non-functional trade-off&Ant- Performance&Pro-Performance&Rise&Drop\\
Type-2&Non-functional trade-off&Pro-Performance&Anti-Performance&Drop&Drop beyond expectation\\
Type-3&Resource allocation&Small&Large&Rise&Drop\\
Type-4&Functionality trade-off&ON&OFF&Rise&Drop\\
Type-4&Functionality trade-off&OFF&ON&Drop&Drop beyond expectation\\
Type-5&Non-influential option&Any&Any&Keep&Drop\\

\bottomrule
\end{tabular}
}
\vspace{-0.2cm}
\end{table*}

However, testing and finding CPBugs are challenging, due primarily to the fact that (1) there are simply too many configurations to examine (\textsc{MySQL} has hundreds of options with more than millions of configurations~\cite{DBLP:conf/sigsoft/XuJFZPT15}); (2) measuring configuration performance is highly expensive, e.g., it can take up to 166 minutes to merely measure one configuration on \textsc{MariaDB}~\cite{DBLP:journals/pacmse/0001L24}; and (3) the test oracle is often unclear, i.e., we do not know when a CPBug occurs. While some tools exist for testing CPBugs~\cite{DBLP:conf/icst/PlazarAPDC19,DBLP:conf/icse/NistorSML13,DBLP:conf/kbse/HeJLXYYWL20}, they are limited in the efficiency of testing and the accuracy of oracle estimation. That is, they have not effectively handled the discriminative importance of the options and their value range with respect to the CPBugs, together with restricted rule-driven oracle inference. Therefore, those tools suffer from issues such as long running times or are ineffective in detecting CPBugs when the testing budget is limited while can be largely mislead by incorrect oracle.

%Performance bugs are triggered under stringent conditions and are often difficult to detect because they rarely exhibit symptoms that lead to system failures. The fundamental reason for their difficulty in detection is the lack of a common test oracle.

%why testing is hard

%Here, we believe that configuration performance (e.g., throughput, latency, execution time, and other performance-related parameters) expectations can serve as a powerful candidate for performance bug detection oracles. There are two main reasons for this. First, previous work has shown that most performance bugs are related to configuration. Second, configuration changes reflect general expectations for performance changes. If the actual performance deviates from the expected design, the configuration-related code segments are likely problematic.

In this paper, we propose to test CPBugs with what we call Neural Dual-level Prioritization, dubbed \approach, a framework that can be paired with different heuristic generators. \approach~leverages neural language models for estimating test oracle and, more importantly, for prioritizing CPBug testing, which is motivated by our observations that only a small proportion of the options are responsible for CPBugs and there are specific ranges of the numeric options' values that are more CPBugs-prone. As such, by leveraging on the probabilities of being CPBug-related to the options, we prioritize the test at two levels: at the options level and at the level of search depth for numeric options, aiming to significantly accelerate the detection of CPBugs. Specifically, our contributions are:

%In this paper, we re-examined and reproduced multiple real-world configuration-related performance bugs (CPBugs) in five representative large-scale software systems. By utilizing more advanced natural language processing methods, we extracted performance attributes related to configurations and ranked the correlation of these attributes with CPBugs to optimize the sampling and identification process of CPBugs. We proposed a series of new sampling methods to improve the exposure process of numerical CPBugs. Additionally, we used an automated performance testing framework with a genetic algorithm to inspect actual configuration options in the system, searching for potential CPBugs.

%In summary, our specific contributions are as follows:
\begin{itemize}
    \item Instead of leveraging on rule mining~\cite{DBLP:conf/kbse/HeJLXYYWL20}, we build a neural language model, i.e., RoBERTa~\cite{DBLP:journals/corr/abs-1907-11692}, that predicts the CPBug type which serves as the oracle in testing.
    \item We fine-tune the other RoBERTa model for estimating the probabilities of an option being CPBug-related, which prioritizes the options to be tested.
    %\item Deriving on the probabilities, we prioritize which options to be tested earlier.
    \item For testing numeric options, we design three search depths that bound an underlying heuristic generator with different search spaces. During the actual testing, those depths (and hence their corresponding search spaces) are prioritized according to the commonality of the CPBug-relatedness and code semantic of an option.
    \item We evaluate \approach~over several widely-used systems with various versions, 2--10 workloads, containing 66 known CPBugs, and against state-of-the-art testing tools.
    % \item We conducted further empirical studies based on previous research, and these findings further improved the oracle and testing process.
    % \item We utilized more advanced natural language processing methods to extract configuration-related performance attributes and ranked their correlation with CPBugs. These configuration-related attributes can optimize the testing and identification process of CPBugs.
    % \item We designed a series of new sampling methods combined with a genetic algorithm, which can more quickly trigger real-world CPBugs and address some deficiencies of previous methods.
\end{itemize}

The results show that, compared with state-of-the-art tools/approaches, \approach~more accurately predicts the oracle of CPBug type in 87\% types/metrics; tests CPBugs with up to 1.73$\times$ speedup by prioritizing the tested options; while finding more CPBugs with from 3.13$\times$ to 88.88$\times$ speedup through prioritizing the search depths of numeric options. All data and source code are publicly accessible via our repository: \textcolor{blue}{\texttt{\url{https://github.com/ideas-labo/ndp}}}.

This paper is organized as: Section~\ref{sec:pre} presents the preliminaries. Section~\ref{sec:char} delineates the known and newly discovered characteristics of CPBugs, which derive our designs. Section~\ref{sec:method} illustrates \approach. Section~\ref{sec:exp} presents the experiment setup followed by the results in Section~\ref{sec:results}. Sections~\ref{sec:dis},~\ref{sec:tov} and~\ref{sec:related} present the discussion, threats to validity, and related work, respectively. Section~\ref{sec:con} concludes the paper.

\section{Preliminaries}
\label{sec:pre}

\subsection{CPBugs in Configurable Systems}

\revision{In general, CPBugs naturally incur from the mismatch between the expected performance (as specified in the documentation) and the actual performance observed by changing a configuration option.} Therefore, measuring the performance deviation between the source and target option value serves as a strong oracle for identifying the CPBugs. In particular, the performance deviations that cause the CPBugs can be mainly observed from a common scenario: \textit{the direction of expected and actual performance changes is different, e.g., the expectation is performance raises but the actual effect is a performance drop\footnote{We follow the de facto standard that a drop is significant only when the change is greater than 5\% on any concerned performance attribute~\cite{DBLP:conf/sigsoft/ChenCWYHM23,DBLP:conf/icse/MuhlbauerSKDAS23}.} under at least one workload,} implying defects in the code segments of the configuration option.

%, leading to completely unanticipated performance changes.

%Here, the expected direction of performance change deviates from the expected change direction,

% \textit{\textbf{Scenario 2:} The direction of expected and actual performance changes is the same, but the magnitude of the changes mismatches with the expectations.} This requires a series of thresholds to indicate abnormal performance change magnitudes. In this work, we follow the methods and values from []:
% \begin{equation}
%     f (\mathbf{c_{s}},w_i) / f (\mathbf{c_{t}},w_i) > t_1
% \end{equation}
% \begin{equation}
%     f (\mathbf{c_{s}},w_i) - f (\mathbf{c_{t}},w_i) > t_1
% \end{equation}
% This means that, under a workload $w_i$, configuration with the source value of an option $c_{s}$ is better than configuration with a target value of the same option $c_{t}$ by more than $t_1$ times while the absolute drop from $c_{s}$ to $c_{t}$ is greater than $t_2$. According to
% [], setting $t_1=3$ and $t_2=5$ (seconds), which have been shown to be the most effective setting for revealing CPBugs.

\subsection{CPBugs Types}

We follow the categorization of the CPBug types proposed by He et al.~\cite{DBLP:conf/kbse/HeJLXYYWL20}, as articulated in Table~\ref{tb:cpbugs-type} and below: 

%Specifically, these types can be discussed as the following:

%Adjusting the configuration values to relax constraints is expected to enhance performance.

\begin{itemize}
\item\textbf{Optimization Switch:} When enabled, an optimization strategy is activated, and the performance is expected to improve. Yet, a performance drop implies a likely CPBug. 
%which would have a significant impact on the CPBug testing. that are used to maintain consistency and correctness
\item\textbf{Non-functional Trade-off:} Configuration options are used to balance the performance and other non-functional needs, such as the ACID properties. Whether the option needs to be increased or decreased is case-dependent. This involves two subtypes (see \textit{Type-2} in Table~\ref{tb:cpbugs-type}). In this work, we do not distinguish these two as the threshold for the ``drop beyond expectation'' is highly subjective. Instead, for an option under this type, there is a CPBug as long as a performance drop is observed\footnote{\revision{Distinguishing the subtypes does not affect the testing designs of the approach. This is because, if an option is classified into either subtype, then what we seek to find during testing is mainly whether there is an actual performance drop for the two subtypes regardless of the expected performance. As such, a CPBug can be found as long as we see a performance drop between configurations with the changes in the option’s value.}}. 
\item\textbf{Resource Allocation:} Options influence resource allocation; more resources are expected to boost performance.
\item\textbf{Functional Switch:} Options control non-performance functionalities but indirectly affect the system's performance. When an option disables a function, system performance usually improves. This also involves two situations (see \textit{Type-4} in Table~\ref{tb:cpbugs-type}). We do not distinguish those two cases due to the same aforementioned reason.
\item\textbf{Non-influential Option:} These options should not affect the system's performance, i.e., performance is expected to remain unchanged after adjusting the values.
\end{itemize}

Each pattern in the above CPBug types determines the oracle of identifying whether there is a CPBug. For example, with \textit{Type-3}, we can interpret it as \textit{``if a resource allocation related option is changed from a small value to a larger value, then we expect the performance to be improved or otherwise there is a CPBug''}. Therefore, it is essential to estimate which CPBug type is the most relevant to a tested option.

%Therefore, it is essential to estimate beforehand which CPBug type an option being tested is most likely to be associated with.

\subsection{Testing CPBugs}

In essence, CPBug testing aims to generate diverse test cases, represented as a pair of configurations (a source $\mathbf{c}_s$ and a target $\mathbf{c}_t$), which differ only on the configuration option to be tested, i.e., $o$ at index 1 (highlighted in red):
\begin{equation}
   \begin{split}
       &\mathbf{c}_s = \{0,\textcolor{red}{10},{15},43,{1}\}\\
       &\mathbf{c}_t = \{0,\textcolor{red}{20},{15},43,{1}\}
   \end{split}  
\end{equation}
An automated CPBug testing tool would perturb the values of the option $o$, such that the performance deviation from the source configuration to the target one under at least one workload matches with a CPBug type from Table~\ref{tb:cpbugs-type}, which serves as the oracle. For example, if changing from $\mathbf{c}_s$ to $\mathbf{c}_t$ (from a smaller value of a resource option to a larger value) under a workload causes a performance drop while in the documentation it should have been a rise, then we find a CPBug of \textit{Type-3}. Yet, due to the large number of options and their values, testing CPBugs is extremely expensive.

%extremely time-consuming, which is a key challenge we address in this work.

\section{Characteristics of CPBugs}
\label{sec:char}

CPBugs naturally come with certain characteristics that can help us design more effective testing. From the systems/versions/options tested in this work, which are taken from a prior study~\cite{DBLP:conf/kbse/HeJLXYYWL20}   \color{black}{and it is worth noting that the number of total options per system we tested has already exceeded what is considered for them.} We have discovered in Table~\ref{tb:cpbugs}\footnote{\revision{Due to the scale of systems and limited resources, for each system, we conducted a preliminary study to identify the most recent designed/discussed options for actual testing instead of examining all options.}} that:

\begin{displayquote}
\textit{\textbf{Characteristic 1:} Overall, only 13.39\% of the configuration options can trigger CPBugs.}
\end{displayquote}

This means that, although CPBugs can be devastating, testing on all unique options to find them is not cost-effective. The numeric configuration options\footnote{Note that the numeric options are often discretized by a set of values, e.g., \texttt{cache\_size} can be 8M, 16M, 32M, etc.} also have known characteristics. For example, He et al.~\cite{DBLP:conf/kbse/HeJLXYYWL20} have shown that:

\begin{displayquote}
\textit{\textbf{Characteristic 2:} The majority of the numeric configuration options studied can trigger CPBugs when they are changed to near one extreme of their values.}
\end{displayquote}

From Table~\ref{tb:cpbugs-numeric}, we have identified a similar pattern from the systems tested: when fixing the source of an option as close to either its maximal or minimal values, all the 18 numerical options can trigger CPBugs when the option in the target is changed to 10\% of the values at the opposed extreme. Further to the above, we have additionally found that:

\begin{displayquote}
\textit{\textbf{Characteristic 3:} 66.67\% (12 out of 18) of the numeric configuration options can trigger CPBugs when they are changed to their middle values.}
\end{displayquote}

\begin{table}[t!]
\centering

\caption{Percentage of unique options that cause CPBugs.}

\label{tb:cpbugs}
\adjustbox{max width=\columnwidth}{
\begin{tabular}{lllll}
 \toprule

\textbf{System}  & \textbf{All Options (Unique)}  & \textbf{CPBugs-related (Unique)} &\%  \\

\midrule

\textsc{MySQL}&139&24&17.27\%\\
\textsc{MariaDB}&127&9&7.09\%\\
\textsc{Apache}&121&6&4.96\%\\
\textsc{Gcc}&38&16&42.11\%\\
\textsc{Clang}&38&7&18.42\%\\

\midrule

Total&  463&62&13.39\%\\

\bottomrule
\end{tabular}
}
\vspace{-0.2cm}
\end{table}

 \begin{table}[t!]
\centering

\caption{Ranges in the numeric options in the target that cause CPBugs when fixing the source of an option as close to either its maximal or minimal values.}

\label{tb:cpbugs-numeric}
\adjustbox{max width=\columnwidth}{
\begin{tabular}{ll|ll}
 \toprule

\textbf{CPBug} & \textbf{Range}  & \textbf{CPBug} & \textbf{Range}  \\

\midrule
MySQL-21727&0\%---1.60\%&MySQL-38511&0\%---0.01\%\\
MySQL-44723&0\%---100\%&MySQL-47529&0\%---0.01\%\\
MySQL-51325&0\%---100\%&MySQL-60074&0\%---100\%\\
MySQL-62478&98.44\%---100\%&MySQL-74325&90\%---100\%\\
MySQL-78262&0\%---100\%&MySQL-80784&5.82E-9\%---100\%\\
MariaDB-145&0\%---0.01\%&MariaDB-8696&3.13\%---100\%\\
MariaDB-12556&0.10\%---100\%&MariaDB-13328&0\%---100\%\\
MariaDB-16283&0.80\%---100\%&Apache-48215&0\%---0.10\%\\
Apache-50002&10\%---100\%&Apache-54852&1.56\%---100\%\\

% MariaDB-15016&0\%---100\%
\bottomrule
\end{tabular}
}
\vspace{-0.2cm}
\end{table}

Table~\ref{tb:cpbugs-numeric} shows that a considerable proportion of the numeric options can trigger CPBugs when their values are changed to be within the middle 80\% and the opposed extreme. The above is because, while most values of numeric options impact the data flow, they usually do not change the control flow. That is, changes in the numeric configuration options need to hit certain critical values that trigger new execution paths, hence more likely to reveal CPBugs.

\revision{To further understand the CPBugs-related numeric options, we manually analyze their source code. Inspired from a recent work~\cite{DBLP:conf/sigsoft/ChenCWYHM23}, we classify those options based on the main type of performance sensitive operations that they can control, i.e., operations related to buffer, memory, network, thread or loops (exclude the other types). From Table~\ref{tb:n-type}, we found that:}

\begin{table}[t!]
\centering

\caption{\revision{Number of CPBugs triggered under extreme and middle value of numeric options based on their categorization.}}

\label{tb:n-type}

\adjustbox{max width=\columnwidth}{
 \begin{threeparttable}
\begin{tabular}{lll}
 \toprule

\textbf{Category of Numeric Option}  & \textbf{Trigger at Extreme}  & \textbf{Trigger at Middle} \\

\midrule

Buffer&12&8\\
Memory&2&1\\
Network&1&1\\
Thread&1&1\\
Loop&2&1\\

\bottomrule
\end{tabular}
 \begin{tablenotes}
      \footnotesize
      \item Different values of an option on the same version might trigger the same CPBug.
    \end{tablenotes}
  \end{threeparttable}
}
\vspace{-0.2cm}
\end{table}

\begin{displayquote}
\revision{\textit{\textbf{Characteristic 4:} Numeric options can trigger CPBugs at extreme and middle values most commonly when they control operations related to buffer.}}
\end{displayquote}

\revision{This is because the buffer operations often cause immediate implication to many parts in the system, hence the corresponding numeric options are highly performance sensitive.}

\revision{While \textbf{\textit{Characteristic 1-2}} have been known, \textbf{\textit{Characteristic 3-4}} are newly discovered information for CPBugs in this work.}

%This pattern indicates that although conventional values of numerical options have limited impact on control flow, they can significantly affect software behavior at certain critical values. These critical values may touch upon sensitive conditions or boundary situations in software design, thereby altering control flow.

\section{Dually Prioritized CPBug Testing with Neural Language Model}
\label{sec:method}

\begin{figure}[t!]
\centering
\includegraphics[width=\columnwidth]{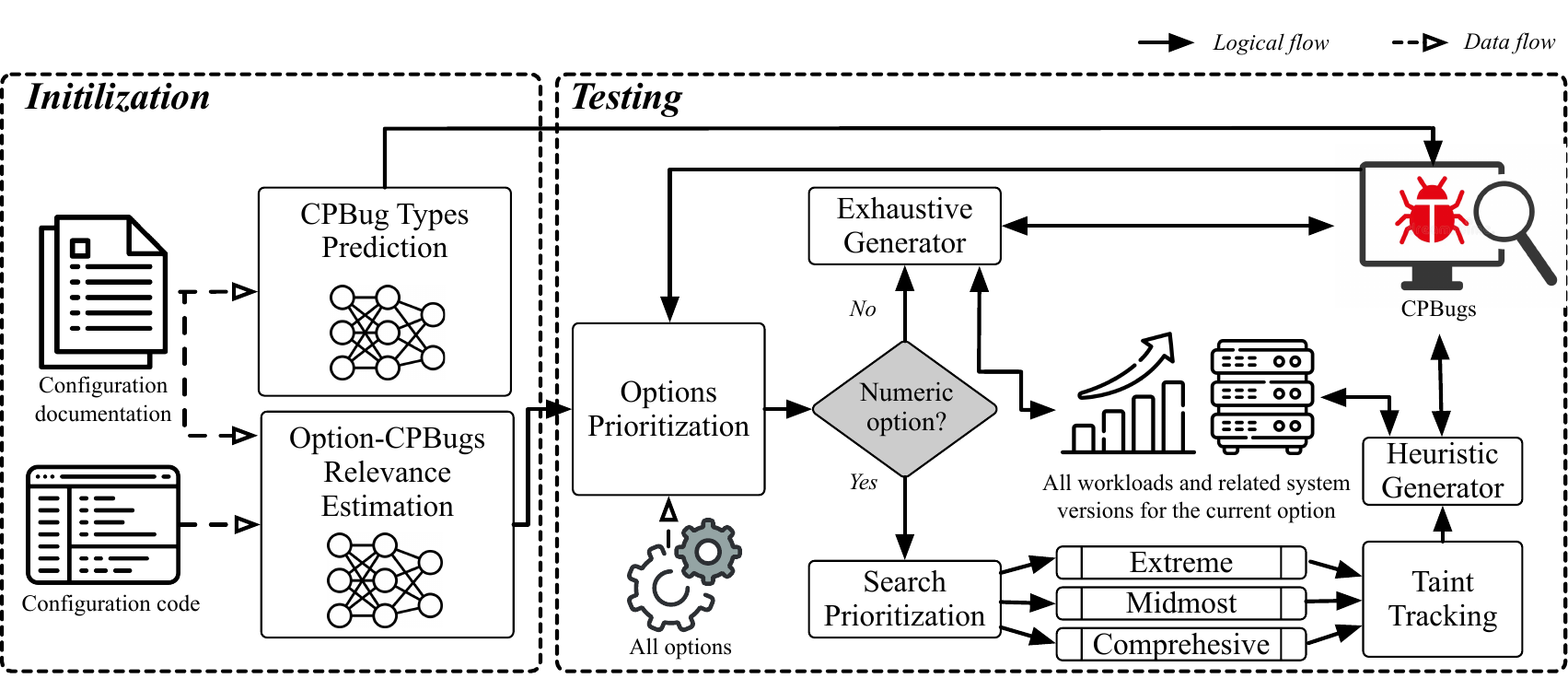}
\caption{Workflow overview of \approach~for CPBug testing.}
\label{fig:overview}
\end{figure}

\approach~seeks to expedite the CPBug testing via dual-prioritization at two levels: the option level that determines which option to test earlier and the search space level of the numeric options, which sets the order of search space to explore. This is supported by two fine-tuned neural language models, one for estimating the probability of an option being CPBug-related and the other for predicting the most relevant CPBug type that determines the test oracle. Similar to the other tools~\cite{DBLP:conf/kbse/HeJLXYYWL20}, \approach~tests the system's configuration option one by one, in which all the combinations of workloads and related versions are also examined in turn. Specifically, we design the two phases in \approach, as shown in Figure~\ref{fig:overview}. For \textit{Initialization}, \approach~focuses on a one-off process that fine-tunes two neural language models using existing data from different systems:

\begin{itemize}
    \item \textbf{CPBug Types Prediction:} Here, the goal is for a neural language model to parse the documentation and predict which CPBug type is most relevant to an option. This then serves as the essential oracle for CPBug testing.

    \item \textbf{Option-CPBugs Relevance Estimation:} We fine-tune another neural language model that takes both the description of options from the documentation and the related code snippet as inputs and estimates the probabilities of those options being CPBug-related. This allows us to handle the identified characteristics of CPBugs.
\end{itemize}

In the \textit{Testing} phase, \approach~contains four components:

\begin{itemize}

    \item \textbf{Options Prioritization (high-level):} The probabilities of whether the options are CPBug-related are used  to prioritize their testing order (due to \textbf{\textit{Characteristic 1}}).

    \item \textbf{Exhaustive Generator:} This is mainly for non-numeric options in which all the possible pairs will be covered under all workloads and related versions considered.

    \item \textbf{Search Prioritization (low-level):} For numeric options, the actual testing would also need to be conducted via a certain search depth. In \approach, we design three search depths, which are prioritized differently depending on the commonality of the probabilities for being CPBug-related. Each of the search depths would trigger an independent run of the heuristic generator, which can be any search algorithm, to generate the test cases. This fits with \textbf{\textit{Characteristic 2}} and \textbf{\textit{Characteristic 3}}. \revision{According to taint analysis of code semantic, buffer-related numeric options are specifically handled given \textbf{\textit{Characteristic 4}}.}

    \item \textbf{Heuristic Generator:} A stochastic search algorithm that samples the values of numeric options in CPBug testing under all workloads and related versions. 
    %\approach~is designed to be compatible with different algorithms.
\end{itemize}

For each option under a version, if its value alteration and the performance change (on any concerned performance attribute) in both configurations of a pair match with the pattern in the predicted CPBug type (which serves as the oracle) under at least one workload, then we found a CPBug; otherwise, we stop testing the option for a version when all pairs/workloads have been explored or a budget has been exhausted. In all cases, \approach~then switches to test the option under the next related version, if any. That is, we only need to find the CPBugs caused by the option on at least one workload under a version but all the related versions would be examined regardless of how many CPBugs are found on the said option. When all related versions have been tested for an option, we move to the next option as prioritized by \approach.

%roberta 1 for classifiying type

% \begin{figure}[t!]
% \centering
% \includegraphics[width=\columnwidth]{fig/eg.fg2.pdf}
% \caption{An example of the texts for an option in the documentation. \textcolor{red}{Youpeng, find this example}}
% \label{fig:doc}
% \end{figure}

 \begin{table}[t!]
\centering

\caption{An exampled option's description from \textsc{MySQL}.}

\label{tb:exp1}
\adjustbox{max width=\columnwidth}{
\begin{tabular}{|p{8.5cm}|}
 \toprule

\textbf{Option:} \verb|innodb_flush_log_at_trx_commit| \\
\midrule
\textbf{Documentation (partial):} The \verb|innodb_flush_log_at_trx_commit| controls the balance between strict {ACID} compliance for commit operations and higher performance that is possible when commit-related I/O operations are rearranged and done in batches. You can achieve better performance by changing the default value......\\
%but then you can lose transactions in a crash

\bottomrule
\end{tabular}
}
\vspace{-0.2cm}
\end{table}

\subsection{Neural CPBug Types Inference}

An essential task in CPBug testing is to determine the oracle, i.e., identifying what types of CPbugs an option is most likely to be associated with, hence triggering the corresponding way to verify whether such an option can trigger CPBugs. To that end, we leverage a single-modal RoBERTa (denoted $\mathcal{M}_s$),  a particular type of neural language model, to predict the CPBug type of a given option to be tested. In \approach, we fine-tune the RoBERTa using the data collected from previous work~\cite{DBLP:conf/kbse/HeJLXYYWL20}, such that the inputs are the natural description of an option from the documentation with a known label from the five CPBug type or a label of ``no CPBug'', \revision{since CPBugs are mainly related to the deviation from the expected performance of an option specified in the documentation} (e.g., Table~\ref{tb:exp1}). Notably, the fine-tuning process is naturally cross-project since the naturalness of the documents ensures its generalization.

% that has been shown to be robust for many software engineering problems~\cite{DBLP:conf/icsm/ZhangXTH0J20}

In particular, RoBERTa is chosen for three reasons:

\begin{itemize}
  \item Compared with the rule mining approach~\cite{DBLP:conf/kbse/HeJLXYYWL20}, it exhibits a stronger generalization ability that can learn hidden information in the documentation in the latent space. In Section~\ref{sec:rq1}, we will experimentally verify this.
  \item In contrast to LLM, RoBERTa fits our problem better: we need a classification of CPBugs type rather than text generation. Further, RoBERTa is more cost-effective~\cite{DBLP:journals/nlpj/RoumeliotisTN24}.
  \item It has been reported that RoBERTa is the generally most promising BERT variant for software engineering~\cite{DBLP:conf/icsm/ZhangXTH0J20}.

\end{itemize}

%The CPBugs data are collected from popular tracking systems (e.g., JIRA and Bugzilla).

%roberta 2 for ranking level 1

\subsection{Neural Multi-Modal Option-CPBugs Relevance Estimation}

From \textbf{\textit{Characteristic 1}}, we note that only a small number of options can potentially be the cause of CPBugs. As a result, a natural idea is to estimate which options are more related to CPBugs. \approach~leverage both documentation and the corresponding code of an option (see Table~\ref{tb:exp2}), together with the corresponding label of whether the option is CPBug related, to fine-tune another multi-modal RoBERTa model, $\mathcal{M}_m$, in a binary classification problem. Our goal here, however, is not to use the model to make a binary prediction but to extract its probability related to the likelihood of an option being CPBug-related, based on which we can rank those options. As such, forming this binary classification to train yet another new RoBERTa model has the benefit of simplicity without producing much noise to fulfill our goal.

%This model specifically aims to produce a probability of an option belong to a CPBug type, based on which allow us to rank their likelihood of being CPBug-related in the testing. 

% \begin{figure}[t!]
% \centering
% \includegraphics[width=\columnwidth]{fig/eg.fg3.pdf}
% \caption{An example of an option's texts from the documentation and the mapped code snippet. \textcolor{red}{Youpeng, find this example}}
% \label{fig:mapping}
% \end{figure}
 \begin{table}[t!]
\centering

\caption{An example of an option’s texts from the documentation and the mapped code snippet from \textsc{MySQL}.}

\label{tb:exp2}
\adjustbox{max width=\columnwidth}{
\begin{tabular}{|p{8.5cm}|}
 \toprule

\textbf{Option:} \verb|innodb_buffer_pool_size| \\
\midrule
\textbf{Documentation (partial):} The \verb|innodb_buffer_pool_size| system variable specifies the size of the buffer pool. If your buffer pool is small and you have sufficient memory, making the pool larger can improve performance by reducing the disk I/O as queries access \texttt{InnoDB} tables......\\

\midrule
\textbf{Code Snippet Mapped (partial):}
\begin{lstlisting}[style=CStyle,numbers=none,frame=none]
if (srv_dedicated_server && sysvar_source_svc != nullptr) {
   static const char *variable_name = "innodb_buffer_pool_size";
   enum enum_variable_source source;
   if (!sysvar_source_svc->get(variable_name, static_cast<unsigned int>(strlen(variable_name)),&source)) {
      if (source == COMPILED) {
         double server_mem = get_sys_mem();
\end{lstlisting}\\
\bottomrule
\end{tabular}
}
\vspace{-0.2cm}
\end{table}

To that end, we locate the code for a corresponding option in the documentation using a pattern matching-based heuristic:

 \begin{itemize}

        \item \textbf{Direct way:} Some systems have a centralized file (or a few files) to maintain all configuration options, such as \textsc{MySQL}. In those cases, we look at the variable with a similar name to those in the documentation and identify the relevant code snippets from the centralized file(s).

        \item \textbf{Indirect way:} Other systems might not have a file(s) that share the same name as those in the documentation. \revision{We consider two cases: (1) there are mechanisms that allow access to those configuration variables via \texttt{setter()} and \texttt{getter()}. In those cases, we can write a script that searches through the relevant files and outputs the similarity of the \texttt{setter()} and \texttt{getter()} to each option in the documentation. We can then manually identify the corresponding code snippets. (2) there are no \texttt{setter()}/\texttt{getter()}, in which case we scan all variables and related functions in the related files.}

    \end{itemize}

\revision{Once the code snippets and the texts of an option are mapped, our heuristic uses the rules below to clean the code:}

 \begin{itemize}

        \item \revision{Remove useless comments, e.g., those with timestep, certain license information, etc.}

        \item \revision{Remove duplicated comments or code snippets.}

        \item \revision{Remove usage examples of the options in the comments. e.g., formats or order of changes.}

        \item \revision{Remove code snippets of incomplete functions extracted.}

    \end{itemize}

%The corresponding code snippets can then be extracted directly. 

In \approach, the texts from documentation and the code snippets of an option are concatenated together (e.g., Table~\ref{tb:exp2}) and we use standard steps such as text cleaning, tokenization, and serialization to parse the data. Those inputs, together with a label of whether the option is CPBug-related, form the data to fine-tune the RoBERTa model. To make RoBERTa work for our simplify binary classification problem, we add a task-specific classification layer with a cross-entropy loss function in the fine-tuning process. Upon predicting a given option, the probability of being a true label (CPBug-related) extracted from the \texttt{softmax} layer is what we are interested in. Again, we use the CPBugs data that has been reported previously~\cite{DBLP:conf/kbse/HeJLXYYWL20}.

\subsection{All Options Prioritization (High-level)}

In \approach, at the high level, we firstly leverage the probabilities of all the options produced by $\mathcal{M}_m$ to determine their order in CPBug testing. This is important as if an option is more likely to cause the CPBugs, then prioritizing it beforehand would help us to identify the bugs quicker. Here, although we do not distinguish the type of options in this prioritization, e.g., numeric and non-numeric ones, their actual testing strategies can be different: for non-numeric options, we generate and test all the combinatorial values of the configurations in the pair using an exhaustive generator, since often those possible values are of limited range~\cite{DBLP:conf/mascots/JamshidiC16}. Notably, all workloads and related versions are considered: we at first pick a version and test all workloads therein in turn; if a CPBug is found for a non-numeric option under a workload, then the remaining untested workloads would be skipped and we switch to the next related version. Finally, \approach~moves to the next option when all related versions have been tested for the current option.

In contrast, for numeric values, we need a stronger way to prioritize their sampling, which we will describe as follows.

% search for numeric level 2

\subsection{Search Prioritization for Numeric Options (Low-level)}

%\subsubsection{Kernel Density-based Prioritization}

\begin{figure}[t!]
\centering
\includegraphics[width=0.7\columnwidth]{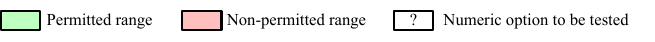}

\subfloat[Extreme]{\includegraphics[width=0.3\columnwidth]{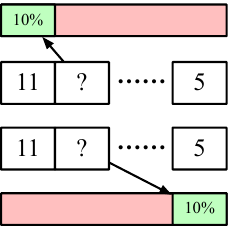}}
~\hfill
\subfloat[Midmost]{\includegraphics[width=0.3\columnwidth]{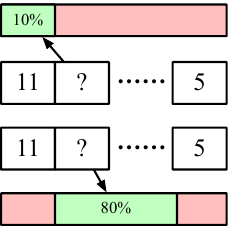}}
~\hfill
\subfloat[Comprehensive]{\includegraphics[width=0.3\columnwidth]{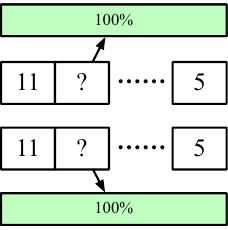}}
\caption{Illustrations of the search space bounded by different search depths for CPBug testing with \approach.}
\label{fig:search}
%\vspace{-0.4cm}
\end{figure}

When the option to be tested is a numeric option, we propose three different search depths that bound the search spaces of the underlying heuristic generator:

\begin{itemize}
    \item \textbf{Extreme search:} As in Figure~\ref{fig:search}a, this is the search with the most restricted search space: the search for test cases happens within $10\%$ of the upper/lower bounds range of values\footnote{\revision{For those options without explicitly defined upper/lower bounds, we set them using our understanding of the domain, e.g., the capacity of our hardware or the extreme values that are commonly set in practice.}} for both configurations in the pair. Yet, we ensure that the two configurations in the pair are searched over the opposed bounds (\textbf{\textit{Characteristic 2}}).
     \item \textbf{Midmost search:} Here, in Figure~\ref{fig:search}b, one configuration in the pair is searched within $10\%$ of the upper or lower bound range values while the other can be changed within the middle $80\%$ of the values (\textbf{\textit{Characteristic 3}}).
      \item \textbf{Comprehensive search:} This is the search that basically means all possible values of the configurations in the pair can be explored (Figure~\ref{fig:search}c).
\end{itemize}

Those search depths differ in terms of the number of tests (for the pairs) required. According to the probabilities produced from $\mathcal{M}_m$ for the numeric options, we can then prioritize how the above three search depths are used on them. In \approach, the idea is to divide the probabilities of all numeric options into three divisions, based on which different prioritization of the search depth is used. To systematically perform such a division, we leverage the Gaussian Kernel Density Estimation (GKDE). In essence, GKDE serves as a one-dimensional binning algorithm that divides the probability density of options being CPBug-related into three bins with no thresholds. This is achieved by dividing the options into the top three peaks\footnote{Given the complexity of configurable systems, we have not seen a case with less than three peaks.}, which are separated by a local trough, based on their closeness of probabilities to those peaks.

\begin{figure}[t!]
\centering
\includegraphics[width=0.5\columnwidth]{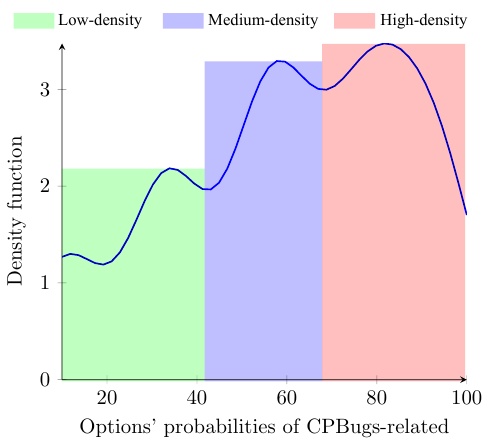}
\caption{Exampled kernel density function on the numeric options' probabilities of being CPBug-related for a system.}
\label{fig:kde}
%\vspace{-0.4cm}
\end{figure}

Figure~\ref{fig:kde} shows an example: we see that the numeric options are divided into three divisions according to their commonality on the probabilities of being CPBug-related. These divisions derive three prioritizations of the search depths:

\begin{itemize}
    \item \textbf{High-density:} For the numeric options belonging to the most frequent bin, we prioritize the more restricted search depth: starting from extreme search, midmost search, and finally comprehensive search. This will speedup testing if CPBugs can be detected within an extreme/middle value.

    \item \textbf{Medium-density:} Here, we prioritize the midmost search first, followed by the extreme search, and then the comprehensive search. The reason is that since the numeric options are of medium density, we start from the midmost search that also assumes a medium size of search space.

    \item \textbf{Low-density:} For the least frequent bin of numeric options, we adopt the {comprehensive search} only.
\end{itemize}

\revision{According to \textit{\textbf{Characteristic 4}}, numeric options that control buffers are most likely to cause CPBugs at their extreme or middle values. Hence, in \approach, we adopt taint tracking\footnote{\revision{The tracking (built on \texttt{LibASTMatchers}~\cite{llvm} for C/C++) is highly efficient: it takes a few seconds to around one minute for a system studied.}} using the option as the source while the buffer-related operations as the sink (e.g., \texttt{release\_sysvar\_source\_service()} for \textsc{MySQL}), hence analyzing the code semantic to identify whether an option controls buffer. These sinks, which are system-dependent, are domain knowledge specified by software engineers. For all buffer-related numeric options, we force their search to follow the depth order of high-density.}

%and \texttt{JavaParser}~\cite{javaparser} for JAVA

For a given numeric option, \approach~follows the steps below:

\begin{enumerate}
    \item Pick a related version of the system under test.
    \item \revision{If the numeric option controls buffer, then make it uses the order of search depths for high-density; otherwise, following the order of the assigned density level.}
    \item Test the option with the order of search depths via the heuristic generator for all workloads (see Section~\ref{sec:hg}).
    \item If a CPBug is found, then jump to \textcolor{black}{5}); otherwise, return to 2) and move to the next search depth.
    \item Repeat from 1) for the next related version, if any; otherwise, move to the next option.
\end{enumerate}

%when either the budget $B$ of the current run of heuristic generator has been exhausted (we set $B=100$ tests) or all possible pairs in the bound has have been explored

%stops immediately once a CPBug has been detected according to the oracle in the predicted CPBug type. In particular, for the high- and medium-density cases, under a search depth, if the entire search space has been explored and no CPBug is found, then \approach~move to the next search depth. Regardless of whether a CPBug has been discovered, the testing of an option terminates whenever the search has reached its pre-defined budget $B$. Under each search depth, an independent budget is given to the underlying heuristic generator. In this work, we set $B=100$ tests. 

% \begin{itemize}
%     \item The entire search space has been explored. For example, in the extreme search, if both sides of the $10\%$ have $16$ combinatorial values in the pair, then the search terminates when all of them have been tested.
%     \item The search has reached its pre-defined budget $B$. Under each search depth, a budget is given which is used by the underlying heuristic tester. In this work, we set $B=100$ tests.
% \end{itemize}

\subsection{Heuristic Generator}
\label{sec:hg}

%random decide which bound and which is middle

\approach~can be paired with any heuristic algorithms for generating test cases for the numeric options. In this work, we use the population-based Genetic Algorithm (GA)~\cite{forrest1996genetic}, but it can be easily replaced by other algorithms. Since \approach~tests one numeric option under a version each time, the solution representation is a pair of values for the tested option. 

%Since the search space is set by the search depth,  

%Depending on the bounded search space, the possible number of pairs might differ, hence we only set a soft population size, i.e., it is possible to have a smaller population.

%that works on a population of the configuration pairs, performing crossover and mutation on the pairs, keeping the most elitist ones in the next generation. 

To determine which pairs to preserve, we use the fitness function below to compare the pairs: 
\begin{equation}
   fitness = \max_{i\in \{1...m\}}  |f (\mathbf{c_{1}},w_i) - f (\mathbf{c_{2}},w_i)|
\end{equation}
whereby $\mathbf{c_{1}}$ and $\mathbf{c_{2}}$ are the configurations in a pair with different values on the numeric option to be tested; $w_i$ denoted the $i$th workloads out of a total of $m$ ones. Since based on the CPBug types, either configurations in the pair can be the source and the options would trigger CPBugs if the performance drops, this fitness reflects the maximum performance deviation between the two paired configurations with different values of the tested numeric option across different workloads---the larger the deviation, the higher possibility of triggering more CPBugs, which should be preserved in testing. 

% as the performance drop in the CPBug types is relative

Under each of the above search depths, the search space of GA is bounded correspondingly. Whether a configuration in the pair starts from the upper or lower sides (for extreme search); or whether it is searched on the middle range (for midmost search) is decided randomly. When GA consumes all of its budget or all pairs within the bound have been explored, \approach~terminates the GA and checks whether a CPBug has been found according to the estimated CPBug type.

%The GA would terminate whenever a particular test case of the pair for the corresponding option has detected a relevant CPBug. To ensure efficiency, we also do not allow GA to generate redundant test cases.

%talk about constraint

\subsection{Handling Dependency}

When changing the tested options, their dependencies need to be complied~\cite{DBLP:conf/icse/LiangChen25}. For example, in \textsc{MySQL}, there is a dependency that option \verb|innodb_buffer_pool_size| (the buffer pool size) must be set as an integer product of that of the option \verb|innodb_buffer_pool_chunk_size| (the granularity of buffer pool resizing) while being greater.

%be no less than that of \verb|innodb_buffer_pool_chunk_size|, and any other values must be an integer product of \verb|innodb_buffer_pool_chunk_size|. 

\approach~leverages \texttt{GPTuner}~\cite{DBLP:journals/pvldb/LaoWLWZCCTW24}---a large language model-based tool that predicts configuration dependency based on the documentation. If, when a value of the tested option violates any dependency, we then (randomly) change the other affected option correspondingly. For example, if we change the value of \verb|innodb_buffer_pool_chunk_size| to 128MB, then we should also set the \verb|innodb_buffer_pool_size| to a value that is an integer product of 128MB, e.g., 128MB, 256MB, or 384MB. Note that, in that case, if either of the two options triggers CPBug, then both are CPBug-related with the same CPBug type. \revision{We chose \texttt{GPTuner} for two reasons:
    \begin{itemize}
    
        \item \revision{it is highly flexibility and can be conveniently used without any fine-tuning.}
        
        \item \revision{thanks to the GPT3.5, it is generalizable to different systems. This is the key advantage compared with other rule-based tools such as \texttt{cDep} \cite{DBLP:conf/sigsoft/ChenWLLX20}.}
    \end{itemize}
 }

\section{Experiments Setup}
\label{sec:exp}

%We now delineate the experiment rigs.

\subsection{Research Questions}

In this work, we answer the following research questions: 
\begin{itemize}

\item\textbf{RQ1:} How well can \approach~estimate the CPBug types?

\item\textbf{RQ2:} How effective dose \approach~in options prioritization against the state-of-the-art tools? 

\item\textbf{RQ3:} How well dose the prioritized search in \approach~perform over the state-of-the-art tools? 

\revision{\item\textbf{RQ4:} Can \approach~discover unknown CPBugs?}

%\item\textbf{RQ4:} What types of CPBugs can be better found by \approach~compared with the others? 

% \item\textbf{RQ1:} Is the method of vectorizing texts using the Transformer architecture to classify option parameter performance properties superior to the method of classifying texts using association rule mining? 
% \item\textbf{RQ2:} Is the method of vectorizing texts using the Transformer architecture to classify option parameter performance properties superior to the method of classifying texts using association rule mining? 
% \item\textbf{RQ3:} Is the performance of the new sampling method superior to the old method? 
% \item\textbf{RQ4:} Can the CPBug oracle discover performance bugs that have not yet occurred under general conditions?
\end{itemize}

%To address the above questions, we plan to validate them through multiple real-world software systems to ensure their broad applicability in detecting configuration-related performance issues. These include databases like MySql and MariaDB, network proxy software like Apache, and compilers like Gcc and Clang. These software systems have had multiple real-world CPBugs in different versions that have been fixed.

\subsection{Systems, Versions, Workloads, and Known CPBugs}

In this work, we use the datasets of 12 systems provided by He et al.~\cite{DBLP:conf/kbse/HeJLXYYWL20} for assessing the CPBug type prediction. For the actual testing, we conduct experiments on five widely used configurable systems therein with known CPBugs, as shown in Table~\ref{tb:software}. The reason is that we have not been able to reproduce the CPBugs for all 12 systems used by He et al.~\cite{DBLP:conf/kbse/HeJLXYYWL20} because, e.g., the related versions are discarded; or the CPBugs have not been documented clearly. Yet, the five systems used for testing are still of diverse domains and scales, including database systems (i.e., \textsc{MySQL} and \textsc{MariaDB}), web servers (i.e., \textsc{Apache}), and compilers (i.e., \textsc{Gcc} and \textsc{Clang}). 

To reproduce the CPBugs in testing, we test the options of each system under various versions. Note that not all the options would go through the same versions, since some do not exist in certain versions, hence \approach~maintains a mapping between the options and related versions. \revision{We selected the related versions that can run successfully, including those that can produce the CPBugs in the ground truth and used previously~\cite{DBLP:conf/kbse/HeJLXYYWL20}; as well as other stable versions that have not been discarded. In practice, we believe that software engineers would come with some domain knowledge about which versions are more likely to have CPBugs, or use all deployable versions that are of interest. Therefore, the selection of versions is case-dependent. \approach~does not make assumptions on the nature and number of versions to be tested.}

%In reality, software engineers can also customize such a mapping based on domain knowledge. 

%Therefore the total number of system instances evaluated is 70. 

%(from 4 to 31 versions)

%4+51+4+40+19

%The selected systems have also been studied previously []. Indeed, [] have reported data related to other systems but we have not been able to replicate the CPBugs for those systems because, e.g., the associated versions are discarded as no longer workable; the information about the CPBug is highly vague; or the CPBug has not been documented clearly.

Each system is tested under different workloads generated by standard benchmarks. For \textsc{MySQL} and \textsc{MariaDB}, we use \textsc{Sysbench}---a powerful multi-threaded benchmark that is frequently employed~\cite{DBLP:conf/icse/MuhlbauerSKDAS23,DBLP:conf/sigsoft/SiegmundGAK15}---to generate 10 workloads that are of various data scales, number of concurrent threads, and test duration. For \textsc{Apache}, we use \textsc{Apachebench} to create 6 workloads of different types. For \textsc{Gcc} and \textsc{Clang}, we use 2 standard programs with different types and scales. All above are important for revealing the CPBugs and have been used in prior work~\cite{DBLP:conf/icse/MuhlbauerSKDAS23,DBLP:conf/sigsoft/SiegmundGAK15,DBLP:conf/icse/SiegmundKKABRS12}. The workloads, combined with the versions, led to a high number of cases in the CPBug testing.

%\textcolor{red}{Youpeng, list the workloads for each system.}(\texttt{Sysbench} is a powerful multi-threaded benchmarking tool that we frequently employ to uncover  CPBugs in \textsc{MySQL} and \textsc{MariaDB}. The \verb|OLTP| test options provided by \texttt{Sysbench} enable the simulation of a variety of workload combinations. By adjusting key parameters such as the scale of the test database, the number of concurrent threads, and the duration of the test, we can construct diverse workloads that reveal performance disparities. For \textsc{Apache}, we similarly use the benchmarking tool \texttt{Apachebench} for performance testing. In the case of \textsc{Gcc} and \textsc{Clang} compilers, the differences in workloads are primarily determined by the type, scale, and complexity of the code being compiled. For test cases that can expose bugs, we increase their scale and complexity without altering the coding type and testing logic. For certain \textsc{CPBugs} in \textsc{MySQL}, \textsc{MariaDB}, and \textsc{Apache}, we adopt a similar approach.)

Derived from prior work~\cite{DBLP:conf/kbse/HeJLXYYWL20}, Table~\ref{tb:software} shows that all systems/versions studied contain various known CPBugs, which are sufficiently complex to challenge CPBug testing tools.

%To verify the generality and effectiveness of our proposed method for the aforementioned issues, we plan to test it on different versions of several large-scale open-source software systems in the real world to ensure its wide applicability in detecting configuration-related performance issues. We selected different categories of software, including database systems (e.g., MySql and MariaDB), network proxy systems (e.g., Apache), and compiler systems (e.g., Gcc and Clang). These software systems have various versions with known, and possibly fixed, types of CPBugs. We successfully reproduced these representative bugs, as shown in Table []. They cover different performance attributes and value types.
%介绍如何选择系统，以及他们的cpbugs数目

\subsection{Compared Approaches}

Our experiments make comparisons with respect to the following state-of-the-art CPBug testing/prediction approaches:

\begin{itemize}
    \item \textbf{\texttt{CP-Detector (CPD)}~\cite{DBLP:conf/kbse/HeJLXYYWL20}}: A state-of-the-art tool that uses rule mining and keyword search to estimate the CPBug types. During the actual testing, it follows the greedy method with a random order of the tested options: fixing each tested option of a configuration in the pair at its minimal value and exponentially increasing the value of the same option for the other configuration.
    \item \textbf{\texttt{Keyword Searching (KS)}}: A baseline that predicts CPBugs type by keyword matching in the documentation.
    \item \textbf{\texttt{Uniform Sampling (US)}~\cite{DBLP:conf/icst/PlazarAPDC19}}: A tool that samples uniformly on the values of the option in a pair with randomly sorted options to be tested. We set the same budget for each option as the GA in \approach, i.e., 100 tests, and the same way as \approach~for CPBug type prediction.
\end{itemize}

\begin{table}[t!]
\centering

\setlength{\tabcolsep}{1mm}
\caption{Configurable software with reproduced CPBugs.}

\label{tb:software}
\adjustbox{max width=\columnwidth}{
 \begin{threeparttable}
\begin{tabular}{lll|llllll|ll}
 \toprule

\textbf{Software} & \textbf{Version} & $\mathcal{W}$& \textbf{$\#$ CPBugs} & \textbf{Type-1}& \textbf{Type-2}& \textbf{Type-3}& \textbf{Type-4}& \textbf{Type-5}& \textbf{$^\neg\mathcal{N}$}& \textbf{$\mathcal{N}$}  \\

\midrule

\textsc{MySQL} & 5.0 - 8.0 & 10 & 30 & 8&7&8&6&1&17&7\\
\textsc{MariaDB} & 5.3 - 10.3 & 10 & 10 & 3&0&4&2&1&5&4\\
\textsc{Apache} & 2.2 - 2.4& 6 &  5 &0&2&1&1&1&3&3\\
\textsc{Gcc} & 3.4 - 7.3& 2 & 15 &1&10&0&2&2&16&0\\
\textsc{Clang} & 3.2 - 5.0& 2 & 6 &0&6&0&0&0&7&0\\

% \textsc{MySQL} & 30 & Type-1=8; Type-2=6; Type-3=8; Type-4=6; Type-5=1\\
% \textsc{MariaDB} & 10 & Type-1=3; Type-2=0; Type-3=4; Type-4=2; Type-5=1\\
% \textsc{Apache} & 5 & Type-1=0; Type-2=1; Type-3=1; Type-4=1; Type-5=1\\
% \textsc{Gcc} & 15 & Type-1=1; Type-2=10; Type-3=0; Type-4=2; Type-5=2\\
% \textsc{Clang} & 6 & Type-1=0; Type-2=6; Type-3=0; Type-4=0; Type-5=0  \\

 \bottomrule
\end{tabular}
 \begin{tablenotes}
      \footnotesize
      \item An option might trigger multiple CPBugs, e.g., different values across different versions; Two options might also lead to a CPBug due to dependency. $^\neg\mathcal{N}$ and $\mathcal{N}$ count the number of non-numeric and numeric CPBug-related options, respectively. $\mathcal{W}$ counts the number of workloads.
    \end{tablenotes}
  \end{threeparttable}
}
    \vspace{-0.2cm}
%\vspace{-0.2cm}
\end{table}

Like \approach, when testing an option under a version, \texttt{CPD} and \texttt{US} also consider all workloads; stop whenever a CPBug is found or cover all pairs/exhaust the budget, then move to the next related version/option. We have omitted some other tools, e.g., \texttt{Toddler}~\cite{DBLP:conf/icse/NistorSML13}, as they have been shown to be significantly inferior to \texttt{CPD}~\cite{DBLP:conf/kbse/HeJLXYYWL20}. For \textbf{RQ1}, we use \texttt{CPD} and \texttt{KS}; while both \texttt{CPD} and \texttt{US} are used for the remaining RQs.

\begin{table*}[t!]
\caption{Effectiveness of estimating CPBugs types. \setlength{\fboxsep}{1.5pt}\colorbox{red!20}{red cells} denote the best-performing approach.}
\label{tb:rq1}
\centering
\footnotesize
\begin{tabular}{llllll|lll|lll}

\toprule

\multirow{2}{*}{\textbf{CPBug Type}}&\multirow{2}{*}{\textbf{Option Purpose}}&\multirow{2}{*}{\textbf{$\#$ Sample Options}}&\multicolumn{3}{c|}{\textbf{Precision}}&\multicolumn{3}{c|}{\textbf{Recall}}&\multicolumn{3}{c}{\textbf{F1 Score}}\\

\cmidrule{4-12}

&&&\approach&\texttt{CPD}&\texttt{KS}&\approach&\texttt{CPD}&\texttt{KS}&\approach&\texttt{CPD}&\texttt{KS}\\

\midrule

Type-1&Optimization& 73&\cellcolor{red!20}73.2\%&69.4\%&27.3\%&\cellcolor{red!20}71.2\%&65.8\%&18.8\%&\cellcolor{red!20}0.72&0.68&0.22\\

Type-2&Tradeoff& 84&\cellcolor{red!20}85.9\%&70.7\%&61.3\%&\cellcolor{red!20}72.6\%&66.9\%&21.4\%&\cellcolor{red!20}0.79&0.69&0.32\\

Type-3&Resource& 143&92.2\%&\cellcolor{red!20}93.9\%&69.4\%&\cellcolor{red!20}99.3\%&92.4\%&93.2\%&\cellcolor{red!20}0.96&0.93&0.80\\

Type-4&Functionality& 100&78.4\%&\cellcolor{red!20}82.2\%&56.2\%&\cellcolor{red!20}80.0\%&55.6\%&39.1\%&\cellcolor{red!20}0.79&0.66&0.46\\

Type-5&Non-influence& 100&\cellcolor{red!20}91.1\%&90.1\%&35.0\%&\cellcolor{red!20}93.0\%&67.1\%&70.0\%&\cellcolor{red!20}0.92&0.77&0.47\\

%\textsc{Trimesh}&Mesh&\textsc{O1:} \# Iteration; \textsc{O2:} Latency&13&239,260\\

%&\makecell[l]{Latency\\Throughput}&3&1,343&\makecell[l]{Apache Storm under the Word \\Count on OpenNebula(1 CPU)}\\

%1.92$\times 10^{10}$

\bottomrule
\end{tabular}
\vspace{-0.2cm}
%}
%\centering
% \begin{tablenotes}
%    \footnotesize
%    %\item The configuration options are preprocessed and thus only the ones (and their ranges) that can affect the performance attributes are considered. 
%    \item $\lvert \mathbfcal{O} \rvert$ denotes number of options. We run all systems under their standard benchmarks. More details can be found at: \href{https://github.com/taochen/mmo-fse-2021}{\texttt{\textcolor{blue}{https://github.com/taochen/mmo-fse-2021}}}.
%    \end{tablenotes}
    % Notably, \textsc{Storm} and \textsc{Keras-DNN} have three benchmarks and two dataset, respectively.
\end{table*}

\revision{Note that, indeed, various sampling methods exist for testing other configuration issues~\cite{DBLP:conf/icse/MedeirosKRGA16}; however, CPBug testing differs from those as it considers different versions and workloads, making it too expensive to be tested by current sampling approaches (which more or less favor diversity).} 

%Further, \texttt{CPD} is a state-of-the-art approach for detecting CPBugs and hence serves as a representative to be compared.}

%In previous studies, the detection of CPBugs was often limited by randomness. This is mainly due to the lack of orderly control over the testing process and precise adjustment of the sampling range. Such randomness makes it difficult to discover a sufficient number of bugs in a short period when detecting known or unknown CPBugs, and detection failures may occur due to misjudgment of performance attributes. To address this issue, our testing method seeks to accelerate CPBugs testing at two levels of priority through NDLP. This method can reveal more bugs faster and reduce detection failures due to performance attribute misjudgments to some extent. This is because we have more opportunities to observe the occurrence of two opposite performance trends during testing, which is usually a strong signal that the configuration option is likely to have a CPBug.
%介绍对比的cpbug测试方法

\subsection{Testing Budgets and Other Settings}

In \approach, when testing numeric options, we set a budget of 100 tests for the heuristic generator (GA in this work). This means that, under a search depth, if the heuristic generator has consumed 100 tests (i.e., testing a pair under a workload for a system version would consume one test), then \approach~would stop testing for the corresponding option. This is the same budget for \texttt{US} but not for \texttt{CPD} since it leverages a greedy search method. In contrast, the testing of non-numeric options is always exhaustive. All other settings of the compared approaches are left as default specified in their work. 

\revision{For setting the GA, we use a mutation and crossover rate of 0.1 and 0.9, respectively, together with a soft population size capped at 10 (i.e., the number of pairs to explore might be less than 10 on the more restricted search depth). The crossover operator is a uniform crossover, i.e., one of the tested option's values in a pair might be swapped with that of the other; the mutation is a random mutation that randomly changes the tested option's value to a different permissible value. All above are standard settings from prior studies~\cite{DBLP:journals/pacmse/0001L24,DBLP:journals/tse/ChenCL24,DBLP:journals/tosem/ChenL23}. As for GKDE, we set all parameters as their default values.}

For training the two neural language models in \approach, we use all the CPBugs data (including CPBug types) that have been previously reported~\cite{DBLP:conf/kbse/HeJLXYYWL20} from different systems, except the system (and its versions) under test; unless otherwise stated. This is the same setup for the rule mining process in \texttt{CPD}~\cite{DBLP:conf/kbse/HeJLXYYWL20}.

\section{Experimental Evaluation}
\label{sec:results}

\subsection{RQ1: CPBugs Type Estimation}
\label{sec:rq1}

\subsubsection{Method}

\revision{To examine oracle prediction via \textbf{RQ1}, we compare \approach~with \texttt{CPD} and \texttt{KS} under the same 500 samples from 12 systems used by He et al.~\cite{DBLP:conf/kbse/HeJLXYYWL20} with no sampling method change, following the same training (fine-tuning)/testing splits for 10-fold cross-validation. He et al.~\cite{DBLP:conf/kbse/HeJLXYYWL20} state that those are randomly sampled from the systems, but they have ensured data quality and representative nature.} The mean recall, precision, and F1 scores for each of the five CPBug types are reported, i.e., a total of 15 types/metrics.

\subsubsection{Results}

From Table~\ref{tb:rq1}, we clearly see that the neural language model in \approach~achieves considerably better results than the others, particularly on the F1 score, leading to superior results on 13 out of 15 types/metrics. In particular, the \texttt{KS} is clearly insufficient due to the limitation of a human-defined keywords sets; \texttt{CPD} is also restricted by the rule mining capability: due to the naturalness, the vast ways of describing the potential CPBugs in the documentation cannot be fully captured by the rules identified. \revision{Indeed, \texttt{CPD} marginally performs better than \approach~on the precision of Type-3 and Type-4. However, this is mainly due to the fact that \texttt{CPD} tends not to include the samples in those two types, as naturally, their descriptions can be more complex. This has led to better precision (better false positive) but serenely comprised recall (worse false negative), which, together have worsened the F1 score in general.} \approach, in contrast, can handle complex cases with good performance over both false positives/negatives, thanks to the reasoning ability in the latent space provided by the neural language model RoBERTa. Overall, we say:

%This has resulted in an acceptable number of false positives (precision) but caused a high number of false negatives (recall) in general. 

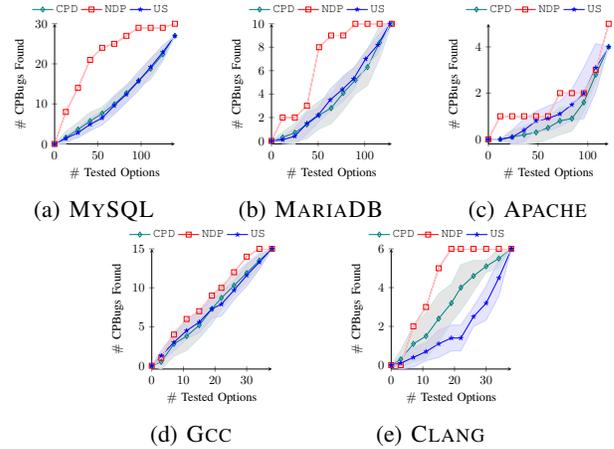
\begin{figure}[t!]
    \centering
    \begin{subfigure}[t!]{0.25\columnwidth}
        \centering
           \begin{tikzpicture}
  \begin{axis}[
  width = 5cm,
  height = 5cm,
  %xtick={1,2,3,4,5},
 % xticklabels={20\%,40\%,60\%,80\%,100\%},
  % xlabel=$p$,
  ylabel={$\#$ CPBugs Found},
  xlabel={$\#$ Tested Options},
  axis x line  = bottom,
  axis y line  = left ,
  legend cell align={left},
   legend columns=3,
 legend style={draw=none,fill=none,at={(1,1.2)}},
   y label style={at={(-0.2,0.5)}}
  ]

    \addplot [mark=diamond,teal,name path=F] plot coordinates {
    (0, 0.00)
(13, 1.60)
(27, 3.60)
(41, 5.70)
(55, 7.60)
(69, 9.90)
(83, 12.70)
(97, 15.80)
(111, 18.80)
(125, 22.40)
(139, 27.00)
 };
 
\addplot [mark=square,red,name path=I] plot coordinates {
% (32, 1)
% (34, 2)
% (35, 3)
% (36, 4)
% (37, 5)
% (40, 6)
% (48, 7)
% (49, 8)
% (58, 9)
% (62, 10)
% (75, 11)
% (88, 12)
% (91, 13)
% (94, 14)
% (97, 15)
% (103, 16)
% (105, 17)
% (107, 18)
% (127, 19)
% (134, 20)
% (208, 21)
% (213, 22)
% (216, 23)
% (299, 24)
% (18.2, 0)
% (32, 1)
% (33.4, 1)
% (34, 2)
% (35, 3)
% (36, 4)
% (37, 5)
% (40, 6)
% (48, 7)
% (48.4, 7)
% (49, 8)
% (58, 9)
% (62, 10)
% (64.7, 10)
% (75, 11)
% (88, 12)
% (91, 13)
% (92.9, 13)
% (94, 14)
% (97, 15)
% (103, 16)
% (104.8, 16)
% (105, 17)
% (107, 18)
% (127, 19)
% (128.6, 19)
% (134, 20)
% (151.6, 20)
% (163.2, 20)
% (179.5, 20)
% (194.2, 20)
% (204.1, 20)
% (208, 21)
% (213, 22)
% (216, 23)
% (220.5, 23)
% (232.3, 23)
% (248.3, 23)
% (260.0, 23)
% (267.9, 23)
% (278.2, 23)
% (288.5, 23)
% (295.1, 23)
% (299, 24)
% (305.9, 24)
% (310.9, 24)
% (318.7, 24)
% (327.0, 24)
(0, 0.00)
(13, 8.00)
(27, 14.00)
(41, 21.00)
(55, 24.00)
(69, 25.00)
(83, 27.00)
(97, 29.00)
(111, 29.00)
(125, 29.00)
(139, 30.00)
 };

    \addplot [mark=star,blue,name path=C] plot coordinates {
% (18.20, 1)
% (33.40, 2)
% (48.40, 3)
% (64.70, 4)
% (92.90, 5)
% (104.80, 6)
% (128.60, 7)
% (151.60, 8)
% (163.20, 9)
% (179.50, 10)
% (194.20, 11)
% (204.10, 12)
% (220.50, 13)
% (232.30, 14)
% (248.30, 15)
% (260.00, 16)
% (267.90, 17)
% (278.20, 18)
% (288.50, 19)
% (295.10, 20)
% (305.90, 21)
% (310.90, 22)
% (318.70, 23)
% (327.00, 24)
% (18.2, 1)
% (32, 1)
% (33.4, 2)
% (34, 2)
% (35, 2)
% (36, 2)
% (37, 2)
% (40, 2)
% (48, 2)
% (48.4, 3)
% (49, 3)
% (58, 3)
% (62, 3)
% (64.7, 4)
% (75, 4)
% (88, 4)
% (91, 4)
% (92.9, 5)
% (94, 5)
% (97, 5)
% (103, 5)
% (104.8, 6)
% (105, 6)
% (107, 6)
% (127, 6)
% (128.6, 7)
% (134, 7)
% (151.6, 8)
% (163.2, 9)
% (179.5, 10)
% (194.2, 11)
% (204.1, 12)
% (208, 12)
% (213, 12)
% (216, 12)
% (220.5, 13)
% (232.3, 14)
% (248.3, 15)
% (260.0, 16)
% (267.9, 17)
% (278.2, 18)
% (288.5, 19)
% (295.1, 20)
% (299, 20)
% (305.9, 21)
% (310.9, 22)
% (318.7, 23)
% (327.0, 24)
(0, 0.00)
(13, 1.40)
(27, 2.80)
(41, 4.90)
(55, 6.50)
(69, 9.60)
(83, 12.40)
(97, 15.70)
(111, 19.20)
(125, 22.90)
(139, 27.00)
 };

\addplot [no marks,teal!20,name path=D] plot coordinates {
(0, 0.00)
(13, 3.10)
(27, 5.34)
(41, 7.89)
(55, 9.16)
(69, 12.11)
(83, 14.75)
(97, 17.58)
(111, 21.44)
(125, 24.14)
(139, 27.00)
 };
\addplot [no marks,teal!20,name path=E] plot coordinates {
(0, 0.00)
(13, 0.10)
(27, 1.86)
(41, 3.51)
(55, 6.04)
(69, 7.69)
(83, 10.65)
(97, 14.02)
(111, 16.16)
(125, 20.66)
(139, 27.00)
 };

\addplot [no marks,red!20,name path=G] plot coordinates {
% (32, 1)
% (34, 2)
% (35, 3)
% (36, 4)
% (37, 5)
% (40, 6)
% (48, 7)
% (49, 8)
% (58, 9)
% (62, 10)
% (75, 11)
% (88, 12)
% (91, 13)
% (94, 14)
% (97, 15)
% (103, 16)
% (105, 17)
% (107, 18)
% (127, 19)
% (134, 20)
% (208, 21)
% (213, 22)
% (216, 23)
% (299, 24)
% (30.49, 0)
% (32, 1)
% (34, 2)
% (35, 3)
% (36, 4)
% (37, 5)
% (40, 6)
% (47.38, 6)
% (48, 7)
% (49, 8)
% (58, 9)
% (62, 10)
% (72.77, 10)
% (75, 11)
% (85.0, 11)
% (88, 12)
% (91, 13)
% (94, 14)
% (97, 15)
% (103, 16)
% (105, 17)
% (107, 18)
% (116.52, 18)
% (127, 19)
% (129.57, 19)
% (134, 20)
% (149.18, 20)
% (180.34, 20)
% (191.17, 20)
% (204.24, 20)
% (208, 21)
% (213, 22)
% (216, 23)
% (218.49, 23)
% (222.94, 23)
% (240.62, 23)
% (255.55, 23)
% (271.48, 23)
% (279.01, 23)
% (284.57, 23)
% (298.36, 23)
% (299, 24)
% (306.32, 24)
% (312.29, 24)
% (316.12, 24)
% (318.79, 24)
% (324.9, 24)
% (330.59, 24)
(0, 0.00)
(13, 8.00)
(27, 14.00)
(41, 21.00)
(55, 24.00)
(69, 25.00)
(83, 27.00)
(97, 29.00)
(111, 29.00)
(125, 29.00)
(139, 30.00)
 };
 \addplot [no marks,red!20,name path=H] plot coordinates {
% (32, 1)
% (34, 2)
% (35, 3)
% (36, 4)
% (37, 5)
% (40, 6)
% (48, 7)
% (49, 8)
% (58, 9)
% (62, 10)
% (75, 11)
% (88, 12)
% (91, 13)
% (94, 14)
% (97, 15)
% (103, 16)
% (105, 17)
% (107, 18)
% (127, 19)
% (134, 20)
% (208, 21)
% (213, 22)
% (216, 23)
% (299, 24)
% (5.91, 0)
% (19.42, 0)
% (24.03, 0)
% (32, 1)
% (34, 2)
% (35, 3)
% (36, 4)
% (37, 5)
% (40, 6)
% (44.4, 6)
% (48, 7)
% (49, 8)
% (58, 9)
% (62, 10)
% (69.28, 10)
% (75, 11)
% (80.03, 11)
% (88, 12)
% (91, 13)
% (94, 14)
% (97, 15)
% (103, 16)
% (105, 17)
% (107, 18)
% (108.02, 18)
% (122.86, 18)
% (127, 19)
% (134, 20)
% (135.23, 20)
% (154.76, 20)
% (169.91, 20)
% (185.26, 20)
% (200.38, 20)
% (208, 21)
% (209.05, 21)
% (213, 22)
% (216, 23)
% (225.12, 23)
% (240.99, 23)
% (251.23, 23)
% (258.04, 23)
% (270.68, 23)
% (277.91, 23)
% (295.68, 23)
% (299, 24)
% (303.01, 24)
% (312.5, 24)
% (323.41, 24)
(0, 0.00)
(13, 8.00)
(27, 14.00)
(41, 21.00)
(55, 24.00)
(69, 25.00)
(83, 27.00)
(97, 29.00)
(111, 29.00)
(125, 29.00)
(139, 30.00)
 }; 

 \addplot [no marks,blue!20,name path=A] plot coordinates {
% (30.49, 1)
% (47.38, 2)
% (72.77, 3)
% (85.00, 4)
% (116.52, 5)
% (129.57, 6)
% (149.18, 7)
% (180.34, 8)
% (191.17, 9)
% (204.24, 10)
% (218.49, 11)
% (222.94, 12)
% (240.62, 13)
% (255.55, 14)
% (271.48, 15)
% (279.01, 16)
% (284.57, 17)
% (298.36, 18)
% (306.32, 19)
% (312.29, 20)
% (316.12, 21)
% (318.79, 22)
% (324.90, 23)
% (330.59, 24)
% (30.49, 1)
% (32, 1)
% (34, 1)
% (35, 1)
% (36, 1)
% (37, 1)
% (40, 1)
% (47.38, 2)
% (48, 2)
% (49, 2)
% (58, 2)
% (62, 2)
% (72.77, 3)
% (75, 3)
% (85.0, 4)
% (88, 4)
% (91, 4)
% (94, 4)
% (97, 4)
% (103, 4)
% (105, 4)
% (107, 4)
% (116.52, 5)
% (127, 5)
% (129.57, 6)
% (134, 6)
% (149.18, 7)
% (180.34, 8)
% (191.17, 9)
% (204.24, 10)
% (208, 10)
% (213, 10)
% (216, 10)
% (218.49, 11)
% (222.94, 12)
% (240.62, 13)
% (255.55, 14)
% (271.48, 15)
% (279.01, 16)
% (284.57, 17)
% (298.36, 18)
% (299, 18)
% (306.32, 19)
% (312.29, 20)
% (316.12, 21)
% (318.79, 22)
% (324.9, 23)
% (330.59, 24)
(0, 0.00)
(13, 2.06)
(27, 4.05)
(41, 6.48)
(55, 8.19)
(69, 11.75)
(83, 14.41)
(97, 18.07)
(111, 21.19)
(125, 24.60)
(139, 27.00)
 };
 \addplot [no marks,blue!20,name path=B] plot coordinates {
% (5.91, 1)
% (19.42, 2)
% (24.03, 3)
% (44.40, 4)
% (69.28, 5)
% (80.03, 6)
% (108.02, 7)
% (122.86, 8)
% (135.23, 9)
% (154.76, 10)
% (169.91, 11)
% (185.26, 12)
% (200.38, 13)
% (209.05, 14)
% (225.12, 15)
% (240.99, 16)
% (251.23, 17)
% (258.04, 18)
% (270.68, 19)
% (277.91, 20)
% (295.68, 21)
% (303.01, 22)
% (312.50, 23)
% (323.41, 24)
% (5.91, 1)
% (19.42, 2)
% (24.03, 3)
% (32, 3)
% (34, 3)
% (35, 3)
% (36, 3)
% (37, 3)
% (40, 3)
% (44.4, 4)
% (48, 4)
% (49, 4)
% (58, 4)
% (62, 4)
% (69.28, 5)
% (75, 5)
% (80.03, 6)
% (88, 6)
% (91, 6)
% (94, 6)
% (97, 6)
% (103, 6)
% (105, 6)
% (107, 6)
% (108.02, 7)
% (122.86, 8)
% (127, 8)
% (134, 8)
% (135.23, 9)
% (154.76, 10)
% (169.91, 11)
% (185.26, 12)
% (200.38, 13)
% (208, 13)
% (209.05, 14)
% (213, 14)
% (216, 14)
% (225.12, 15)
% (240.99, 16)
% (251.23, 17)
% (258.04, 18)
% (270.68, 19)
% (277.91, 20)
% (295.68, 21)
% (299, 21)
% (303.01, 22)
% (312.5, 23)
% (323.41, 24)
(0, 0.00)
(13, 0.74)
(27, 1.55)
(41, 3.32)
(55, 4.81)
(69, 7.45)
(83, 10.39)
(97, 13.33)
(111, 17.21)
(125, 21.20)
(139, 27.00)
 };
 
   % filling
\addplot[blue!20,opacity=0.5] fill between[of=B and C];
\addplot[blue!20,opacity=0.5] fill between[of=A and C];

\addplot[black!20,opacity=0.5] fill between[of=D and F];
\addplot[black!20,opacity=0.5] fill between[of=E and F];    
    
           \addplot[red!20,opacity=0.5] fill between[of=H and I];
    \addplot[red!20,opacity=0.5] fill between[of=G and I];

%\addlegendentry{\textsc{Flash}$_\textsc{mmo}$}
%\addlegendentry{\textsc{Flash}}
\addlegendentry{\texttt{CPD}}
\addlegendentry{\approach}
\addlegendentry{\texttt{US}}

  \end{axis}
\end{tikzpicture}
        \caption{\textsc{MySQL}}
    \end{subfigure}
     ~\hspace{0.3cm}
    \begin{subfigure}[t!]{0.25\columnwidth}
        \centering        
           \begin{tikzpicture}
  \begin{axis}[
  width = 5cm,
  height = 5cm,
  %xtick={1,2,3,4,5},
 % xticklabels={20\%,40\%,60\%,80\%,100\%},
  % xlabel=$p$,
  ylabel={$\#$ CPBugs Found},
  xlabel={$\#$ Tested Options},
  axis x line  = bottom,
  axis y line  = left ,
  legend cell align={left},
  legend columns=3,
 legend style={draw=none,fill=none,at={(1,1.2)}},
   y label style={at={(-0.2,0.5)}}
  ]

    \addplot [mark=diamond,teal,name path=F] plot coordinates {
    (0, 0.00)
(12, 0.30)
(25, 0.70)
(38, 1.40)
(51, 2.20)
(64, 2.80)
(77, 4.10)
(90, 5.20)
(103, 6.30)
(116, 8.40)
(129, 10.00)
 }; 
\addplot [mark=square,red,name path=I] plot coordinates {
% (16, 1)
% (43, 2)
% (75, 3)
% (80, 4)
% (81, 5)
% (87, 6)
% (103, 7)
% (118, 8)
% (235, 9)
(0, 0.00)
(12, 2.00)
(25, 2.00)
(38, 3.00)
(51, 8.00)
(64, 9.00)
(77, 9.00)
(90, 10.00)
(103, 10.00)
(116, 10.00)
(129, 10.00)
 };

    \addplot [mark=star,blue,name path=C] plot coordinates {
% (79.80, 1)
% (122.40, 2)
% (164.50, 3)
% (205.40, 4)
% (236.60, 5)
% (262.80, 6)
% (282.10, 7)
% (298.80, 8)
% (314.70, 9)
% (16, 0)
% (43, 0)
% (75, 0)
% (79.8, 1)
% (80, 1)
% (81, 1)
% (87, 1)
% (103, 1)
% (118, 1)
% (122.4, 2)
% (164.5, 3)
% (205.4, 4)
% (235, 4)
% (236.6, 5)
% (262.8, 6)
% (282.1, 7)
% (298.8, 8)
% (314.7, 9)
(0, 0.00)
(12, 0.10)
(25, 0.40)
(38, 1.50)
(50, 2.20)
(63, 3.50)
(76, 4.40)
(88, 5.30)
(101, 7.00)
(114, 8.20)
(127, 10.00)
 };

\addplot [no marks,teal!20,name path=D] plot coordinates {
(0, 0.00)
(12, 0.76)
(25, 1.48)
(38, 2.60)
(51, 3.37)
(64, 4.13)
(77, 5.55)
(90, 6.37)
(103, 7.85)
(116, 9.76)
(129, 10.00)
 };
\addplot [no marks,teal!20,name path=E] plot coordinates {
(0, 0.00)
(12, -0.16)
(25, -0.08)
(38, 0.20)
(51, 1.03)
(64, 1.47)
(77, 2.65)
(90, 4.03)
(103, 4.75)
(116, 7.04)
(129, 10.00)
 };

\addplot [no marks,red!20,name path=G] plot coordinates {
% (16, 1)
% (43, 2)
% (75, 3)
% (80, 4)
% (81, 5)
% (87, 6)
% (103, 7)
% (118, 8)
% (235, 9)
(0, 0.00)
(12, 2.00)
(25, 2.00)
(38, 3.00)
(51, 8.00)
(64, 9.00)
(77, 9.00)
(90, 10.00)
(103, 10.00)
(116, 10.00)
(129, 10.00)
 };
 \addplot [no marks,red!20,name path=H] plot coordinates {
% (16, 1)
% (43, 2)
% (75, 3)
% (80, 4)
% (81, 5)
% (87, 6)
% (103, 7)
% (118, 8)
% (235, 9)
(0, 0.00)
(12, 2.00)
(25, 2.00)
(38, 3.00)
(51, 8.00)
(64, 9.00)
(77, 9.00)
(90, 10.00)
(103, 10.00)
(116, 10.00)
(129, 10.00)
 }; 

 \addplot [no marks,blue!20,name path=A] plot coordinates {
% (135.45, 1)
% (161.14, 2)
% (199.10, 3)
% (248.12, 4)
% (270.53, 5)
% (290.17, 6)
% (307.74, 7)
% (320.88, 8)
% (332.00, 9)
(0, 0.00)
(12, 0.40)
(25, 1.06)
(38, 2.42)
(50, 2.80)
(63, 4.42)
(76, 5.42)
(88, 6.20)
(101, 7.63)
(114, 8.95)
(127, 10.00)
 };
 \addplot [no marks,blue!20,name path=B] plot coordinates {
% (24.15, 1)
% (83.66, 2)
% (129.90, 3)
% (162.68, 4)
% (202.67, 5)
% (235.43, 6)
% (256.46, 7)
% (276.72, 8)
% (297.40, 9)
(0, 0.00)
(12, -0.20)
(25, -0.26)
(38, 0.58)
(50, 1.60)
(63, 2.58)
(76, 3.38)
(88, 4.40)
(101, 6.37)
(114, 7.45)
(127, 10.00)
 };
 
   % filling
\addplot[blue!20,opacity=0.5] fill between[of=B and C];
\addplot[blue!20,opacity=0.5] fill between[of=A and C];

\addplot[black!20,opacity=0.5] fill between[of=D and F];
\addplot[black!20,opacity=0.5] fill between[of=E and F];

           \addplot[red!20,opacity=0.5] fill between[of=H and I];
    \addplot[red!20,opacity=0.5] fill between[of=G and I];

%\addlegendentry{\textsc{Flash}$_\textsc{mmo}$}
%\addlegendentry{\textsc{Flash}}
\addlegendentry{\texttt{CPD}}
\addlegendentry{\approach}
\addlegendentry{\texttt{US}}

  \end{axis}
\end{tikzpicture}
         \caption{\textsc{MariaDB}}
    \end{subfigure}
    ~\hspace{0.3cm}
     \begin{subfigure}[t!]{0.25\columnwidth}
        \centering        
           \begin{tikzpicture}
  \begin{axis}[
  width = 5cm,
  height = 5cm,
  %xtick={1,2,3,4,5},
 % xticklabels={20\%,40\%,60\%,80\%,100\%},
  % xlabel=$p$,
  ylabel={$\#$ CPBugs Found},
  xlabel={$\#$ Tested Options},
  axis x line  = bottom,
  axis y line  = left ,
  legend cell align={left},
  legend columns=3,
  legend style={draw=none,fill=none,at={(1,1.2)}},
  y label style={at={(-0.2,0.5)}}
  ]

    \addplot [mark=diamond,teal,name path=F] plot coordinates {
(0, 0.00)
(12, 0.00)
(24, 0.10)
(36, 0.20)
(48, 0.30)
(60, 0.50)
(72, 0.80)
(84, 0.90)
(96, 1.60)
(108, 2.80)
(121, 4.00)
 }; 
\addplot [mark=square,red,name path=I] plot coordinates {
% (15, 1)
% (197, 2)
% (272, 3)
% (303, 4)
% (315, 5)
% (15, 1)
% (122.10, 1)
% (197, 2)
% (222.90, 2)
% (269.90, 2)
% (272, 3)
% (285.1, 3)
% (303, 4)
% (311.90, 4)
% (315, 5)
(0, 0.00)
(12, 1.00)
(24, 1.00)
(36, 1.00)
(48, 1.00)
(60, 1.00)
(72, 2.00)
(84, 2.00)
(96, 2.00)
(108, 3.00)
(121, 5.00)
 };

    \addplot [mark=star,blue,name path=C] plot coordinates {
% (122.10, 1)
% (222.90, 2)
% (269.90, 3)
% (285.10, 4)
% (311.90, 5)
% (15, 0)
% (122.10, 1)
% (197, 1)
% (222.90, 2)
% (269.90, 3)
% (272, 3)
% (285.10, 4)
% (303, 4)
% (311.90, 5)
% (315, 5)
(0, 0.00)
(12, 0.00)
(24, 0.10)
(36, 0.40)
(48, 0.80)
(60, 0.90)
(72, 1.10)
(84, 1.50)
(96, 2.00)
(108, 3.10)
(121, 4.00)
 };

\addplot [no marks,teal!20,name path=D] plot coordinates {
(0, 0.00)
(12, 0.00)
(24, 0.40)
(36, 0.60)
(48, 0.76)
(60, 1.00)
(72, 1.40)
(84, 1.44)
(96, 2.09)
(108, 3.20)
(121, 4.00)
 };
\addplot [no marks,teal!20,name path=E] plot coordinates {
(0, 0.00)
(12, 0.00)
(24, -0.20)
(36, -0.20)
(48, -0.16)
(60, 0.00)
(72, 0.20)
(84, 0.36)
(96, 1.11)
(108, 2.40)
(121, 4.00)
 };

\addplot [no marks,red!20,name path=G] plot coordinates {
% (15, 1)
% (197, 2)
% (272, 3)
% (303, 4)
% (315, 5)
% (15, 1)
% (169.80, 1)
% (197, 2)
% (248.81, 2)
% (272, 3)
% (283.31, 3)
% (295.50, 3)
% (303, 4)
% (315, 5)
% (325.60, 5)
(0, 0.00)
(12, 1.00)
(24, 1.00)
(36, 1.00)
(48, 1.00)
(60, 1.00)
(72, 2.00)
(84, 2.00)
(96, 2.00)
(108, 3.00)
(121, 5.00)
 };
 \addplot [no marks,red!20,name path=H] plot coordinates {
% (15, 1)
% (197, 2)
% (272, 3)
% (303, 4)
% (315, 5)
% (15, 1)
% (74.40, 1)
% (196.99, 1)
% (197, 2)
% (256.49, 2)
% (272, 3)
% (274.70, 3)
% (298.20, 3)
% (303, 4)
(0, 0.00)
(12, 1.00)
(24, 1.00)
(36, 1.00)
(48, 1.00)
(60, 1.00)
(72, 2.00)
(84, 2.00)
(96, 2.00)
(108, 3.00)
(121, 5.00)
 }; 

 \addplot [no marks,blue!20,name path=A] plot coordinates {
% (169.80, 1)
% (248.81, 2)
% (283.31, 3)
% (295.50, 4)
% (325.60, 5)
% (15, 0)
% (169.80, 1)
% (197, 1)
% (248.81, 2)
% (272, 2)
% (283.31, 3)
% (295.50, 4)
% (303, 4)
% (315, 4)
% (325.60, 5)
(0, 0.00)
(12, 0.00)
(24, 0.40)
(36, 0.89)
(48, 1.20)
(60, 1.44)
(72, 1.64)
(84, 2.17)
(96, 3.10)
(108, 4.14)
(121, 4.00)
 };
 \addplot [no marks,blue!20,name path=B] plot coordinates {
% (74.40, 1)
% (196.99, 2)
% (256.49, 3)
% (274.70, 4)
% (298.20, 5)
% (15, 0)
% (74.4, 1)
% (196.99, 2)
% (197, 2)
% (256.49, 3)
% (272, 3)
% (274.7, 4)
% (298.2, 5)
% (303, 5)
% (315, 5)
(0, 0.00)
(12, 0.00)
(24, -0.20)
(36, -0.09)
(48, 0.40)
(60, 0.36)
(72, 0.56)
(84, 0.83)
(96, 0.90)
(108, 2.06)
(121, 4.00)
 };
 
   % filling
\addplot[blue!20,opacity=0.5] fill between[of=B and C];
\addplot[blue!20,opacity=0.5] fill between[of=A and C];
    
\addplot[black!20,opacity=0.5] fill between[of=D and F];
\addplot[black!20,opacity=0.5] fill between[of=E and F];

           \addplot[red!20,opacity=0.5] fill between[of=H and I];
    \addplot[red!20,opacity=0.5] fill between[of=G and I];

%\addlegendentry{\textsc{Flash}$_\textsc{mmo}$}
%\addlegendentry{\textsc{Flash}}

\addlegendentry{\texttt{CPD}}
\addlegendentry{\approach}
\addlegendentry{\texttt{US}}

  \end{axis}
\end{tikzpicture}
         \caption{\textsc{Apache}}   
    \end{subfigure}

        \begin{subfigure}[t!]{0.25\columnwidth}
        \centering        
           \begin{tikzpicture}
  \begin{axis}[
   width = 5cm,
  height = 5cm,
  %xtick={1,2,3,4,5},
 % xticklabels={20\%,40\%,60\%,80\%,100\%},
  % xlabel=$p$,
  ylabel={$\#$ CPBugs Found},
  xlabel={$\#$ Tested Options},
  axis x line  = bottom,
  axis y line  = left ,
  legend cell align={left},
  legend columns=3,
 legend style={draw=none,fill=none,at={(1,1.2)}},
    y label style={at={(-0.2,0.5)}}
  ]

    \addplot [mark=diamond,teal,name path=F] plot coordinates {
    (0, 0.00)
(3, 0.50)
(7, 2.80)
(11, 3.80)
(15, 5.20)
(19, 7.30)
(22, 8.70)
(26, 10.30)
(30, 11.90)
(34, 13.50)
(38, 15.00)
 }; 
\addplot [mark=square,red,name path=I] plot coordinates {
% (72, 1)
% (82, 2)
% (84, 3)
% (89, 4)
% (98, 5)
% (101, 6)
% (108, 7)
% (113, 8)
% (116, 9)
% (124, 10)
% (125, 11)
% (147, 12)
% (166, 13)
% (170, 14)
% (197, 15)
% (239, 16)
% (23.6, 0)
% (57.2, 0)
% (72, 1)
% (81.5, 1)
% (82, 2)
% (84, 3)
% (89, 4)
% (98, 5)
% (99.8, 5)
% (101, 6)
% (108, 7)
% (113, 8)
% (116, 9)
% (124, 10)
% (125, 11)
% (135.8, 11)
% (147, 12)
% (166, 13)
% (170, 14)
% (177.5, 14)
% (187.1, 14)
% (197, 15)
% (213.9, 15)
% (233.5, 15)
% (239, 16)
% (246.5, 16)
% (264.5, 16)
% (276.4, 16)
% (285.7, 16)
% (298.7, 16)
% (311.0, 16)
% (321.5, 16)
(0, 0.00)
(3, 1.00)
(7, 4.00)
(11, 6.00)
(15, 7.00)
(19, 9.00)
(22, 10.00)
(26, 12.00)
(30, 14.00)
(34, 15.00)
(38, 15.00)
 };

    \addplot [mark=star,blue,name path=C] plot coordinates {
% (23.60, 1)
% (57.20, 2)
% (81.50, 3)
% (99.80, 4)
% (135.80, 5)
% (177.50, 6)
% (187.10, 7)
% (213.90, 8)
% (233.50, 9)
% (246.50, 10)
% (264.50, 11)
% (276.40, 12)
% (285.70, 13)
% (298.70, 14)
% (311.00, 15)
% (321.50, 16)
% (23.6, 1)
% (57.2, 2)
% (72, 2)
% (81.5, 3)
% (82, 3)
% (84, 3)
% (89, 3)
% (98, 3)
% (99.8, 4)
% (101, 4)
% (108, 4)
% (113, 4)
% (116, 4)
% (124, 4)
% (125, 4)
% (135.8, 5)
% (147, 5)
% (166, 5)
% (170, 5)
% (177.5, 6)
% (187.1, 7)
% (197, 7)
% (213.9, 8)
% (233.5, 9)
% (239, 9)
% (246.5, 10)
% (264.5, 11)
% (276.4, 12)
% (285.7, 13)
% (298.7, 14)
% (311.0, 15)
% (321.5, 16)
(0, 0.00)
(3, 1.30)
(7, 3.00)
(11, 4.50)
(15, 5.60)
(19, 7.30)
(22, 7.90)
(26, 9.70)
(30, 11.60)
(34, 13.30)
(38, 15.00)
 };

\addplot [no marks,teal!20,name path=D] plot coordinates {
(0, 0.00)
(3, 1.42)
(7, 4.40)
(11, 5.63)
(15, 6.86)
(19, 9.20)
(22, 10.25)
(26, 11.65)
(30, 13.20)
(34, 14.42)
(38, 15.00)
 };
\addplot [no marks,teal!20,name path=E] plot coordinates {
(0, 0.00)
(3, -0.42)
(7, 1.20)
(11, 1.97)
(15, 3.54)
(19, 5.40)
(22, 7.15)
(26, 8.95)
(30, 10.60)
(34, 12.58)
(38, 15.00)
 };

\addplot [no marks,red!20,name path=G] plot coordinates {
% (72, 1)
% (82, 2)
% (84, 3)
% (89, 4)
% (98, 5)
% (101, 6)
% (108, 7)
% (113, 8)
% (116, 9)
% (124, 10)
% (125, 11)
% (147, 12)
% (166, 13)
% (170, 14)
% (197, 15)
% (239, 16)
% (39.38, 0)
% (72, 1)
% (80.77, 1)
% (82, 2)
% (84, 3)
% (89, 4)
% (98, 5)
% (99.6, 5)
% (101, 6)
% (108, 7)
% (113, 8)
% (116, 9)
% (120.12, 9)
% (124, 10)
% (125, 11)
% (147, 12)
% (162.69, 12)
% (166, 13)
% (170, 14)
% (197, 15)
% (200.11, 15)
% (209.67, 15)
% (236.42, 15)
% (239, 16)
% (249.2, 16)
% (260.91, 16)
% (288.25, 16)
% (294.73, 16)
% (304.18, 16)
% (310.21, 16)
% (322.06, 16)
% (329.79, 16)
(0, 0.00)
(3, 1.00)
(7, 4.00)
(11, 6.00)
(15, 7.00)
(19, 9.00)
(22, 10.00)
(26, 12.00)
(30, 14.00)
(34, 15.00)
(38, 15.00)
 };
 \addplot [no marks,red!20,name path=H] plot coordinates {
% (72, 1)
% (82, 2)
% (84, 3)
% (89, 4)
% (98, 5)
% (101, 6)
% (108, 7)
% (113, 8)
% (116, 9)
% (124, 10)
% (125, 11)
% (147, 12)
% (166, 13)
% (170, 14)
% (197, 15)
% (239, 16)
% (7.82, 0)
% (33.63, 0)
% (63.4, 0)
% (72, 1)
% (79.48, 1)
% (82, 2)
% (84, 3)
% (89, 4)
% (98, 5)
% (101, 6)
% (108, 7)
% (108.91, 7)
% (113, 8)
% (116, 9)
% (124, 10)
% (125, 11)
% (147, 12)
% (154.89, 12)
% (164.53, 12)
% (166, 13)
% (170, 14)
% (191.38, 14)
% (197, 15)
% (217.8, 15)
% (232.09, 15)
% (239, 16)
% (240.75, 16)
% (258.07, 16)
% (267.22, 16)
% (287.19, 16)
% (299.94, 16)
% (313.21, 16)
(0, 0.00)
(3, 1.00)
(7, 4.00)
(11, 6.00)
(15, 7.00)
(19, 9.00)
(22, 10.00)
(26, 12.00)
(30, 14.00)
(34, 15.00)
(38, 15.00)
 }; 

 \addplot [no marks,blue!20,name path=A] plot coordinates {
% (39.38, 1)
% (80.77, 2)
% (99.60, 3)
% (120.12, 4)
% (162.69, 5)
% (200.11, 6)
% (209.67, 7)
% (236.42, 8)
% (249.20, 9)
% (260.91, 10)
% (288.25, 11)
% (294.73, 12)
% (304.18, 13)
% (310.21, 14)
% (322.06, 15)
% (329.79, 16)
% (39.38, 1)
% (72, 1)
% (80.77, 2)
% (82, 2)
% (84, 2)
% (89, 2)
% (98, 2)
% (99.6, 3)
% (101, 3)
% (108, 3)
% (113, 3)
% (116, 3)
% (120.12, 4)
% (124, 4)
% (125, 4)
% (147, 4)
% (162.69, 5)
% (166, 5)
% (170, 5)
% (197, 5)
% (200.11, 6)
% (209.67, 7)
% (236.42, 8)
% (239, 8)
% (249.2, 9)
% (260.91, 10)
% (288.25, 11)
% (294.73, 12)
% (304.18, 13)
% (310.21, 14)
% (322.06, 15)
% (329.79, 16)
(0, 0.00)
(3, 1.94)
(7, 4.10)
(11, 5.52)
(15, 6.96)
(19, 8.49)
(22, 9.41)
(26, 10.89)
(30, 12.96)
(34, 14.20)
(38, 15.00)
 };
 \addplot [no marks,blue!20,name path=B] plot coordinates {
% (7.82, 1)
% (33.63, 2)
% (63.40, 3)
% (79.48, 4)
% (108.91, 5)
% (154.89, 6)
% (164.53, 7)
% (191.38, 8)
% (217.80, 9)
% (232.09, 10)
% (240.75, 11)
% (258.07, 12)
% (267.22, 13)
% (287.19, 14)
% (299.94, 15)
% (313.21, 16)
% (7.82, 1)
% (33.63, 2)
% (63.4, 3)
% (72, 3)
% (79.48, 4)
% (82, 4)
% (84, 4)
% (89, 4)
% (98, 4)
% (101, 4)
% (108, 4)
% (108.91, 5)
% (113, 5)
% (116, 5)
% (124, 5)
% (125, 5)
% (147, 5)
% (154.89, 6)
% (164.53, 7)
% (166, 7)
% (170, 7)
% (191.38, 8)
% (197, 8)
% (217.8, 9)
% (232.09, 10)
% (239, 10)
% (240.75, 11)
% (258.07, 12)
% (267.22, 13)
% (287.19, 14)
% (299.94, 15)
% (313.21, 16)
(0, 0.00)
(3, 0.66)
(7, 1.90)
(11, 3.48)
(15, 4.24)
(19, 6.11)
(22, 6.39)
(26, 8.51)
(30, 10.24)
(34, 12.40)
(38, 15.00)
 };
 
   % filling
\addplot[blue!20,opacity=0.5] fill between[of=B and C];
\addplot[blue!20,opacity=0.5] fill between[of=A and C];

\addplot[black!20,opacity=0.5] fill between[of=D and F];
\addplot[black!20,opacity=0.5] fill between[of=E and F];

           \addplot[red!20,opacity=0.5] fill between[of=H and I];
    \addplot[red!20,opacity=0.5] fill between[of=G and I];

%\addlegendentry{\textsc{Flash}$_\textsc{mmo}$}
%\addlegendentry{\textsc{Flash}}

\addlegendentry{\texttt{CPD}}
\addlegendentry{\approach}
\addlegendentry{\texttt{US}}

  \end{axis}
\end{tikzpicture}
         \caption{\textsc{Gcc}}   
    \end{subfigure}
    ~\hspace{0.6cm}
         \begin{subfigure}[t!]{0.25\columnwidth}
        \centering        
           \begin{tikzpicture}
  \begin{axis}[
  width = 5cm,
  height = 5cm,
  %xtick={1,2,3,4,5},
 % xticklabels={20\%,40\%,60\%,80\%,100\%},
  % xlabel=$p$,
  ylabel={$\#$ CPBugs Found},
  xlabel={$\#$ Tested Options},
  axis x line  = bottom,
  axis y line  = left ,
  legend cell align={left},
  legend columns=3,
 legend style={draw=none,fill=none,at={(1,1.2)}},
  y label style={at={(-0.2,0.5)}}
  ]

    \addplot [mark=diamond,teal,name path=F] plot coordinates {
    (0, 0.00)
(3, 0.30)
(7, 1.10)
(11, 1.50)
(15, 2.40)
(19, 3.20)
(22, 4.00)
(26, 4.60)
(30, 5.10)
(34, 5.50)
(38, 6.00)
 }; 
\addplot [mark=square,red,name path=I] plot coordinates {
% (84, 1)
% (89, 2)
% (92, 3)
% (94, 4)
% (108, 5)
% (113, 6)
% (124, 7)
% (84, 1)
% (89, 2)
% (92, 3)
% (94, 4)
% (95.8, 4)
% (108, 5)
% (113, 6)
% (124, 7)
% (170.4, 7)
% (203.8, 7)
% (236.2, 7)
% (266.0, 7)
% (290.2, 7)
% (309.5, 7)
(0, 0.00)
(3, 0.00)
(7, 2.00)
(11, 3.00)
(15, 5.00)
(19, 6.00)
(22, 6.00)
(26, 6.00)
(30, 6.00)
(34, 6.00)
(38, 6.00)
 };

    \addplot [mark=star,blue,name path=C] plot coordinates {
% (95.80, 1)
% (170.40, 2)
% (203.80, 3)
% (236.20, 4)
% (266.00, 5)
% (290.20, 6)
% (309.50, 7)
% (84, 0)
% (89, 0)
% (92, 0)
% (94, 0)
% (95.8, 1)
% (108, 1)
% (113, 1)
% (124, 1)
% (170.4, 2)
% (203.8, 3)
% (236.2, 4)
% (266.0, 5)
% (290.2, 6)
% (309.5, 7)
(0, 0.00)
(3, 0.10)
(7, 0.40)
(11, 0.70)
(15, 1.10)
(19, 1.40)
(22, 1.40)
(26, 2.50)
(30, 3.20)
(34, 4.50)
(38, 6.00)
 };

\addplot [no marks,teal!20,name path=D] plot coordinates {
(0, 0.00)
(3, 0.76)
(7, 2.04)
(11, 2.70)
(15, 3.68)
(19, 4.18)
(22, 5.18)
(26, 5.40)
(30, 5.40)
(34, 6.00)
(38, 6.00)
 };
\addplot [no marks,teal!20,name path=E] plot coordinates {
(0, 0.00)
(3, -0.16)
(7, 0.16)
(11, 0.30)
(15, 1.12)
(19, 2.22)
(22, 2.82)
(26, 3.80)
(30, 4.80)
(34, 5.00)
(38, 6.00)
 };

\addplot [no marks,red!20,name path=G] plot coordinates {
% (84, 1)
% (89, 2)
% (92, 3)
% (94, 4)
% (108, 5)
% (113, 6)
% (124, 7)
% (84, 1)
% (89, 2)
% (92, 3)
% (94, 4)
% (108, 5)
% (113, 6)
% (124, 7)
% (149.05, 7)
% (201.08, 7)
% (224.9, 7)
% (268.08, 7)
% (293.46, 7)
% (317.71, 7)
% (334.27, 7)
(0, 0.00)
(3, 0.00)
(7, 2.00)
(11, 3.00)
(15, 5.00)
(19, 6.00)
(22, 6.00)
(26, 6.00)
(30, 6.00)
(34, 6.00)
(38, 6.00)
 };
 \addplot [no marks,red!20,name path=H] plot coordinates {
% (84, 1)
% (89, 2)
% (92, 3)
% (94, 4)
% (108, 5)
% (113, 6)
% (124, 7)
% (42.55, 0)
% (84, 1)
% (89, 2)
% (92, 3)
% (94, 4)
% (108, 5)
% (113, 6)
% (124, 7)
% (139.72, 7)
% (182.7, 7)
% (204.32, 7)
% (238.54, 7)
% (262.69, 7)
% (284.73, 7)
(0, 0.00)
(3, 0.00)
(7, 2.00)
(11, 3.00)
(15, 5.00)
(19, 6.00)
(22, 6.00)
(26, 6.00)
(30, 6.00)
(34, 6.00)
(38, 6.00)
 }; 

 \addplot [no marks,blue!20,name path=A] plot coordinates {
% (149.05, 1)
% (201.08, 2)
% (224.90, 3)
% (268.08, 4)
% (293.46, 5)
% (317.71, 6)
% (334.27, 7)
% (84, 0)
% (89, 0)
% (92, 0)
% (94, 0)
% (108, 0)
% (113, 0)
% (124, 0)
% (149.05, 1)
% (201.08, 2)
% (224.9, 3)
% (268.08, 4)
% (293.46, 5)
% (317.71, 6)
% (334.27, 7)
(0, 0.00)
(3, 0.40)
(7, 0.89)
(11, 1.16)
(15, 1.80)
(19, 2.06)
(22, 2.06)
(26, 3.00)
(30, 4.07)
(34, 5.42)
(38, 6.00)
 };
 \addplot [no marks,blue!20,name path=B] plot coordinates {
% (42.55, 1)
% (139.72, 2)
% (182.70, 3)
% (204.32, 4)
% (238.54, 5)
% (262.69, 6)
% (284.73, 7)
% (42.55, 1)
% (84, 1)
% (89, 1)
% (92, 1)
% (94, 1)
% (108, 1)
% (113, 1)
% (124, 1)
% (139.72, 2)
% (182.7, 3)
% (204.32, 4)
% (238.54, 5)
% (262.69, 6)
% (284.73, 7)
(0, 0.00)
(3, -0.20)
(7, -0.09)
(11, 0.24)
(15, 0.40)
(19, 0.74)
(22, 0.74)
(26, 2.00)
(30, 2.33)
(34, 3.58)
(38, 6.00)
 };
 
   % filling
\addplot[blue!20,opacity=0.5] fill between[of=B and C];
\addplot[blue!20,opacity=0.5] fill between[of=A and C];

\addplot[black!20,opacity=0.5] fill between[of=D and F];
\addplot[black!20,opacity=0.5] fill between[of=E and F];

           \addplot[red!20,opacity=0.5] fill between[of=H and I];
    \addplot[red!20,opacity=0.5] fill between[of=G and I];

%\addlegendentry{\textsc{Flash}$_\textsc{mmo}$}
%\addlegendentry{\textsc{Flash}}

\addlegendentry{\texttt{CPD}}
\addlegendentry{\approach}
\addlegendentry{\texttt{US}}

  \end{axis}
\end{tikzpicture}
         \caption{\textsc{Clang}}   
    \end{subfigure}
    \caption{Effectiveness of testing CPBugs over all options.}
    \label{fig:rq2}
    \vspace{-0.2cm}
\end{figure}

\begin{rbox}
   \textit{\approach~better predicts CPBugs type than the state-of-the-art approaches over 87\% (13/15) types/metrics.}
\end{rbox}

\subsection{RQ2: Tested Option Prioritization}

\subsubsection{Method}

\revision{To verify high-level option prioritization in \textbf{RQ2},} we use five systems (and their versions) for which we have successfully reproduced the CPBugs. We compare \approach~against \texttt{CPD} and \texttt{US}, which test the options in random order. The mean/deviation of the cumulative number of CPBugs found with respect to the number of options tested for each system over 10 runs are reported. We also calculate the speedup of \approach~via $o \over o'$, where $o$ is the number of options tested to find the most CPBugs by the other tool and $o'$ is the number of tested options tested for \approach~to achieve the same.

\subsubsection{Results}

\revision{As can be seen in Figure~\ref{fig:rq2}, \approach~can reveal more CPBugs for \textsc{MySQL} and \textsc{Apache} due to the prioritization in testing numeric options (if we compare the last point),} which we will evaluate in \textbf{RQ3}. More importantly, it runs CPBug testing with much better efficiency: we see that for all systems, \approach~exhibits much stepper slops, meaning that many more CPBugs are discovered in an earlier stage of the testing---a significant contribution made by the prioritization at the tested options level. In particular, \approach~achieves speedup range from 1.11$\times$ to 1.73$\times$ against both \texttt{CPD} and \texttt{US}.

\revision{Notably, testing a single option can be rather expensive, i.e., 2 hours on average; some can be days (since we need to go through many versions/workloads), hence there will be significant savings if we can find the same (or more) CPBugs by testing even slightly fewer options. It can be seen from the Table~\ref{tb:rq2}, which shows the least tested options/clock time required to find per CPBug, that  \approach~only needs to test 1.6--12 options (18--536.2 minutes) against the 2.5--30.3 tested options (19.4--1351.6 minutes) for the state-of-the-art tools.}

\revision{The reduced improvement of \approach~for \textsc{Gcc} is due to its larger CPBugs ratio: 16 out of 38 options can trigger CPBugs. Indeed, since \approach~speedups testing by prioritizing the options, clearly a high ratio of CPBugs can blur the benefits.}

\begin{table}[t!]
%\caption{\revision{The best efficiency of each approach by computing how many tested options/much clock time is required to find per CPBug.}}
\caption{\revision{The least (average) tested options/clock time required to find per CPBug; \setlength{\fboxsep}{1.5pt}\colorbox{red!20}{red cells} denote the best.}}
\label{tb:rq2}
\centering
\begin{tabular}{llll|lll}
\hline
\toprule
\multirow{2}{*}{\textbf{System}}&\multicolumn{3}{c|}{\textbf{$\#$ Tested Options}}&\multicolumn{3}{c}{\textbf{Time (min)}}\\

\cmidrule{2-7}

&\approach&\texttt{CPD}&\texttt{US}&\approach&\texttt{CPD}&\texttt{US}\\

\midrule

\textsc{MySQL} &\cellcolor{red!20}1.6&5.1&5.1 &\cellcolor{red!20}119.1&380.2&380.2 \\
\textsc{MariaDB} &\cellcolor{red!20}6.0&12.7&12.7 &\cellcolor{red!20}153.0&324.7&324.7 \\
\textsc{Apache} &\cellcolor{red!20}12.0&30.3&30.3  &\cellcolor{red!20}536.2&1351.6&1351.6 \\
\textsc{Gcc} &\cellcolor{red!20}2.1&2.5&2.3  &\cellcolor{red!20}18.0&21.0&19.4 \\
\textsc{Clang} &\cellcolor{red!20}3.0&5.5&6.3  &\cellcolor{red!20}25.2&46.2&53.2 \\
\bottomrule
\end{tabular}
\vspace{-0.2cm}
\end{table}

All those results suggest that:

\begin{rbox}
   \textit{\approach~produces significantly better efficiency than state-of-the-art tools by prioritizing the order of tested options on all systems, achieving up to 1.73$\times$ speedup.}
\end{rbox}

\subsection{RQ3: Search Prioritization for Numeric Options}

\subsubsection{Method}

\revision{In \textbf{RQ3}, we examine the search prioritization when testing numeric options on the systems/versions from \textbf{RQ2}} (only \textsc{MySQL}, \textsc{MariaDB}, and \textsc{Apache} contain numeric options) against others. Since the number of tests---testing a pair of configurations is one test---is crucial for testing numeric options, for all systems, we report on the mean/deviation of the cumulative number of CPBugs found along with the number of tests for numeric options across 10 runs. All tools follow the same order of testing the numeric options prioritized by \approach: among all the numeric options of each system, \approach~remarkably prioritizes the CPBug-related ones before the others. We measure the cumulative CPBugs found for every 10\% tests (rounded) when the number of tests required is greater than 10; otherwise, we report every test. We calculate the speedup of \approach~via the same way as for \textbf{RQ2}.

%We calculate the speedup of \approach~via $t \over t'$, where $t$ is the number of tests to find the most CPBugs by the other tool and $t'$ is the number of tests required for \approach~to achieve the same.

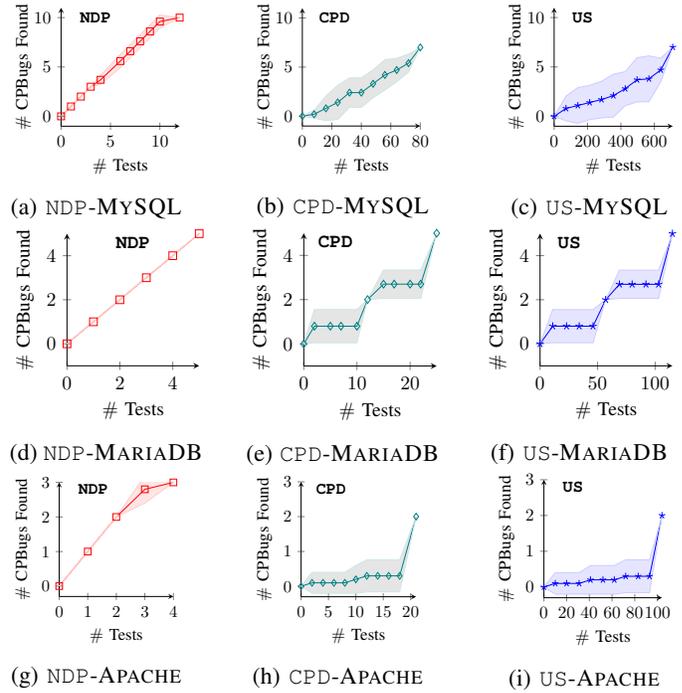
\begin{figure}[t!]
    \centering
    \begin{subfigure}[t!]{0.26\columnwidth}
        \centering
           \begin{tikzpicture}
  \begin{axis}[
   width = 4cm,
  height = 4cm,
  %xtick={1,2,3,4,5},
 % xticklabels={20\%,40\%,60\%,80\%,100\%},
  % xlabel=$p$,
  ylabel={$\#$ CPBugs Found},
  xlabel={$\#$ Tests},
  axis x line  = bottom,
  axis y line  = left ,
  legend cell align={left},
  legend columns=2,
  legend style={draw=none,fill=none,at={(1.08,1.05)}},
   ymin=-1, 
  ymax=11,
   y label style={at={(-0.2,0.5)}}
  ]
 
\addplot [mark=square,red,name path=I] plot coordinates {
% (1.00, 1)
% (2.00, 2)
% (3.00, 3)
% (4.20, 4)
% (5.20, 5)
% (6.20, 6)
% (7.20, 7)
% (8.20, 8)
% (9.70, 9)
% (10.70, 10)
% (1.0, 1)
% (2.0, 2)
% (3.0, 3)
% (4.2, 4)
% (5.2, 5)
% (6.2, 6)
% (7.2, 7)
% (8.2, 8)
% (9.7, 9)
% (10.7, 10)
% (41.0, 10)
% (54.5, 10)
% (78.9, 10)
% (99.8, 10)
% (127.0, 10)
% (511.1, 10)
% (600.1, 10)
% (834.1, 10)
% (1069.2, 10)
% (1340.0, 10)
% (1, 1.00)
% (2, 2.00)
% (3, 3.00)
% (5, 4.80)
% (6, 5.80)
% (7, 6.80)
% (9, 8.60)
% (10, 9.40)
% (11, 9.70)
% (13, 10.00)
(0, 0.00)
(1, 1.00)
(2, 2.00)
(3, 3.00)
(4, 3.70)
(6, 5.60)
(7, 6.60)
(8, 7.60)
(9, 8.60)
(10, 9.60)
(12, 10.00)
 };

\addplot [no marks,red!20,name path=G] plot coordinates {
% (1.00, 1)
% (2.00, 2)
% (3.00, 3)
% (4.60, 4)
% (5.60, 5)
% (6.60, 6)
% (7.60, 7)
% (8.60, 8)
% (10.70, 9)
% (11.70, 10)
% (1.0, 1)
% (2.0, 2)
% (3.0, 3)
% (4.6, 4)
% (5.6, 5)
% (6.6, 6)
% (7.6, 7)
% (8.6, 8)
% (10.7, 9)
% (11.7, 10)
% (66.44, 10)
% (86.63, 10)
% (114.42, 10)
% (123.23, 10)
% (127.0, 10)
% (861.76, 10)
% (1012.41, 10)
% (1278.7, 10)
% (1340.04, 10)
% (1340.0, 10)
% (1, 1.00)
% (2, 2.00)
% (3, 3.00)
% (5, 5.20)
% (6, 6.20)
% (7, 7.20)
% (9, 9.09)
% (10, 10.20)
% (11, 10.34)
% (13, 10.00)
(1, 1.00)
(2, 2.00)
(3, 3.00)
(4, 4.16)
(6, 6.26)
(7, 7.26)
(8, 8.26)
(9, 9.26)
(10, 10.26)
(12, 10.00)
 };
 \addplot [no marks,red!20,name path=H] plot coordinates {
% (1.00, 1)
% (2.00, 2)
% (3.00, 3)
% (3.80, 4)
% (4.80, 5)
% (5.80, 6)
% (6.80, 7)
% (7.80, 8)
% (8.70, 9)
% (9.70, 10)
% (1.0, 1)
% (2.0, 2)
% (3.0, 3)
% (3.8, 4)
% (4.8, 5)
% (5.8, 6)
% (6.8, 7)
% (7.8, 8)
% (8.7, 9)
% (9.7, 10)
% (15.56, 10)
% (22.37, 10)
% (43.38, 10)
% (76.37, 10)
% (127.0, 10)
% (160.44, 10)
% (187.79, 10)
% (389.5, 10)
% (798.36, 10)
% (1340.0, 10)
(0, 0.00)
(1, 1.00)
(2, 2.00)
(3, 3.00)
(4, 3.24)
(6, 4.94)
(7, 5.94)
(8, 6.94)
(9, 7.94)
(10, 8.94)
(12, 10.00)
 };

   % filling
    
\addplot[red!20,opacity=0.5] fill between[of=H and I];
\addplot[red!20,opacity=0.5] fill between[of=G and I];
    
   \node at (3.5,10) {\textbf{\texttt{NDP}}}; 
%\addlegendentry{\textsc{Flash}$_\textsc{mmo}$}
%\addlegendentry{\textsc{Flash}}

%\addlegendentry{NDP}

  \end{axis}
\end{tikzpicture}
        \caption{\approach-\textsc{MySQL}}
    \end{subfigure}
    ~\hfill%\hspace{-0.2cm}
    \begin{subfigure}[t!]{0.275\columnwidth}
        \centering        
           \begin{tikzpicture}
  \begin{axis}[
  width = 4cm,
  height = 4cm,
  ylabel={$\#$ CPBugs Found},
  xlabel={$\#$ Tests},
  axis x line  = bottom,
  axis y line  = left ,
  legend cell align={left},
  legend columns=2,
  legend style={draw=none,fill=none,at={(1.08,1.05)}},
    ymin=-1, 
  ymax=11,
   y label style={at={(-0.2,0.5)}}
  ]
   \addplot [mark=diamond,teal,name path=F] plot coordinates {
% (41.00, 1)
% (54.50, 2)
% (78.90, 3)
% (99.80, 4)
% (127.00, 5)
% (1.0, 0)
% (2.0, 0)
% (3.0, 0)
% (4.2, 0)
% (5.2, 0)
% (6.2, 0)
% (7.2, 0)
% (8.2, 0)
% (9.7, 0)
% (10.7, 0)
% (41.0, 1)
% (54.5, 2)
% (78.9, 3)
% (99.8, 4)
% (127.0, 5)
% (511.1, 5)
% (600.1, 5)
% (834.1, 5)
% (1069.2, 5)
% (1340.0, 5)
% (12, 0.40)
% (25, 0.60)
% (38, 1.00)
% (50, 1.30)
% (63, 1.80)
% (76, 2.40)
% (88, 2.50)
% (101, 3.10)
% (114, 3.20)
% (127, 5.00)
(0, 0.00)
(8, 0.20)
(16, 0.80)
(24, 1.40)
(32, 2.40)
(40, 2.40)
(48, 3.30)
(56, 4.20)
(64, 4.70)
(72, 5.40)
(80, 7.00)
 };

\addplot [no marks,teal!20,name path=D] plot coordinates {
% (66.44, 1)
% (86.63, 2)
% (114.42, 3)
% (123.23, 4)
% (127.00, 5)
% (1.0, 0)
% (2.0, 0)
% (3.0, 0)
% (4.6, 0)
% (5.6, 0)
% (6.6, 0)
% (7.6, 0)
% (8.6, 0)
% (10.7, 0)
% (11.7, 0)
% (66.44, 1)
% (86.63, 2)
% (114.42, 3)
% (123.23, 4)
% (127.0, 5)
% (861.76, 5)
% (1012.41, 5)
% (1278.7, 5)
% (1340.04, 5)
% (1340.0, 5)
% (12, 1.20)
% (25, 1.62)
% (38, 2.26)
% (50, 2.79)
% (63, 3.05)
% (76, 3.42)
% (88, 3.42)
% (101, 4.14)
% (114, 4.28)
% (127, 5.00)
(0, 0.00)
(8, 0.60)
(16, 2.05)
(24, 3.14)
(32, 3.90)
(40, 3.90)
(48, 4.72)
(56, 5.74)
(64, 5.89)
(72, 6.32)
(80, 7.00)
 };
\addplot [no marks,teal!20,name path=E] plot coordinates {
% (15.56, 1)
% (22.37, 2)
% (43.38, 3)
% (76.37, 4)
% (127.00, 5)
% (1.0, 0)
% (2.0, 0)
% (3.0, 0)
% (3.8, 0)
% (4.8, 0)
% (5.8, 0)
% (6.8, 0)
% (7.8, 0)
% (8.7, 0)
% (9.7, 0)
% (15.56, 1)
% (22.37, 2)
% (43.38, 3)
% (76.37, 4)
% (127.0, 5)
% (160.44, 5)
% (187.79, 5)
% (389.5, 5)
% (798.36, 5)
% (1340.0, 5)
% (12, -0.40)
% (25, -0.42)
% (38, -0.26)
% (50, -0.19)
% (63, 0.55)
% (76, 1.38)
% (88, 1.58)
% (101, 2.06)
% (114, 2.12)
% (127, 5.00)
(0, 0.00)
(8, -0.20)
(16, -0.45)
(24, -0.34)
(32, 0.90)
(40, 0.90)
(48, 1.88)
(56, 2.66)
(64, 3.51)
(72, 4.48)
(80, 7.00)
 };

 \addplot[black!20,opacity=0.5] fill between[of=D and F];
 \addplot[black!20,opacity=0.5] fill between[of=E and F];

       \node at (22,10) {\textbf{\texttt{CPD}}}; 
%\addlegendentry{\textsc{Flash}$_\textsc{mmo}$}
%\addlegendentry{\textsc{Flash}}
%\addlegendentry{CPD}

  \end{axis}
\end{tikzpicture}
         \caption{\texttt{CPD}-\textsc{MySQL}}
    \end{subfigure}
    ~\hfill%\hspace{-0.2cm}
     \begin{subfigure}[t!]{0.26\columnwidth}
        \centering        
           \begin{tikzpicture}
  \begin{axis}[
  width = 4cm,
  height = 4cm,
  %xtick={1,2,3,4,5},
 % xticklabels={20\%,40\%,60\%,80\%,100\%},
  % xlabel=$p$,
  ylabel={$\#$ CPBugs Found},
  xlabel={$\#$ Tests},
  axis x line  = bottom,
  axis y line  = left ,
  legend cell align={left},
  legend columns=2,
  legend style={draw=none,fill=none,at={(1.08,1.05)}},
  ymin=-1, 
  ymax=11,
   y label style={at={(-0.2,0.5)}},
  ]

    \addplot [mark=star,blue,name path=C] plot coordinates {
% (511.10, 1)
% (600.10, 2)
% (834.10, 3)
% (1069.20, 4)
% (1340.00, 5)
% (1.0, 0)
% (2.0, 0)
% (3.0, 0)
% (4.2, 0)
% (5.2, 0)
% (6.2, 0)
% (7.2, 0)
% (8.2, 0)
% (9.7, 0)
% (10.7, 0)
% (41.0, 0)
% (54.5, 0)
% (78.9, 0)
% (99.8, 0)
% (127.0, 0)
% (511.1, 1)
% (600.1, 2)
% (834.1, 3)
% (1069.2, 4)
% (1340.0, 5)
% (134, 0.50)
% (268, 0.80)
% (402, 1.10)
% (536, 1.40)
% (670, 1.40)
% (804, 2.00)
% (938, 2.20)
% (1072, 2.90)
% (1206, 3.20)
% (1340, 5.00)
(0, 0.00)
(71, 0.80)
(142, 1.10)
(213, 1.40)
(284, 1.70)
(355, 2.10)
(426, 2.80)
(497, 3.70)
(568, 3.80)
(639, 4.70)
(711, 7.00)
 };

 \addplot [no marks,blue!20,name path=A] plot coordinates {
% (861.76, 1)
% (1012.41, 2)
% (1278.70, 3)
% (1340.04, 4)
% (1340.00, 5)
% (1.0, 0)
% (2.0, 0)
% (3.0, 0)
% (4.6, 0)
% (5.6, 0)
% (6.6, 0)
% (7.6, 0)
% (8.6, 0)
% (10.7, 0)
% (11.7, 0)
% (66.44, 0)
% (86.63, 0)
% (114.42, 0)
% (123.23, 0)
% (127.0, 0)
% (861.76, 1)
% (1012.41, 2)
% (1278.7, 3)
% (1340.04, 4)
% (1340.0, 5)
% (134, 1.52)
% (268, 2.05)
% (402, 2.47)
% (536, 2.90)
% (670, 2.90)
% (804, 3.41)
% (938, 3.60)
% (1072, 3.94)
% (1206, 4.28)
% (1340, 5.00)
(0, 0.00)
(71, 2.05)
(142, 2.91)
(213, 3.14)
(284, 3.55)
(355, 4.27)
(426, 4.52)
(497, 5.94)
(568, 6.12)
(639, 5.89)
(711, 7.00)
 };
 \addplot [no marks,blue!20,name path=B] plot coordinates {
% (160.44, 1)
% (187.79, 2)
% (389.50, 3)
% (798.36, 4)
% (1340.00, 5)
% (1.0, 0)
% (2.0, 0)
% (3.0, 0)
% (3.8, 0)
% (4.8, 0)
% (5.8, 0)
% (6.8, 0)
% (7.8, 0)
% (8.7, 0)
% (9.7, 0)
% (15.56, 0)
% (22.37, 0)
% (43.38, 0)
% (76.37, 0)
% (127.0, 0)
% (160.44, 1)
% (187.79, 2)
% (389.5, 3)
% (798.36, 4)
% (1340.0, 5)
% (134, -0.52)
% (268, -0.45)
% (402, -0.27)
% (536, -0.10)
% (670, -0.10)
% (804, 0.59)
% (938, 0.80)
% (1072, 1.86)
% (1206, 2.12)
% (1340, 5.00)
(0, 0.00)
(71, -0.45)
(142, -0.71)
(213, -0.34)
(284, -0.15)
(355, -0.07)
(426, 1.08)
(497, 1.46)
(568, 1.48)
(639, 3.51)
(711, 7.00)
 };
 
   % filling
\addplot[blue!20,opacity=0.5] fill between[of=B and C];
\addplot[blue!20,opacity=0.5] fill between[of=A and C];

     \node at (170,10) {\textbf{\texttt{US}}};
%\addlegendentry{\textsc{Flash}$_\textsc{mmo}$}
%\addlegendentry{\textsc{Flash}}

%\addlegendentry{US}

  \end{axis}
\end{tikzpicture}
    \caption{\texttt{US}-\textsc{MySQL}} 
    \end{subfigure}

\begin{subfigure}[t!]{0.29\columnwidth}
        \centering
           \begin{tikzpicture}
  \begin{axis}[
  width = 4cm,
  height = 4cm,
  %xtick={1,2,3,4,5},
 % xticklabels={20\%,40\%,60\%,80\%,100\%},
  % xlabel=$p$,
   ylabel={$\#$ CPBugs Found},
  xlabel={$\#$ Tests},
  axis x line  = bottom,
  axis y line  = left ,
  legend cell align={left},
  legend columns=2,
  legend style={draw=none,fill=none,at={(1.08,1.05)}},
  ymin=-1,
  ymax=5,
   y label style={at={(-0.2,0.5)}}
  ]

\addplot [mark=square,red,name path=I] plot coordinates {
% (1.00, 1)
% (2.00, 2)
% (3.00, 3)
% (4.30, 4)
% (5.30, 5)
% (1.0, 1)
% (2.0, 2)
% (3.0, 3)
% (4.3, 4)
% (5.3, 5)
% (14.1, 5)
% (21.8, 5)
% (36.2, 5)
% (51.0, 5)
% (141.9, 5)
% (202.7, 5)
% (323.3, 5)
% (492.0, 5)
% (1, 1.00)
% (2, 2.00)
% (3, 3.00)
% (4, 3.70)
% (5, 4.70)
% (6, 5.00)
(0, 0.00)
(1, 1.00)
(2, 2.00)
(3, 3.00)
(4, 4.00)
(5, 5.00)
 };

\addplot [no marks,red!20,name path=G] plot coordinates {
% (1.00, 1)
% (2.00, 2)
% (3.00, 3)
% (4.76, 4)
% (5.76, 5)
% (1.0, 1)
% (2.0, 2)
% (3.0, 3)
% (4.76, 4)
% (5.76, 5)
% (28.42, 5)
% (36.14, 5)
% (51.66, 5)
% (51.0, 5)
% (323.12, 5)
% (393.04, 5)
% (518.89, 5)
% (492.0, 5)
% (1, 1.00)
% (2, 2.00)
% (3, 3.00)
% (4, 4.16)
% (5, 5.16)
% (6, 5.00)
(0, 0.00)
(1, 1.00)
(2, 2.00)
(3, 3.00)
(4, 4.00)
(5, 5.00)
 };
\addplot [no marks,red!20,name path=H] plot coordinates {
% (1.00, 1)
% (2.00, 2)
% (3.00, 3)
% (3.84, 4)
% (4.84, 5)
% (-39.32, 0)
% (-0.22, 0)
% (1.0, 1)
% (2.0, 2)
% (3.0, 3)
% (3.84, 4)
% (4.84, 5)
% (7.46, 5)
% (12.36, 5)
% (20.74, 5)
% (51.0, 5)
% (127.71, 5)
% (492.0, 5)
% (1, 1.00)
% (2, 2.00)
% (3, 3.00)
% (4, 3.24)
% (5, 4.24)
% (6, 5.00)
(0, 0.00)
(1, 1.00)
(2, 2.00)
(3, 3.00)
(4, 4.00)
(5, 5.00)
 };

   % filling

           \addplot[red!20,opacity=0.5] fill between[of=H and I];
    \addplot[red!20,opacity=0.5] fill between[of=G and I];
    
    \node at (2.5,4.6) {\textbf{\texttt{NDP}}};
%\addlegendentry{\textsc{Flash}$_\textsc{mmo}$}
%\addlegendentry{\textsc{Flash}}
%\addlegendentry{NDP}

  \end{axis}
\end{tikzpicture}
        \caption{\approach-\textsc{MariaDB}}
    \end{subfigure}
   ~\hfill%\hspace{-0.3cm}
\begin{subfigure}[t!]{0.29\columnwidth}
        \centering        
           \begin{tikzpicture}
  \begin{axis}[
  width = 4cm,
  height = 4cm,
  %xtick={1,2,3,4,5},
 % xticklabels={20\%,40\%,60\%,80\%,100\%},
  % xlabel=$p$,
  ylabel={$\#$ CPBugs Found},
  xlabel={$\#$ Tests},
  axis x line  = bottom,
  axis y line  = left ,
  legend cell align={left},
  legend columns=2,
  legend style={draw=none,fill=none,at={(1.08,1.05)}},
  ymin=-1,
  ymax=5,
   y label style={at={(-0.2,0.5)}}
  ]
  \addplot [mark=diamond,teal,name path=F] plot coordinates {
% (14.10, 1)
% (21.80, 2)
% (36.20, 3)
% (51.00, 4)
% (1.0, 0)
% (2.0, 0)
% (3.0, 0)
% (4.3, 0)
% (5.3, 0)
% (14.1, 1)
% (21.8, 2)
% (36.2, 3)
% (51.0, 4)
% (141.9, 4)
% (202.7, 4)
% (323.3, 4)
% (492.0, 4)
% (5, 0.50)
% (10, 0.50)
% (15, 1.50)
% (20, 1.50)
% (25, 1.70)
% (30, 1.80)
% (35, 1.80)
% (40, 2.50)
% (45, 2.50)
% (51, 4.00)
(0, 0.00)
(2, 0.80)
(5, 0.80)
(7, 0.80)
(10, 0.80)
(12, 2.00)
(15, 2.70)
(17, 2.70)
(20, 2.70)
(22, 2.70)
(25, 5.00)
 };

\addplot [no marks,teal!20,name path=D] plot coordinates {

% (28.42, 1)
% (36.14, 2)
% (51.66, 3)
% (51.00, 4)
% (1.0, 0)
% (2.0, 0)
% (3.0, 0)
% (4.76, 0)
% (5.76, 0)
% (28.42, 1)
% (36.14, 2)
% (51.66, 3)
% (51.0, 4)
% (323.12, 4)
% (393.04, 4)
% (518.89, 4)
% (492.0, 4)
% (5, 1.17)
% (10, 1.17)
% (15, 2.62)
% (20, 2.62)
% (25, 2.97)
% (30, 2.97)
% (35, 2.97)
% (40, 3.00)
% (45, 3.00)
% (51, 4.00)
(0, 0.00)
(2, 1.55)
(5, 1.55)
(7, 1.55)
(10, 1.55)
(12, 2.00)
(15, 3.34)
(17, 3.34)
(20, 3.34)
(22, 3.34)
(25, 5.00)
 };
\addplot [no marks,teal!20,name path=E] plot coordinates {
% (-0.22, 1)
% (7.46, 2)
% (20.74, 3)
% (51.00, 4)
% (-39.32, 0)
% (-0.22, 1)
% (1.0, 1)
% (2.0, 1)
% (3.0, 1)
% (3.84, 1)
% (4.84, 1)
% (7.46, 2)
% (12.36, 2)
% (20.74, 3)
% (51.0, 4)
% (127.71, 4)
% (492.0, 4)
% (5, -0.17)
% (10, -0.17)
% (15, 0.38)
% (20, 0.38)
% (25, 0.43)
% (30, 0.63)
% (35, 0.63)
% (40, 2.00)
% (45, 2.00)
% (51, 4.00)
(0, 0.00)
(2, 0.05)
(5, 0.05)
(7, 0.05)
(10, 0.05)
(12, 2.00)
(15, 2.06)
(17, 2.06)
(20, 2.06)
(22, 2.06)
(25, 5.00)
 };

 \addplot[black!20,opacity=0.5] fill between[of=D and F];
 \addplot[black!20,opacity=0.5] fill between[of=E and F];

     \node at (6,4.6) {\textbf{\texttt{CPD}}};
    
%\addlegendentry{\textsc{Flash}$_\textsc{mmo}$}
%\addlegendentry{\textsc{Flash}}
%\addlegendentry{CPD}

  \end{axis}
\end{tikzpicture}
       \caption{\texttt{CPD}-\textsc{MariaDB}}
    \end{subfigure}
    ~\hfill%\hspace{-0.3cm}
 \begin{subfigure}[t!]{0.29\columnwidth}
        \centering        
           \begin{tikzpicture}
  \begin{axis}[
  width = 4cm,
  height = 4cm,
  %xtick={1,2,3,4,5},
 % xticklabels={20\%,40\%,60\%,80\%,100\%},
  % xlabel=$p$,
 ylabel={$\#$ CPBugs Found},
  xlabel={$\#$ Tests},
  axis x line  = bottom,
  axis y line  = left ,
  legend cell align={left},
  legend columns=2,
  legend style={draw=none,fill=none,at={(1.08,1.05)}},
  ymin=-1,
  ymax=5,
   y label style={at={(-0.2,0.5)}}
  ]

\addplot [mark=star,blue,name path=C] plot coordinates {
% (141.90, 1)
% (202.70, 2)
% (323.30, 3)
% (492.00, 4)
% (1.0, 0)
% (2.0, 0)
% (3.0, 0)
% (4.3, 0)
% (5.3, 0)
% (14.1, 0)
% (21.8, 0)
% (36.2, 0)
% (51.0, 0)
% (141.9, 1)
% (202.7, 2)
% (323.3, 3)
% (492.0, 4)
% (49, 0.50)
% (98, 1.50)
% (147, 1.70)
% (196, 1.70)
% (246, 1.70)
% (295, 1.70)
% (344, 1.70)
% (393, 1.80)
% (442, 2.50)
% (492, 4.00)
(0, 0.00)
(11, 0.80)
(23, 0.80)
(34, 0.80)
(46, 0.80)
(57, 2.00)
(69, 2.70)
(80, 2.70)
(92, 2.70)
(103, 2.70)
(115, 5.00)
 };

\addplot [no marks,blue!20,name path=A] plot coordinates {
% (323.12, 1)
% (393.04, 2)
% (518.89, 3)
% (492.00, 4)
% (1.0, 0)
% (2.0, 0)
% (3.0, 0)
% (4.76, 0)
% (5.76, 0)
% (28.42, 0)
% (36.14, 0)
% (51.66, 0)
% (51.0, 0)
% (323.12, 1)
% (393.04, 2)
% (518.89, 3)
% (492.0, 4)
% (49, 1.17)
% (98, 2.62)
% (147, 2.97)
% (196, 2.97)
% (246, 2.97)
% (295, 2.97)
% (344, 2.97)
% (393, 2.97)
% (442, 3.00)
% (492, 4.00)
(0, 0.00)
(11, 1.55)
(23, 1.55)
(34, 1.55)
(46, 1.55)
(57, 2.00)
(69, 3.34)
(80, 3.34)
(92, 3.34)
(103, 3.34)
(115, 5.00)
 };
\addplot [no marks,blue!20,name path=B] plot coordinates {
% (-39.32, 1)
% (12.36, 2)
% (127.71, 3)
% (492.00, 4)
% (-39.32, 1)
% (-0.22, 1)
% (1.0, 1)
% (2.0, 1)
% (3.0, 1)
% (3.84, 1)
% (4.84, 1)
% (7.46, 1)
% (12.36, 2)
% (20.74, 2)
% (51.0, 2)
% (127.71, 3)
% (492.0, 4)
% (49, -0.17)
% (98, 0.38)
% (147, 0.43)
% (196, 0.43)
% (246, 0.43)
% (295, 0.43)
% (344, 0.43)
% (393, 0.63)
% (442, 2.00)
% (492, 4.00)
(0, 0.00)
(11, 0.05)
(23, 0.05)
(34, 0.05)
(46, 0.05)
(57, 2.00)
(69, 2.06)
(80, 2.06)
(92, 2.06)
(103, 2.06)
(115, 5.00)
 };

   % filling
\addplot[blue!20,opacity=0.5] fill between[of=B and C];
\addplot[blue!20,opacity=0.5] fill between[of=A and C];

     \node at (25.5,4.6) {\textbf{\texttt{US}}};
%\addlegendentry{\textsc{Flash}$_\textsc{mmo}$}
%\addlegendentry{\textsc{Flash}}

%\addlegendentry{US}

  \end{axis}
\end{tikzpicture}
          \caption{\texttt{US}-\textsc{MariaDB}}
    \end{subfigure}

      \begin{subfigure}[t!]{0.26\columnwidth}
        \centering
           \begin{tikzpicture}
  \begin{axis}[
  width = 4cm,
  height = 4cm,
  %xtick={1,2,3,4,5},
 % xticklabels={20\%,40\%,60\%,80\%,100\%},
  % xlabel=$p$,
  ylabel={$\#$ CPBugs Found},
  xlabel={$\#$ Tests},
  axis x line  = bottom,
  axis y line  = left ,
  legend cell align={left},
  legend columns=2,
  legend style={draw=none,fill=none,at={(1.08,1.05)}},
  ymin=-0.3,
  ymax=3,
   y label style={at={(-0.2,0.5)}}
  ]

\addplot [mark=square,red,name path=I] plot coordinates {
% (1.00, 1)
% (2.00, 2)
% (3.20, 3)
% (1.0, 1)
% (2.0, 2)
% (3.2, 3)
% (16.2, 3)
% (21.0, 3)
% (82.6, 3)
% (104.0, 3)
(0, 0.00)
(1, 1.00)
(2, 2.00)
(3, 2.80)
(4, 3.00)
 };

\addplot [no marks,red!20,name path=G] plot coordinates {
% (1.00, 1)
% (2.00, 2)
% (3.60, 3)
% (1.0, 1)
% (2.0, 2)
% (3.6, 3)
% (22.54, 3)
% (21.0, 3)
% (117.09, 3)
% (104.0, 3)
(0, 0.00)
(1, 1.00)
(2, 2.00)
(3, 3.20)
(4, 3.00)
 };
 \addplot [no marks,red!20,name path=H] plot coordinates {
% (1.00, 1)
% (2.00, 2)
% (2.80, 3)
% (1.0, 1)
% (2.0, 2)
% (2.8, 3)
% (9.86, 3)
% (21.0, 3)
% (48.11, 3)
% (104.0, 3)
(0, 0.00)
(1, 1.00)
(2, 2.00)
(3, 2.40)
(4, 3.00)
 };

   % filling

           \addplot[red!20,opacity=0.5] fill between[of=H and I];
    \addplot[red!20,opacity=0.5] fill between[of=G and I];
    
    \node at (1.2,2.8) {\textbf{\texttt{NDP}}};
%\addlegendentry{\textsc{Flash}$_\textsc{mmo}$}
%\addlegendentry{\textsc{Flash}}

%\addlegendentry{NDP}

  \end{axis}
\end{tikzpicture}
           \caption{\approach-\textsc{Apache}}
    \end{subfigure}
    ~\hfill%\hspace{-0.2cm}
 \begin{subfigure}[t!]{0.26\columnwidth}
        \centering        
           \begin{tikzpicture}
  \begin{axis}[
  width = 4cm,
  height = 4cm,
  %xtick={1,2,3,4,5},
 % xticklabels={20\%,40\%,60\%,80\%,100\%},
  % xlabel=$p$,
 ylabel={$\#$ CPBugs Found},
  xlabel={$\#$ Tests},
  axis x line  = bottom,
  axis y line  = left ,
  legend cell align={left},
  legend columns=2,
  legend style={draw=none,fill=none,at={(1.08,1.05)}},
   ymin=-0.3,
  ymax=3,
   y label style={at={(-0.2,0.5)}}
  ]
   \addplot [mark=diamond,teal,name path=F] plot coordinates {
% (16.20, 1)
% (21.00, 2)
% (1.0, 0)
% (2.0, 0)
% (3.2, 0)
% (16.2, 1)
% (21.0, 2)
% (82.6, 2)
% (104.0, 2)
(0, 0.00)
(2, 0.10)
(4, 0.10)
(6, 0.10)
(8, 0.10)
(10, 0.20)
(12, 0.30)
(14, 0.30)
(16, 0.30)
(18, 0.30)
(21, 2.00)
 };

\addplot [no marks,teal!20,name path=D] plot coordinates {

% (22.54, 1)
% (21.00, 2)
% (1.0, 0)
% (2.0, 0)
% (3.6, 0)
% (22.54, 1)
% (21.0, 2)
% (117.09, 2)
% (104.0, 2)
(0, 0.00)
(2, 0.40)
(4, 0.40)
(6, 0.40)
(8, 0.40)
(10, 0.60)
(12, 0.76)
(14, 0.76)
(16, 0.76)
(18, 0.76)
(21, 2.00)
 };
\addplot [no marks,teal!20,name path=E] plot coordinates {
% (9.86, 1)
% (21.00, 2)
% (1.0, 0)
% (2.0, 0)
% (2.8, 0)
% (9.86, 1)
% (21.0, 2)
% (48.11, 2)
% (104.0, 2)
(0, 0.00)
(2, -0.20)
(4, -0.20)
(6, -0.20)
(8, -0.20)
(10, -0.20)
(12, -0.16)
(14, -0.16)
(16, -0.16)
(18, -0.16)
(21, 2.00)
 };

   % filling
\
    
 \addplot[black!20,opacity=0.5] fill between[of=D and F];
 \addplot[black!20,opacity=0.5] fill between[of=E and F];

%\addlegendentry{\textsc{Flash}$_\textsc{mmo}$}
%\addlegendentry{\textsc{Flash}}
%\addlegendentry{CPD}
 \node at (5.5,2.8) {\textbf{\texttt{CPD}}};

  \end{axis}
\end{tikzpicture}
           \caption{\texttt{CPD}-\textsc{Apache}}
    \end{subfigure}
    ~\hfill%\hspace{-0.2cm}
\begin{subfigure}[t!]{0.275\columnwidth}
        \centering        
           \begin{tikzpicture}
  \begin{axis}[
  width = 4cm,
  height = 4cm,
  %xtick={1,2,3,4,5},
 % xticklabels={20\%,40\%,60\%,80\%,100\%},
  % xlabel=$p$,
 ylabel={$\#$ CPBugs Found},
   xlabel={$\#$ Tests},
  axis x line  = bottom,
  axis y line  = left ,
  legend cell align={left},
  legend columns=2,
  legend style={draw=none,fill=none,at={(1.08,1.05)}},
   ymin=-0.3,
  ymax=3,
   y label style={at={(-0.2,0.5)}}
  ]

    \addplot [mark=star,blue,name path=C] plot coordinates {
% (82.60, 1)
% (104.00, 2)
% (1.0, 0)
% (2.0, 0)
% (3.2, 0)
% (16.2, 0)
% (21.0, 0)
% (82.6, 1)
% (104.0, 2)
(0, 0.00)
(10, 0.10)
(20, 0.10)
(31, 0.10)
(41, 0.20)
(52, 0.20)
(62, 0.20)
(72, 0.30)
(83, 0.30)
(93, 0.30)
(104, 2.00)
 };

 \addplot [no marks,blue!20,name path=A] plot coordinates {

% (117.09, 1)
% (104.00, 2)
% (1.0, 0)
% (2.0, 0)
% (3.6, 0)
% (22.54, 0)
% (21.0, 0)
% (117.09, 1)
% (104.0, 2)
(0, 0.00)
(10, 0.40)
(20, 0.40)
(31, 0.40)
(41, 0.60)
(52, 0.60)
(62, 0.60)
(72, 0.76)
(83, 0.76)
(93, 0.76)
(104, 2.00)
 };
 \addplot [no marks,blue!20,name path=B] plot coordinates {

% (48.11, 1)
% (104.00, 2)
% (1.0, 0)
% (2.0, 0)
% (2.8, 0)
% (9.86, 0)
% (21.0, 0)
% (48.11, 1)
% (104.0, 2)
(0, 0.00)
(10, -0.20)
(20, -0.20)
(31, -0.20)
(41, -0.20)
(52, -0.20)
(62, -0.20)
(72, -0.16)
(83, -0.16)
(93, -0.16)
(104, 2.00)
 };
 
   % filling
\addplot[blue!20,opacity=0.5] fill between[of=B and C];
\addplot[blue!20,opacity=0.5] fill between[of=A and C];

%\addlegendentry{\textsc{Flash}$_\textsc{mmo}$}
%\addlegendentry{\textsc{Flash}}
 \node at (25.5,2.8) {\textbf{\texttt{US}}};
%\addlegendentry{US}

  \end{axis}
\end{tikzpicture}
        \caption{\texttt{US}-\textsc{Apache}} 
    \end{subfigure}

    %    \begin{subfigure}[t!]{0.26\columnwidth}
    %     \centering
    %     \includestandalone[width=\columnwidth]{fig/rq3-mysql-ndp}
    %     \caption{\approach-\textsc{MySQL}}
    % \end{subfigure}

    %    \begin{subfigure}[t!]{0.26\columnwidth}
    %     \centering
    %     \includestandalone[width=\columnwidth]{fig/rq3-mariadb-ndp}
    %     \caption{\approach-\textsc{MariaDB}}
    % \end{subfigure}
    \caption{Testing CPBugs over numeric options.}
    \label{fig:rq3}
    \vspace{-0.2cm}
\end{figure}

\subsubsection{Results}

Figure~\ref{fig:rq3} shows the traces of testing numeric options. Clearly, we see that \approach~exhibits remarkably better results compared with the others: \revision{it discovers the same (\textsc{MariaDB}) or more numeric options-related CPBugs (\textsc{MySQL} and \textsc{Apache}) than \texttt{CPD} and \texttt{US}, e.g., 10 for \approach~while the other two can only find 7 CPBugs on \textsc{MySQL}.} In particular, \approach~achieves such with significantly less number of tests---for all three systems, \approach~does so with as few as 4--13 tests while \texttt{CPD} and \texttt{US} need 21--80 tests and 104--711 tests to reach their maximum number of CPBugs, respectively. Notably, to find the same maximum number of CPBugs as achieved by others, \approach~has 3.13$\times$ to 10.50$\times$ and 14.38$\times$ to 88.88$\times$ speedup over \texttt{CPD} and \texttt{US}, respectively. 

\revision{Since a single test run is highly expensive, the saving is thereby significant. Table~\ref{tb:rq3} shows the least number of tests/clock time required to find per CPBug: \approach~only needs 1 test runs (as small as 1.1--4.4 minutes) against the 2.5--88.8 test runs ( 4.25--432.4 minutes) for the state-of-the-art tools.}

All above demonstrate the effectiveness of the search level prioritization for numeric options in \approach. Thus, we conclude:

%\approach~achieves such with significantly less number of tests, e.g., for \textsc{MariadDB}, it needs as few as 6 tests to reveal the same CPBugs that can only achieved by 51 tests for \texttt{CPD} and 492 tests for \texttt{US};

\begin{rbox}
   \textit{For all systems, \approach~finds considerably more numeric options-related CPBugs than the state-of-the-art tools with 3.13$\times$ to 88.88$\times$ speedup by prioritizing the search.}
\end{rbox}

\begin{table}[t!]
%\caption{\revision{The best efficiency of each approach by computing how many test counts/much clock time is required to find per CPBug on numeric options.}}
\caption{\revision{The least (average) test counts/clock time required to find per CPBug on numeric options; \setlength{\fboxsep}{1.5pt}\colorbox{red!20}{red cells} are the best.}}
\label{tb:rq3}
\centering
\footnotesize
\begin{tabular}{llll|lll}
\hline

\toprule
\multirow{2}{*}{\textbf{System}}&\multicolumn{3}{c|}{\textbf{$\#$ Test Counts}}&\multicolumn{3}{c}{\textbf{Time (min)}}\\

\cmidrule{2-7}

&\approach&\texttt{CPD}&\texttt{US}&\approach&\texttt{CPD}&\texttt{US}\\

\midrule

\textsc{MySQL}&\cellcolor{red!20}1.0&7.6&88.8&\cellcolor{red!20}4.4&37.0&432.4 \\
\textsc{MariaDB}&\cellcolor{red!20}1.0&2.5&23.0&\cellcolor{red!20}1.1&4.25&39.1 \\
\textsc{Apache}&\cellcolor{red!20}1.0&10.5&52.0&\cellcolor{red!20}4.2&44.1&218.4 \\

\bottomrule
\end{tabular}
\vspace{-0.2cm}
\end{table}

\subsection{RQ4: Detecting New CPBugs}

\subsubsection{Method}
\revision{To verify whether \approach~can reveal unknown CPBugs, we apply \approach~to further extended sets of versions compared with those used for \textbf{RQ1}--\textbf{RQ3}, and left the testing runs. For any CPBugs discovered, we also compute the measured performance drop by setting the target option value against the performance obtained via the source option value.}

%To quantify the possible performance degradation caused by these bugs, we conducted calculation and analysis by comparing with the maximum and reasonable possible test performance value that conforms to the original design intention of configuration options. For example, in GCC, the optimization level of -O3 is higher than that of -O2, so the possible test performance value of the -O3 option should not be less than the actual test performance value of the -O2 option.

%Through the analysis of test results, \approach~reports multiple test results of CPBugs related to GCC (the same configuration option may be caused by the same reason for CPBug). These CPBugs cause configuration options to deviate from the original design intention in multiple aspects, specifically manifested in abnormal performance of compilation time and execution time in GCC. 

\subsubsection{Results}

\revision{From Table~\ref{tb:rq4}, we see that \approach~has successfully discovered 12 CPBugs that are previously unknown on \textsc{Gcc} and \textsc{Clang}, which we have reported. These CPBugs can lead to significant performance impact, e.g., ranging between 1.06$\times$--2.75$\times$ and 1.06$\times$--1.17$\times$ degradations on the execution time and compiling time, respectively. As such, we say that:}

\begin{rbox}
   \textit{\approach~can discover previously unknown CPBugs given sufficient resources and versions/workloads.}
\end{rbox}

%After further verification, we found seven CPBugs triggered by different configuration options. The results show that these CPBugs have a significant impact on the software system (1.13$\times$-1.18$\times$ execution time, 1.12$\times$-1.59$\times$ compilation time), further revealing potential incorrect software designs and ultimately leading to a decline in user experience. See the table~\ref{tb:rq4} for details. It is worth noting that the improvement of various indicators of \approach~for predicting CPBug Type further reduces the possibility of false positives.

%\input{table/rq4}
\begin{table}[t!]
\caption{\revision{New CPBugs discovered by NDP.}}
\label{tb:rq4}
\centering
\adjustbox{max width=\columnwidth}{
  \begin{threeparttable}
\begin{tabular}{lll|l|l}
\toprule
\textbf{CPBug} & \textbf{System} & \textbf{Version}  & \textbf{Performance Degradation} & \textbf{CPBug Type} \\
\midrule
$\#$Pending (\textcolor{blue}{\href{https://github.com/ideas-labo/ndp/blob/main/new_bug/bug1.md}{link}}) & \textsc{Gcc} & v12 & 1.12$\times$ Execution Time & Type-4 \\
$\#$Pending (\textcolor{blue}{\href{https://github.com/ideas-labo/ndp/blob/main/new_bug/bug2.md}{link}}) & \textsc{Gcc} & v9 & 2.75$\times$ Execution Time & Type-2 \\
$\#$Pending (\textcolor{blue}{\href{https://github.com/ideas-labo/ndp/blob/main/new_bug/bug2.md}{link}}) & \textsc{Gcc} & v12 & 1.06$\times$ Execution Time & Type-2 \\
$\#$Pending (\textcolor{blue}{\href{https://github.com/ideas-labo/ndp/blob/main/new_bug/bug3.md}{link}}) & \textsc{Gcc} & v12 & 1.10$\times$ Compiling Time & Type-2 \\
$\#$Pending (\textcolor{blue}{\href{https://github.com/ideas-labo/ndp/blob/main/new_bug/bug4.md}{link}}) & \textsc{Gcc} & v12 & 1.17$\times$ Execution Time & Type-2 \\
$\#$Pending (\textcolor{blue}{\href{https://github.com/ideas-labo/ndp/blob/main/new_bug/bug5.md}{link}}) & \textsc{Gcc} & v9 & 1.14$\times$ Compiling Time & Type-2 \\
$\#$Pending (\textcolor{blue}{\href{https://github.com/ideas-labo/ndp/blob/main/new_bug/bug6.md}{link}}) & \textsc{Gcc} & v12 & 1.17$\times$ Compiling Time & Type-2 \\
$\#$Pending (\textcolor{blue}{\href{https://github.com/ideas-labo/ndp/blob/main/new_bug/bug7.md}{link}}) & \textsc{Gcc} & v9,v12 & 1.06--1.10$\times$ Execution Time & Type-2 \\
$\#$Pending (\textcolor{blue}{\href{https://github.com/ideas-labo/ndp/blob/main/new_bug/bug8.md}{link}}) & \textsc{Gcc} & v9,v12 & 1.31$\times$ File Size & Type-2 \\
$\#$117992(1) & \textsc{Clang} & v14 & 1.09$\times$ Compiling Time & Type-2 \\
$\#$117992(2) & \textsc{Clang} & v9,v14 & 1.06$\times$ Compiling Time & Type-2 \\
$\#$117993 & \textsc{Clang} & v9,v14 & 1.06--1.13$\times$ Execution Time & Type-2 \\

 \bottomrule
\end{tabular}
 \begin{tablenotes}
      \footnotesize
      \item We published the new \textsc{Gcc} bugs on our repository since \textsc{Gcc} has stopped bug reporting.
    \end{tablenotes}
  \end{threeparttable}
}
\vspace{-0.2cm}
\end{table}

\section{Discussion: Why \approach~Works?}
\label{sec:dis}
%\subsection{RQ4: Why \approach~Works?}

\revision{\textbf{RQ1}--\textbf{RQ3} serve as the ablation analysis of \approach. Here, we further explain why \approach~work with a qualitative analysis.}

\subsection{Predicting CPBugs Types}

A key benefit of the neural language model in \approach~is the significant reduction of false negatives compared with \texttt{CPD} and \texttt{KS}. For example, option \verb|innodb_fill_factor| for \textsc{MySQL} has the description of ``\textit{\texttt{innodb\_fill\_factor} defines the percentage of space on each B-tree page that is filled during a sorted index build, with the remaining space reserved for future index growth. For example, setting \texttt{innodb\_fill\_factor} to 80 reserves 20 percent of the space on each B-tree page for future index growth...}''. This option should belong to \textit{Type-2} since both a too small or a too large value could downgrade the performance as the former creates many recursions while the latter processes too many pages. Yet, for consistency, a smaller value is preferred since fewer pages need to be maintained, hence there is a trade-off. However, the description has no clear pattern to indicate such, hence both \texttt{CPD} and \texttt{KS} have wrongly classified it as \textit{Type-3} due to the presence of the word ``space''. \approach, in contrast, has correctly estimated the option since it has learned that texts with ``B-tree'' and ``page'' are likely to be related to trade-offs.

%\textcolor{red}{Youpeng, give an example of CPBug type predicting where we predict correctly but the other cause false negative, showing the natural language part of the example} (\verb|innodb_fill_factor| defines the percentage of space on each B-tree page that is filled during a sorted index build, with the remaining space reserved for future index growth. In different scenarios, we can observe varying performance tendencies. Inappropriate values can lead to different degrees of performance degradation. We consider this type of option as a more complex trade-off compared to regular configuration options, as its performance characteristics are much richer. In NDP, we highly value such configuration options with multiple performance tendencies. The chaotic performance tendencies can cause it to deviate from its original design intention, potentially serving as a condition for us to expose CPBugs.).

% \begin{figure}[t!]
%     \centering
%     \begin{subfigure}[t!]{0.35\columnwidth}
%         \centering
%         \includestandalone[width=\columnwidth]{fig/rank1}
%         \caption{Prioritization in \approach}
%     \end{subfigure}
%      ~\hspace{0.8cm}
%     \begin{subfigure}[t!]{0.4\columnwidth}
%         \centering        
%         \includestandalone[width=\columnwidth]{fig/rank2}
%          \caption{Random in \texttt{CPD}}
%     \end{subfigure}
%     \caption{Order of testing options for \textsc{MySQL} (left to right).}
%     \label{fig:rank}
%       \vspace{-0.2cm}
% \end{figure}

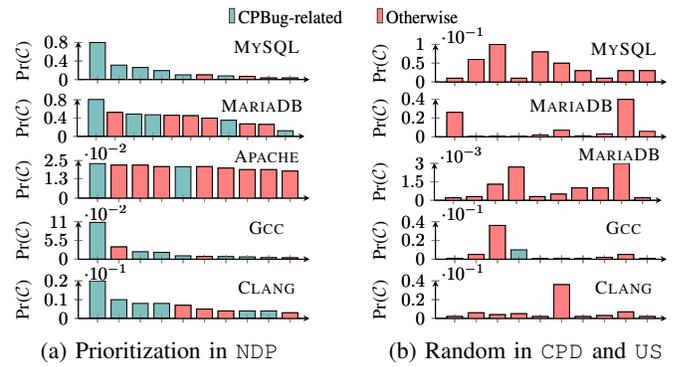
\begin{figure}[t!]
    \centering

\includegraphics[width=0.35\columnwidth]{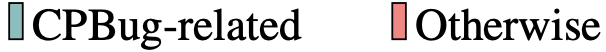}
%\vspace{-0.1cm}
    
    \begin{subfigure}[t!]{0.45\columnwidth}
        \centering
        % \begin{figure}
%     \centering
    \begin{tikzpicture}
        \begin{axis}[
            ybar,
            axis x line  = bottom,
            axis y line  = left  ,
             reverse legend,
            bar width=.3cm,
            width=6cm,
            height=2.3cm,
               enlarge x limits=0.1,
            ylabel={Pr($\mathcal{C}$)},
            symbolic x coords={1,2,3,4,5,6,7,8,9,10,11,12,13,14,15,16,17,18,19,20,21,22,23,24,25,26,27,28,29,30},
            xtick=data,
            xticklabels=\empty,
           ytick={0,0.4,0.8},
            ymax=0.81,
            ymin=0,
            legend style={at={(0.1,1.9)},draw=none,fill=none,anchor=north,legend columns=1},
            legend cell align={left},
           % y tick label style={rotate=45,anchor=east},
            % legend style={at={(0.9,1)},
            %     anchor=north,legend columns=-1},
            % enlarge x limits={abs=0.65cm},
            % nodes near coords,
            % point meta=explicit symbolic
        ]
        \addplot+[fill=teal!50,draw=black
            ] plot coordinates {
             % (1,99.99)
             % (2,99.99)
             % (3,99.99)
             % (4,99.99)
             % (5,99.99) 
             % (6,99.99)
             % (7,80.01)%1
             % (8,31.06)%1
             % (9,26.07)%1
             % (10,19.28)%1
             % (11,10.12)%1
             % (12,9.98)
             % (13,7.85)%1 
             % (14,6.81)
             % (15,3.29)
             % (16,3.27) 
             % (17,2.81)%1
             % (18,2.73)%1
             % (19,2.69)
             % (20,2.55) 
             % (21,2.51)
             % (22,1.63)
             % (23,1.56)%1
             % (24,1.16)%1
             % (25,0.90)
             % (26,0.74)
             % (27,0.73)
             % (28,0.40) 
             % (29,0.39)%1
             % (30,0.22)
            (1,0.8001)%1
            (2,0.3106)%1
            (3,0.2607)%1
            (4,0.1928)%1
            (5,0.1012)%1
            (6,0.0998)
            (7,0.0785)%1
            (8,0.0681)
            (9,0.0329)
            (10,0.0327)
            % (11,2.81)%1
            % (12,2.73)%1
            % (13,2.69)
            % (14,2.55)
            % (15,2.51)
            % (16,1.63)
            % (17,1.56)%1
            % (18,1.16)%1
            % (19,0.90)
            % (20,0.74)
            % (21,0.73)
            % (22,0.40)
            % (23,0.39)%1
            % (24,0.23)
            % (25,0.18)
            % (26,0.18)
            % (27,0.17)%1
            % (28,0.15)%1
            % (29,0.15)
            % (30,0.14)%1
        };
%\addlegendentry{\texttt{FTCP}}

%\node at (150,0.08) {$\log_{10} 10=1$};  
%\addlegendentry{\texttt{FairRF}}
       %  \legend{CPBug-related}
        \end{axis}
\node at (3.6,0.6) {\textsc{MySQL}};  

               \begin{axis}[
            ybar,
            axis x line  = bottom,
            axis y line  = left  ,
             reverse legend,
             bar width=.3cm,
            width=6cm,
            height=2.3cm,
               enlarge x limits=0.1,
            symbolic x coords={1,2,3,4,5,6,7,8,9,10,11,12,13,14,15,16,17,18,19,20,21,22,23,24,25,26,27,28,29,30},
            xtick=data,
            xticklabels=\empty,
            ytick={0,0.4,0.8},
          ymax=0.81,
            ymin=0,
            legend style={at={(0.7,1.9)},draw=none,fill=none,anchor=north,legend columns=1},
            legend cell align={left},
           % y tick label style={rotate=45,anchor=east},
            % legend style={at={(0.9,1)},
            %     anchor=north,legend columns=-1},
            % enlarge x limits={abs=0.65cm},
            % nodes near coords,
            % point meta=explicit symbolic
        ]
        \addplot+[fill=red!50,draw=black
            ] plot coordinates {
             % (1,99.99)
             % (2,99.99)
             % (3,99.99)
             % (4,99.99)
             % (5,99.99) 
             % (6,99.99)
             
             % (12,9.98)

             % (14,6.81)
             % (15,3.29)
             % (16,3.27)

             % (19,2.69)
             % (20,2.55) 
             % (21,2.51)
             % (22,1.63)
            
             % (25,0.90)
             % (26,0.74)
             % (27,0.73)
             % (28,0.40)

             % (30,0.22)
             
             % (6,9.98)

             % (8,6.81)
             % (9,3.29)
             % (10,3.27)

             % (13,2.69)
             % (14,2.55)
             % (15,2.51)
             % (16,1.63)

             % (19,0.90)
             % (20,0.74)
             % (21,0.73)
             % (22,0.40)

             % (24,0.23)
             % (25,0.18)
             % (26,0.18)

             % (29,0.15)
            (1,0)
            (2,0)
            (3,0)
            (4,0)
            (5,0)
            (6,0.0998)
            (7,0)
            (8,0.0681)
            (9,0.0329)
            (10,0.0327)
            % (11,0)
            % (12,0)
            % (13,2.69)
            % (14,2.55)
            % (15,2.51)
            % (16,1.63)
            % (17,0)
            % (18,0)
            % (19,0.90)
            % (20,0.74)
            % (21,0.73)
            % (22,0.40)
            % (23,0)
            % (24,0.23)
            % (25,0.18)
            % (26,0.18)
            % (27,0) 
            % (28,0)
            % (29,0.15)
            % (30,0)
        };
%\addlegendentry{\texttt{FTCP}}

%\addlegendentry{\texttt{FairRF}}
        % \legend{Otherwise}
        \end{axis}

    \end{tikzpicture}
    % \caption{The runtime in all dataset on Causal Graph.}
    % \label{fig:runtime}
% \end{figure}
       % \caption{Prioritization in \approach}
    \end{subfigure}
     ~\hfill
    \begin{subfigure}[t!]{0.45\columnwidth}
        \centering        
           % \begin{figure}
%     \centering
    \begin{tikzpicture}
        \begin{axis}[
            ybar,
            axis x line  = bottom,
            axis y line  = left  ,
             reverse legend,
            bar width=.3cm,
            width=6cm,
            height=2.3cm,
               enlarge x limits=0.1,
            ylabel={Pr($\mathcal{C}$)},
            symbolic x coords={1,2,3,4,5,6,7,8,9,10,11,12,13,14,15,16,17,18,19,20,21,22,23,24,25,26,27,28,29,30},
            xtick=data,
            xticklabels=\empty,
            ytick={0,0.05,0.1},
              scaled y ticks={base 10:1},
            ymin=0,
            legend style={at={(0.1,1.9)},draw=none,fill=none,anchor=north,legend columns=1},
            legend cell align={left},
           % y tick label style={rotate=45,anchor=east},
            % legend style={at={(0.9,1)},
            %     anchor=north,legend columns=-1},
            % enlarge x limits={abs=0.65cm},
            % nodes near coords,
            % point meta=explicit symbolic
        ]
        \addplot+[fill=teal!50,draw=black
            ] plot coordinates {
             % (1,99.99)
             % (2,99.99)
             % (3,99.99)
             % (4,99.99)
             % (5,99.99) 
             % (6,99.99)
             % (7,80.01)%1
             % (8,31.06)%1
             % (9,26.07)%1
             % (10,19.28)%1
             % (11,10.12)%1
             % (12,9.98)
             % (13,7.85)%1 
             % (14,6.81)
             % (15,3.29)
             % (16,3.27) 
             % (17,2.81)%1
             % (18,2.73)%1
             % (19,2.69)
             % (20,2.55) 
             % (21,2.51)
             % (22,1.63)
             % (23,1.56)%1
             % (24,1.16)%1
             % (25,0.90)
             % (26,0.74)
             % (27,0.73)
             % (28,0.40) 
             % (29,0.39)%1
             % (30,0.22)
              (1,0.01)
             (2,0.06)
             (3,0.10)
             (4,0.01)
             (5,0.08) 
             (6,0.05)
             (7,0.03)
             (8,0.01)
             (9,0.03)
             (10,0.03)
            % (11,2.81)%1
            % (12,2.73)%1
            % (13,2.69)
            % (14,2.55)
            % (15,2.51)
            % (16,1.63)
            % (17,1.56)%1
            % (18,1.16)%1
            % (19,0.90)
            % (20,0.74)
            % (21,0.73)
            % (22,0.40)
            % (23,0.39)%1
            % (24,0.23)
            % (25,0.18)
            % (26,0.18)
            % (27,0.17)%1
            % (28,0.15)%1
            % (29,0.15)
            % (30,0.14)%1
        };
%\addlegendentry{\texttt{FTCP}}

%\addlegendentry{\texttt{FairRF}}
        % \legend{CPBug-related}
        \end{axis}

\node at (3.6,0.6) {\textsc{MySQL}}; 
               \begin{axis}[
            ybar,
            axis x line  = bottom,
            axis y line  = left  ,
             reverse legend,
             bar width=.3cm,
            width=6cm,
            height=2.3cm,
               enlarge x limits=0.1,
            symbolic x coords={1,2,3,4,5,6,7,8,9,10,11,12,13,14,15,16,17,18,19,20,21,22,23,24,25,26,27,28,29,30},
            xtick=data,
            xticklabels=\empty,
            ytick={0,0.05,0.1},
              scaled y ticks={base 10:1},
            ymin=0,
            legend style={at={(0.7,1.9)},draw=none,fill=none,anchor=north,legend columns=1},
            legend cell align={left},
           % y tick label style={rotate=45,anchor=east},
            % legend style={at={(0.9,1)},
            %     anchor=north,legend columns=-1},
            % enlarge x limits={abs=0.65cm},
            % nodes near coords,
            % point meta=explicit symbolic
        ]
        \addplot+[fill=red!50,draw=black
            ] plot coordinates {
             % (1,99.99)
             % (2,99.99)
             % (3,99.99)
             % (4,99.99)
             % (5,99.99) 
             % (6,99.99)
             
             % (12,9.98)

             % (14,6.81)
             % (15,3.29)
             % (16,3.27)

             % (19,2.69)
             % (20,2.55) 
             % (21,2.51)
             % (22,1.63)
            
             % (25,0.90)
             % (26,0.74)
             % (27,0.73)
             % (28,0.40)

             % (30,0.22)
             
             % (6,9.98)

             % (8,6.81)
             % (9,3.29)
             % (10,3.27)

             % (13,2.69)
             % (14,2.55)
             % (15,2.51)
             % (16,1.63)

             % (19,0.90)
             % (20,0.74)
             % (21,0.73)
             % (22,0.40)

             % (24,0.23)
             % (25,0.18)
             % (26,0.18)

             % (29,0.15)
           (1,0.01)
             (2,0.06)
             (3,0.10)
             (4,0.01)
             (5,0.08) 
             (6,0.05)
             (7,0.03)
             (8,0.01)
             (9,0.03)
             (10,0.03)
            % (11,0)
            % (12,0)
            % (13,2.69)
            % (14,2.55)
            % (15,2.51)
            % (16,1.63)
            % (17,0)
            % (18,0)
            % (19,0.90)
            % (20,0.74)
            % (21,0.73)
            % (22,0.40)
            % (23,0)
            % (24,0.23)
            % (25,0.18)
            % (26,0.18)
            % (27,0) 
            % (28,0)
            % (29,0.15)
            % (30,0)
        };
%\addlegendentry{\texttt{FTCP}}

%\addlegendentry{\texttt{FairRF}}
        % \legend{Otherwise}
        \end{axis}

    \end{tikzpicture}
    % \caption{The runtime in all dataset on Causal Graph.}
    % \label{fig:runtime}
% \end{figure}
         %\caption{Random in \texttt{CPD}}
    \end{subfigure}
\vspace{-0.15cm}

      \begin{subfigure}[t!]{0.45\columnwidth}
        \centering
        % \begin{figure}
%     \centering
    \begin{tikzpicture}
        \begin{axis}[
            ybar,
            axis x line  = bottom,
            axis y line  = left  ,
             reverse legend,
         bar width=.3cm,
            width=6cm,
            height=2.3cm,
               enlarge x limits=0.1,
               ylabel={Pr($\mathcal{C}$)},
            symbolic x coords={39, 40, 41, 42, 43, 44, 45, 46, 47, 48, 49},
            xtick=data,
            xticklabels=\empty,
                 ytick={0,0.4,0.8},
            ymax=0.805,
            ymin=0,
            legend style={at={(0.1,1.15)},draw=none,fill=none,anchor=north,legend columns=1},
            legend cell align={left},
           % y tick label style={rotate=45,anchor=east},
            % legend style={at={(0.9,1)},
            %     anchor=north,legend columns=-1},
            % enlarge x limits={abs=0.65cm},
            % nodes near coords,
            % point meta=explicit symbolic
        ]
        \addplot+[fill=teal!50,draw=black
            ] plot coordinates {
             % (1,39.37)
             % (2,30.27)
             % (3,20.27)
             % (4,20.27)
             % (5,20.27) 
             % (6,20.27)
             % (7,20.27)
             % (8,20.27) 
             % (9,39.37)
             % (10,30.27)
             % (11,20.27)
             % (12,20.27)
             % (13,20.27) 
             % (14,20.27)
             % (15,20.27)
             % (16,20.27) 
             % (17,30.27)
             % (18,20.27)
             % (19,20.27)
             % (20,20.27) 
             % (21,20.27)
             % (22,20.27)
             % (23,20.27) 
             % (24,20.27) 
             % (25,30.27)
             % (26,20.27)
             % (27,20.27)
             % (28,20.27) 
             % (29,20.27)
             % (30,20.27)
             (39,0.805)%1
             (40,0.523)
             (41,0.487)%1
             (42,0.471)%1
             (43,0.460)
             (44,0.451)
             (45,0.399)
             (46,0.353)%1
             (47,0.272)
             (48,0.260)
             (49,0.123)%1
             % (11,10.12)
             % (12,9.98)
             % (13,7.85)
             % (14,6.81)
             % (15,3.29)
             % (16,3.27) 
             % (17,2.81)
             % (18,2.73)
             % (19,2.69)
             % (20,2.55) 
             % (21,2.51)
             % (22,1.63)
             % (23,1.56)
             % (24,1.16)
             % (25,0.90)
             % (26,0.74)
             % (27,0.73)
             % (28,0.40) 
             % (29,0.39)
             % (30,0.22)
        };
%\addlegendentry{\texttt{FTCP}}

%\addlegendentry{\texttt{FairRF}}
        % \legend{CPBug-related}
        \end{axis}

\node at (3.6,0.6) {\textsc{MariaDB}}; 
               \begin{axis}[
            ybar,
            axis x line  = bottom,
            axis y line  = left  ,
             reverse legend,
           bar width=.3cm,
            width=6cm,
            height=2.3cm,
               enlarge x limits=0.1,
            symbolic x coords={39, 40, 41, 42, 43, 44, 45, 46, 47, 48, 49},
            xtick=data,
            xticklabels=\empty,
              ytick={0,0.4,0.8},
            ymax=0.805,
            ymin=0,
            legend style={at={(0.7,1.15)},draw=none,fill=none,anchor=north,legend columns=1},
            legend cell align={left},
           % y tick label style={rotate=45,anchor=east},
            % legend style={at={(0.9,1)},
            %     anchor=north,legend columns=-1},
            % enlarge x limits={abs=0.65cm},
            % nodes near coords,
            % point meta=explicit symbolic
        ]
        \addplot+[fill=red!50,draw=black
            ] plot coordinates {
             % (1,39.37)
             % (2,30.27)
             % (3,20.27)
             % (4,20.27)
             % (5,20.27) 
             % (6,20.27)
             
             % (12,20.27)
             % (13,20.27) 
             % (14,20.27)
             % (15,20.27)
             % (16,20.27) 
             % (17,30.27)
             % (18,20.27)
           
             % (26,20.27)
             % (27,20.27)
             % (28,20.27) 
             % (29,20.27)
             % (30,20.27)
             % (1,99.99)

             % (2,99.99)
             % (3,99.99)
             % (4,99.99)
             % (5,99.99) 
             % (6,99.99)
             % (7,80.01)
             % (8,31.06)
             % (9,26.07)
             % (10,19.28)
             % (11,10.12)
             % (12,9.98)

             % (14,6.81)
             % (15,3.29)

             % (17,2.81)
             % (18,2.73)
             % (19,2.69)
             % (20,2.55)

             % (22,1.63)
             % (23,1.56)
             % (24,1.16)
             % (25,0.90)
             % (26,0.74)
             % (27,0.73)
             % (28,0.40) 
             % (29,0.39)
             % (30,0.22)
             (39,0)%1
             (40,0.523)
             (41,0)%1
             (42,0)%1
             (43,0.460)
             (44,0.451)
             (45,0.399)
             (46,0)%1
             (47,0.272)
             (48,0.260)
             (49,0)%1
        };
%\addlegendentry{\texttt{FTCP}}

%\addlegendentry{\texttt{FairRF}}
        % \legend{Otherwise}
        \end{axis}

    \end{tikzpicture}
    % \caption{The runtime in all dataset on Causal Graph.}
    % \label{fig:runtime}
% \end{figure}
    \end{subfigure}
     ~\hfill
    \begin{subfigure}[t!]{0.45\columnwidth}
        \centering        
        % \begin{figure}
%     \centering
    \begin{tikzpicture}
        \begin{axis}[
            ybar,
            axis x line  = bottom,
            axis y line  = left  ,
             reverse legend,
         bar width=.3cm,
            width=6cm,
            height=2.3cm,
               enlarge x limits=0.1,
          ylabel={Pr($\mathcal{C}$)},
            symbolic x coords={39, 40, 41, 42, 43, 44, 45, 46, 47, 48, 49},
            xtick=data,
            xticklabels=\empty,
               ytick={0,0.2,0.4},
            ymax=0.4,
            ymin=0,
            legend style={at={(0.1,1.15)},draw=none,fill=none,anchor=north,legend columns=1},
            legend cell align={left},
           % y tick label style={rotate=45,anchor=east},
            % legend style={at={(0.9,1)},
            %     anchor=north,legend columns=-1},
            % enlarge x limits={abs=0.65cm},
            % nodes near coords,
            % point meta=explicit symbolic
        ]
        \addplot+[fill=teal!50,draw=black
            ] plot coordinates {
             % (1,39.37)
             % (2,30.27)
             % (3,20.27)
             % (4,20.27)
             % (5,20.27) 
             % (6,20.27)
             % (7,20.27)
             % (8,20.27) 
             % (9,39.37)
             % (10,30.27)
             % (11,20.27)
             % (12,20.27)
             % (13,20.27) 
             % (14,20.27)
             % (15,20.27)
             % (16,20.27) 
             % (17,30.27)
             % (18,20.27)
             % (19,20.27)
             % (20,20.27) 
             % (21,20.27)
             % (22,20.27)
             % (23,20.27) 
             % (24,20.27) 
             % (25,30.27)
             % (26,20.27)
             % (27,20.27)
             % (28,20.27) 
             % (29,20.27)
             % (30,20.27)
             (39,0.260)%1
             (40,0.00003)
             (41,0.00004)%1
             (42,0.00003)%1
             (43,0.0002)
             (44,0.0007)
             (45,0.0001)
             (46,0.0003)%1
             (47,0.398)
             (48,0.0058)
             % (49,0.01)%1
             % (11,10.12)
             % (12,9.98)
             % (13,7.85)
             % (14,6.81)
             % (15,3.29)
             % (16,3.27) 
             % (17,2.81)
             % (18,2.73)
             % (19,2.69)
             % (20,2.55) 
             % (21,2.51)
             % (22,1.63)
             % (23,1.56)
             % (24,1.16)
             % (25,0.90)
             % (26,0.74)
             % (27,0.73)
             % (28,0.40) 
             % (29,0.39)
             % (30,0.22)
        };
%\addlegendentry{\texttt{FTCP}}

%\addlegendentry{\texttt{FairRF}}
         %\legend{CPBug-related}
        \end{axis}
\node at (2.6,0.6) {\textsc{MariaDB}}; 

               \begin{axis}[
            ybar,
            axis x line  = bottom,
            axis y line  = left  ,
             reverse legend,
           bar width=.3cm,
            width=6cm,
            height=2.3cm,
               enlarge x limits=0.1,
            symbolic x coords={39, 40, 41, 42, 43, 44, 45, 46, 47, 48, 49},
            xtick=data,
            xticklabels=\empty,
            ytick={0,0.2,0.4},
            ymax=0.4,
            ymin=0,
            legend style={at={(0.7,1.15)},draw=none,fill=none,anchor=north,legend columns=1},
            legend cell align={left},
           % y tick label style={rotate=45,anchor=east},
            % legend style={at={(0.9,1)},
            %     anchor=north,legend columns=-1},
            % enlarge x limits={abs=0.65cm},
            % nodes near coords,
            % point meta=explicit symbolic
        ]
        \addplot+[fill=red!50,draw=black
            ] plot coordinates {
             % (1,39.37)
             % (2,30.27)
             % (3,20.27)
             % (4,20.27)
             % (5,20.27) 
             % (6,20.27)
             
             % (12,20.27)
             % (13,20.27) 
             % (14,20.27)
             % (15,20.27)
             % (16,20.27) 
             % (17,30.27)
             % (18,20.27)
           
             % (26,20.27)
             % (27,20.27)
             % (28,20.27) 
             % (29,20.27)
             % (30,20.27)
             % (1,99.99)

             % (2,99.99)
             % (3,99.99)
             % (4,99.99)
             % (5,99.99) 
             % (6,99.99)
             % (7,80.01)
             % (8,31.06)
             % (9,26.07)
             % (10,19.28)
             % (11,10.12)
             % (12,9.98)

             % (14,6.81)
             % (15,3.29)

             % (17,2.81)
             % (18,2.73)
             % (19,2.69)
             % (20,2.55)

             % (22,1.63)
             % (23,1.56)
             % (24,1.16)
             % (25,0.90)
             % (26,0.74)
             % (27,0.73)
             % (28,0.40) 
             % (29,0.39)
             % (30,0.22)
             (39,0.260)
             (40,0.003)
             (41,0.004)
             (42,0.003)
             (43,0.02)
             (44,0.07)
             (45,0.01)
             (46,0.03)
             (47,0.398)
             (48,0.058)
             % (49,0.01)%1
        };
%\addlegendentry{\texttt{FTCP}}

%\addlegendentry{\texttt{FairRF}}
        % \legend{Otherwise}
        \end{axis}

    \end{tikzpicture}
    % \caption{The runtime in all dataset on Causal Graph.}
    % \label{fig:runtime}
% \end{figure}
    \end{subfigure}
\vspace{-0.15cm}

        \begin{subfigure}[t!]{0.45\columnwidth}
        \centering
        % \begin{figure}
%     \centering
    \begin{tikzpicture}
        \begin{axis}[
            ybar,
            axis x line  = bottom,
            axis y line  = left  ,
             reverse legend,
         bar width=.3cm,
            width=6cm,
            height=2.3cm,
               enlarge x limits=0.1,    
               ylabel={Pr($\mathcal{C}$)},
            symbolic x coords={109, 110, 111, 112, 113, 114, 115, 116, 117, 118},
            xtick=data,
            xticklabels=\empty,
             ytick={0,0.013,0.025},
            ymax=0.025,
            ymin=0,
            legend style={at={(0.1,1.15)},draw=none,fill=none,anchor=north,legend columns=1},
            legend cell align={left},
           % y tick label style={rotate=45,anchor=east},
            % legend style={at={(0.9,1)},
            %     anchor=north,legend columns=-1},
            % enlarge x limits={abs=0.65cm},
            % nodes near coords,
            % point meta=explicit symbolic
        ]
        \addplot+[fill=teal!50,draw=black
            ] plot coordinates {
             % (1,39.37)
             % (2,30.27)
             % (3,20.27)
             % (4,20.27)
             % (5,20.27) 
             % (6,20.27)
             % (7,20.27)
             % (8,20.27) 
             % (9,39.37)
             % (10,30.27)
             % (11,20.27)
             % (12,20.27)
             % (13,20.27) 
             % (14,20.27)
             % (15,20.27)
             % (16,20.27) 
             % (17,30.27)
             % (18,20.27)
             % (19,20.27)
             % (20,20.27) 
             % (21,20.27)
             % (22,20.27)
             % (23,20.27) 
             % (24,20.27) 
             % (25,30.27)
             % (26,20.27)
             % (27,20.27)
             % (28,20.27) 
             % (29,20.27)
             % (30,20.27)
             (109,0.023)%1
             (110,0.022)
             (111,0.022)
             (112,0.021)
             (113,0.021)%1
             (114,0.021)
             (115,0.020)
             (116,0.019)
             (117,0.019)
             (118,0.018)
             % (11,10.12)
             % (12,9.98)
             % (13,7.85)
             % (14,6.81)
             % (15,3.29)
             % (16,3.27) 
             % (17,2.81)
             % (18,2.73)
             % (19,2.69)
             % (20,2.55) 
             % (21,2.51)
             % (22,1.63)
             % (23,1.56)
             % (24,1.16)
             % (25,0.90)
             % (26,0.74)
             % (27,0.73)
             % (28,0.40) 
             % (29,0.39)
             % (30,0.22)
        };
%\addlegendentry{\texttt{FTCP}}

%\addlegendentry{\texttt{FairRF}}
       %  \legend{CPBug-related}
        \end{axis}

\node at (3.6,0.8) {\textsc{Apache}}; 
               \begin{axis}[
            ybar,
            axis x line  = bottom,
            axis y line  = left  ,
             reverse legend,
           bar width=.3cm,
            width=6cm,
            height=2.3cm,
               enlarge x limits=0.1,
            symbolic x coords={109, 110, 111, 112, 113, 114, 115, 116, 117, 118, 119},
            xtick=data,
            xticklabels=\empty,
              ytick={0,0.013,0.025},
            ymax=0.025,
            ymin=0,
            legend style={at={(0.7,1.15)},draw=none,fill=none,anchor=north,legend columns=1},
            legend cell align={left},
           % y tick label style={rotate=45,anchor=east},
            % legend style={at={(0.9,1)},
            %     anchor=north,legend columns=-1},
            % enlarge x limits={abs=0.65cm},
            % nodes near coords,
            % point meta=explicit symbolic
        ]
        \addplot+[fill=red!50,draw=black
            ] plot coordinates {
             % (1,39.37)
             % (2,30.27)
             % (3,20.27)
             % (4,20.27)
             % (5,20.27) 
             % (6,20.27)
             
             % (12,20.27)
             % (13,20.27) 
             % (14,20.27)
             % (15,20.27)
             % (16,20.27) 
             % (17,30.27)
             % (18,20.27)
           
             % (26,20.27)
             % (27,20.27)
             % (28,20.27) 
             % (29,20.27)
             % (30,20.27)
             % (1,99.99)

             % (2,99.99)
             % (3,99.99)
             % (4,99.99)
             % (5,99.99) 
             % (6,99.99)
             % (7,80.01)
             % (8,31.06)
             % (9,26.07)
             % (10,19.28)
             % (11,10.12)
             % (12,9.98)

             % (14,6.81)
             % (15,3.29)

             % (17,2.81)
             % (18,2.73)
             % (19,2.69)
             % (20,2.55)

             % (22,1.63)
             % (23,1.56)
             % (24,1.16)
             % (25,0.90)
             % (26,0.74)
             % (27,0.73)
             % (28,0.40) 
             % (29,0.39)
             % (30,0.22)
             (109,0)%1
             (110,0.022)
             (111,0.022)
             (112,0.021)
             (113,0)%1
             (114,0.021)
             (115,0.020)
             (116,0.019)
             (117,0.019)
             (118,0.018)
        };
%\addlegendentry{\texttt{FTCP}}

%\addlegendentry{\texttt{FairRF}}
       %  \legend{Otherwise}
        \end{axis}

    \end{tikzpicture}
    % \caption{The runtime in all dataset on Causal Graph.}
    % \label{fig:runtime}
% \end{figure}
    \end{subfigure}
     ~\hfill
    \begin{subfigure}[t!]{0.45\columnwidth}
        \centering        
        % \begin{figure}
%     \centering
    \begin{tikzpicture}
        \begin{axis}[
            ybar,
            axis x line  = bottom,
            axis y line  = left  ,
             reverse legend,
         bar width=.3cm,
            width=6cm,
            height=2.3cm,
               enlarge x limits=0.1,
                ylabel={Pr($\mathcal{C}$)},
            symbolic x coords={109, 110, 111, 112, 113, 114, 115, 116, 117, 118},
            xtick=data,
            xticklabels=\empty,
              ytick={0,0.0015,0.003},
            ymax=0.003,
            ymin=0,
            legend style={at={(0.1,1.15)},draw=none,fill=none,anchor=north,legend columns=1},
            legend cell align={left},
           % y tick label style={rotate=45,anchor=east},
            % legend style={at={(0.9,1)},
            %     anchor=north,legend columns=-1},
            % enlarge x limits={abs=0.65cm},
            % nodes near coords,
            % point meta=explicit symbolic
        ]
        \addplot+[fill=teal!50,draw=black
            ] plot coordinates {
             % (1,39.37)
             % (2,30.27)
             % (3,20.27)
             % (4,20.27)
             % (5,20.27) 
             % (6,20.27)
             % (7,20.27)
             % (8,20.27) 
             % (9,39.37)
             % (10,30.27)
             % (11,20.27)
             % (12,20.27)
             % (13,20.27) 
             % (14,20.27)
             % (15,20.27)
             % (16,20.27) 
             % (17,30.27)
             % (18,20.27)
             % (19,20.27)
             % (20,20.27) 
             % (21,20.27)
             % (22,20.27)
             % (23,20.27) 
             % (24,20.27) 
             % (25,30.27)
             % (26,20.27)
             % (27,20.27)
             % (28,20.27) 
             % (29,20.27)
             % (30,20.27)
             % (109,0.003)%1
             % (110,0.0003)
             % (111,0.027)
             % (112,0.0005)
             % (113,0.938)%1
             % (114,0.0005)
             % (115,0.002)
             % (116,0.0003)
             % (117,0.0018)
             % (118,0.0014)
             % (119,0.0004)
             (109,0.0002)
             (110,0.0003)
             (111,0.0013)
             (112,0.0027)
             (113,0.0003)
             (114,0.0005)
             (115,0.001)
             (116,0.001)
             (117,0.003)
             (118,0.0002)
             % (11,10.12)
             % (12,9.98)
             % (13,7.85)
             % (14,6.81)
             % (15,3.29)
             % (16,3.27) 
             % (17,2.81)
             % (18,2.73)
             % (19,2.69)
             % (20,2.55) 
             % (21,2.51)
             % (22,1.63)
             % (23,1.56)
             % (24,1.16)
             % (25,0.90)
             % (26,0.74)
             % (27,0.73)
             % (28,0.40) 
             % (29,0.39)
             % (30,0.22)
        };
%\addlegendentry{\texttt{FTCP}}

%\addlegendentry{\texttt{FairRF}}
       %  \legend{CPBug-related}
        \end{axis}

\node at (3.6,0.9) {\textsc{MariaDB}}; 
               \begin{axis}[
            ybar,
            axis x line  = bottom,
            axis y line  = left  ,
             reverse legend,
           bar width=.3cm,
            width=6cm,
            height=2.3cm,
               enlarge x limits=0.1,
            symbolic x coords={109, 110, 111, 112, 113, 114, 115, 116, 117, 118},
            xtick=data,
            xticklabels=\empty,
                     ytick={0,0.0015,0.003},
            ymax=0.003,
            ymin=0,
            legend style={at={(0.7,1.15)},draw=none,fill=none,anchor=north,legend columns=1},
            legend cell align={left},
           % y tick label style={rotate=45,anchor=east},
            % legend style={at={(0.9,1)},
            %     anchor=north,legend columns=-1},
            % enlarge x limits={abs=0.65cm},
            % nodes near coords,
            % point meta=explicit symbolic
        ]
        \addplot+[fill=red!50,draw=black
            ] plot coordinates {
             % (1,39.37)
             % (2,30.27)
             % (3,20.27)
             % (4,20.27)
             % (5,20.27) 
             % (6,20.27)
             
             % (12,20.27)
             % (13,20.27) 
             % (14,20.27)
             % (15,20.27)
             % (16,20.27) 
             % (17,30.27)
             % (18,20.27)
           
             % (26,20.27)
             % (27,20.27)
             % (28,20.27) 
             % (29,20.27)
             % (30,20.27)
             % (1,99.99)

             % (2,99.99)
             % (3,99.99)
             % (4,99.99)
             % (5,99.99) 
             % (6,99.99)
             % (7,80.01)
             % (8,31.06)
             % (9,26.07)
             % (10,19.28)
             % (11,10.12)
             % (12,9.98)

             % (14,6.81)
             % (15,3.29)

             % (17,2.81)
             % (18,2.73)
             % (19,2.69)
             % (20,2.55)

             % (22,1.63)
             % (23,1.56)
             % (24,1.16)
             % (25,0.90)
             % (26,0.74)
             % (27,0.73)
             % (28,0.40) 
             % (29,0.39)
             % (30,0.22)
             (109,0.0002)
             (110,0.0003)
             (111,0.0013)
             (112,0.0027)
             (113,0.0003)
             (114,0.0005)
             (115,0.001)
             (116,0.001)
             (117,0.003)
             (118,0.0002)
        };
%\addlegendentry{\texttt{FTCP}}

%\addlegendentry{\texttt{FairRF}}
        % \legend{Otherwise}
        \end{axis}

    \end{tikzpicture}
    % \caption{The runtime in all dataset on Causal Graph.}
    % \label{fig:runtime}
% \end{figure}
    \end{subfigure}
\vspace{-0.15cm}

      \begin{subfigure}[t!]{0.45\columnwidth}
        \centering
        % \begin{figure}
%     \centering
    \begin{tikzpicture}
        \begin{axis}[
            ybar,
            axis x line  = bottom,
            axis y line  = left  ,
             reverse legend,
         bar width=.3cm,
            width=6cm,
            height=2.3cm,
               enlarge x limits=0.1,
            ylabel={Pr($\mathcal{C}$)},
            symbolic x coords={3, 4, 5, 6, 7, 8, 9, 10, 11, 12, 13, 14},
            xtick=data,
            xticklabels=\empty,
               ytick={0,0.055,0.11},
            scaled y ticks={base 10:2},
            ymax=0.11,
            ymin=0,
            legend style={at={(0.1,1.15)},draw=none,fill=none,anchor=north,legend columns=1},
            legend cell align={left},
           % y tick label style={rotate=45,anchor=east},
            % legend style={at={(0.9,1)},
            %     anchor=north,legend columns=-1},
            % enlarge x limits={abs=0.65cm},
            % nodes near coords,
            % point meta=explicit symbolic
        ]
        \addplot+[fill=teal!50,draw=black
            ] plot coordinates {
             % (1,39.37)
             % (2,30.27)
             % (3,20.27)
             % (4,20.27)
             % (5,20.27) 
             % (6,20.27)
             % (7,20.27)
             % (8,20.27) 
             % (9,39.37)
             % (10,30.27)
             % (11,20.27)
             % (12,20.27)
             % (13,20.27) 
             % (14,20.27)
             % (15,20.27)
             % (16,20.27) 
             % (17,30.27)
             % (18,20.27)
             % (19,20.27)
             % (20,20.27) 
             % (21,20.27)
             % (22,20.27)
             % (23,20.27) 
             % (24,20.27) 
             % (25,30.27)
             % (26,20.27)
             % (27,20.27)
             % (28,20.27) 
             % (29,20.27)
             % (30,20.27)
             (3,0.108)%1
             (4,0.036)
             (5, 0.022)%1
             (6, 0.020)%1
             (7, 0.010)%1
             (8,0.008)%1
             (9,0.008)%1
             (10,0.007)%1
             (11,0.005)%1
             (12,0.004)
             % (13,0.004)%1
             % (14,0.004)%1
             % (11,10.12)
             % (12,9.98)
             % (13,7.85)
             % (14,6.81)
             % (15,3.29)
             % (16,3.27) 
             % (17,2.81)
             % (18,2.73)
             % (19,2.69)
             % (20,2.55) 
             % (21,2.51)
             % (22,1.63)
             % (23,1.56)
             % (24,1.16)
             % (25,0.90)
             % (26,0.74)
             % (27,0.73)
             % (28,0.40) 
             % (29,0.39)
             % (30,0.22)
        };
%\addlegendentry{\texttt{FTCP}}

%\addlegendentry{\texttt{FairRF}}
        % \legend{CPBug-related}
        \end{axis}

\node at (3.6,0.6) {\textsc{Gcc}}; 
               \begin{axis}[
            ybar,
            axis x line  = bottom,
            axis y line  = left  ,
             reverse legend,
           bar width=.3cm,
            width=6cm,
            height=2.3cm,
               enlarge x limits=0.1,
            symbolic x coords={3, 4, 5, 6, 7, 8, 9, 10, 11, 12, 13, 14},
            xtick=data,
            xticklabels=\empty,
              ytick={0,0.055,0.11},
            scaled y ticks={base 10:2},
            ymax=0.11,
            ymin=0,
            legend style={at={(0.7,1.15)},draw=none,fill=none,anchor=north,legend columns=1},
            legend cell align={left},
           % y tick label style={rotate=45,anchor=east},
            % legend style={at={(0.9,1)},
            %     anchor=north,legend columns=-1},
            % enlarge x limits={abs=0.65cm},
            % nodes near coords,
            % point meta=explicit symbolic
        ]
        \addplot+[fill=red!50,draw=black
            ] plot coordinates {
             % (1,39.37)
             % (2,30.27)
             % (3,20.27)
             % (4,20.27)
             % (5,20.27) 
             % (6,20.27)
             
             % (12,20.27)
             % (13,20.27) 
             % (14,20.27)
             % (15,20.27)
             % (16,20.27) 
             % (17,30.27)
             % (18,20.27)
           
             % (26,20.27)
             % (27,20.27)
             % (28,20.27) 
             % (29,20.27)
             % (30,20.27)
             % (1,99.99)

             % (2,99.99)
             % (3,99.99)
             % (4,99.99)
             % (5,99.99) 
             % (6,99.99)
             % (7,80.01)
             % (8,31.06)
             % (9,26.07)
             % (10,19.28)
             % (11,10.12)
             % (12,9.98)

             % (14,6.81)
             % (15,3.29)

             % (17,2.81)
             % (18,2.73)
             % (19,2.69)
             % (20,2.55)

             % (22,1.63)
             % (23,1.56)
             % (24,1.16)
             % (25,0.90)
             % (26,0.74)
             % (27,0.73)
             % (28,0.40) 
             % (29,0.39)
             % (30,0.22)
             (3,0)%1
             (4,0.036)
             (5, 0)%1
             (6, 0)%1
             (7, 0)%1
             (8,0.008)
             (9,0)%1
             (10,0)%1
             (11,0)%1
             (12,0.004)
             % (13,0)%1
             % (14,0)%1
        };
%\addlegendentry{\texttt{FTCP}}

%\addlegendentry{\texttt{FairRF}}
        % \legend{Otherwise}
        \end{axis}

    \end{tikzpicture}
    % \caption{The runtime in all dataset on Causal Graph.}
    % \label{fig:runtime}
% \end{figure}
    \end{subfigure}
     ~\hfill
    \begin{subfigure}[t!]{0.45\columnwidth}
        \centering        
        % \begin{figure}
%     \centering
    \begin{tikzpicture}
        \begin{axis}[
            ybar,
            axis x line  = bottom,
            axis y line  = left  ,
             reverse legend,
         bar width=.3cm,
            width=6cm,
            height=2.3cm,
               enlarge x limits=0.1,
            ylabel={Pr($\mathcal{C}$)},
            symbolic x coords={3, 4, 5, 6, 7, 8, 9, 10, 11, 12, 13, 14},
            xtick=data,
            xticklabels=\empty,
                 ytick={0,0.02,0.04},
                 scaled y ticks={base 10:1},
            ymax=0.04,
            ymin=0,
            legend style={at={(0.1,1.15)},draw=none,fill=none,anchor=north,legend columns=1},
            legend cell align={left},
           % y tick label style={rotate=45,anchor=east},
            % legend style={at={(0.9,1)},
            %     anchor=north,legend columns=-1},
            % enlarge x limits={abs=0.65cm},
            % nodes near coords,
            % point meta=explicit symbolic
        ]
        \addplot+[fill=teal!50,draw=black
            ] plot coordinates {
             % (1,39.37)
             % (2,30.27)
             % (3,20.27)
             % (4,20.27)
             % (5,20.27) 
             % (6,20.27)
             % (7,20.27)
             % (8,20.27) 
             % (9,39.37)
             % (10,30.27)
             % (11,20.27)
             % (12,20.27)
             % (13,20.27) 
             % (14,20.27)
             % (15,20.27)
             % (16,20.27) 
             % (17,30.27)
             % (18,20.27)
             % (19,20.27)
             % (20,20.27) 
             % (21,20.27)
             % (22,20.27)
             % (23,20.27) 
             % (24,20.27) 
             % (25,30.27)
             % (26,20.27)
             % (27,20.27)
             % (28,20.27) 
             % (29,20.27)
             % (30,20.27)
             (3,0.001)
             (4,0.005)
             (5, 0.036)
             (6, 0.010)%1
             (7, 0.001)%1
             (8,0.001)
             (9,0.001)
             (10,0.002)
             (11,0.005)
             (12,0.001)
             % (13,0.003)%1
             % (14,0.004)%1
             % (11,10.12)
             % (12,9.98)
             % (13,7.85)
             % (14,6.81)
             % (15,3.29)
             % (16,3.27) 
             % (17,2.81)
             % (18,2.73)
             % (19,2.69)
             % (20,2.55) 
             % (21,2.51)
             % (22,1.63)
             % (23,1.56)
             % (24,1.16)
             % (25,0.90)
             % (26,0.74)
             % (27,0.73)
             % (28,0.40) 
             % (29,0.39)
             % (30,0.22)
        };
%\addlegendentry{\texttt{FTCP}}

%\addlegendentry{\texttt{FairRF}}
        % \legend{CPBug-related}
        \end{axis}

\node at (3.6,0.6) {\textsc{Gcc}}; 
               \begin{axis}[
            ybar,
            axis x line  = bottom,
            axis y line  = left  ,
             reverse legend,
           bar width=.3cm,
            width=6cm,
            height=2.3cm,
               enlarge x limits=0.1,
            symbolic x coords={3, 4, 5, 6, 7, 8, 9, 10, 11, 12, 13, 14},
            xtick=data,
            xticklabels=\empty,
               ytick={0,0.02,0.04},
               scaled y ticks={base 10:1},
            ymax=0.04,
            ymin=0,
            legend style={at={(0.7,1.15)},draw=none,fill=none,anchor=north,legend columns=1},
            legend cell align={left},
           % y tick label style={rotate=45,anchor=east},
            % legend style={at={(0.9,1)},
            %     anchor=north,legend columns=-1},
            % enlarge x limits={abs=0.65cm},
            % nodes near coords,
            % point meta=explicit symbolic
        ]
        \addplot+[fill=red!50,draw=black
            ] plot coordinates {
             % (1,39.37)
             % (2,30.27)
             % (3,20.27)
             % (4,20.27)
             % (5,20.27) 
             % (6,20.27)
             
             % (12,20.27)
             % (13,20.27) 
             % (14,20.27)
             % (15,20.27)
             % (16,20.27) 
             % (17,30.27)
             % (18,20.27)
           
             % (26,20.27)
             % (27,20.27)
             % (28,20.27) 
             % (29,20.27)
             % (30,20.27)
             % (1,99.99)

             % (2,99.99)
             % (3,99.99)
             % (4,99.99)
             % (5,99.99) 
             % (6,99.99)
             % (7,80.01)
             % (8,31.06)
             % (9,26.07)
             % (10,19.28)
             % (11,10.12)
             % (12,9.98)

             % (14,6.81)
             % (15,3.29)

             % (17,2.81)
             % (18,2.73)
             % (19,2.69)
             % (20,2.55)

             % (22,1.63)
             % (23,1.56)
             % (24,1.16)
             % (25,0.90)
             % (26,0.74)
             % (27,0.73)
             % (28,0.40) 
             % (29,0.39)
             % (30,0.22)
             (3,0.001)
             (4,0.005)
             (5, 0.036)
             (6, 0)%1
             (7, 0)%1
             (8,0.001)
             (9,0.001)
             (10,0.002)
             (11,0.005)
             (12,0.001)
             % (13,0)%1
             % (14,0)%1
        };
%\addlegendentry{\texttt{FTCP}}

%\addlegendentry{\texttt{FairRF}}
       %  \legend{Otherwise}
        \end{axis}

    \end{tikzpicture}
    % \caption{The runtime in all dataset on Causal Graph.}
    % \label{fig:runtime}
% \end{figure}
    \end{subfigure}
\vspace{-0.15cm}

         \begin{subfigure}[t!]{0.45\columnwidth}
        \centering
        % \begin{figure}
%     \centering
    \begin{tikzpicture}
        \begin{axis}[
            ybar,
            axis x line  = bottom,
            axis y line  = left  ,
             reverse legend,
         bar width=.3cm,
            width=6cm,
            height=2.3cm,
               enlarge x limits=0.1,
                ylabel={Pr($\mathcal{C}$)},
            symbolic x coords={6, 7, 8, 9, 10, 11, 12, 13, 14, 15},
            xtick=data,
            xticklabels=\empty,
             ytick={0,0.01,0.02},
              scaled y ticks={base 10:1},
            ymax=0.02,
            ymin=0,
            legend style={at={(0.1,1.15)},draw=none,fill=none,anchor=north,legend columns=1},
            legend cell align={left},
           % y tick label style={rotate=45,anchor=east},
            % legend style={at={(0.9,1)},
            %     anchor=north,legend columns=-1},
            % enlarge x limits={abs=0.65cm},
            % nodes near coords,
            % point meta=explicit symbolic
        ]
        \addplot+[fill=teal!50,draw=black
            ] plot coordinates {
             % (1,39.37)
             % (2,30.27)
             % (3,20.27)
             % (4,20.27)
             % (5,20.27) 
             % (6,20.27)
             % (7,20.27)
             % (8,20.27) 
             % (9,39.37)
             % (10,30.27)
             % (11,20.27)
             % (12,20.27)
             % (13,20.27) 
             % (14,20.27)
             % (15,20.27)
             % (16,20.27) 
             % (17,30.27)
             % (18,20.27)
             % (19,20.27)
             % (20,20.27) 
             % (21,20.27)
             % (22,20.27)
             % (23,20.27) 
             % (24,20.27) 
             % (25,30.27)
             % (26,20.27)
             % (27,20.27)
             % (28,20.27) 
             % (29,20.27)
             % (30,20.27)
             (6, 0.020)%1
             (7, 0.010)%1
             (8,0.008)%1
             (9,0.008)%1
             (10,0.007)
             (11,0.005)
             (12,0.004)
             (13,0.004)%1
             (14,0.004)%1
             (15,0.003)
             % (11,10.12)
             % (12,9.98)
             % (13,7.85)
             % (14,6.81)
             % (15,3.29)
             % (16,3.27) 
             % (17,2.81)
             % (18,2.73)
             % (19,2.69)
             % (20,2.55) 
             % (21,2.51)
             % (22,1.63)
             % (23,1.56)
             % (24,1.16)
             % (25,0.90)
             % (26,0.74)
             % (27,0.73)
             % (28,0.40) 
             % (29,0.39)
             % (30,0.22)
        };
%\addlegendentry{\texttt{FTCP}}

%\addlegendentry{\texttt{FairRF}}
         %\legend{CPBug-related}
        \end{axis}
\node at (3.6,0.6) {\textsc{Clang}}; 

               \begin{axis}[
            ybar,
            axis x line  = bottom,
            axis y line  = left  ,
             reverse legend,
           bar width=.3cm,
            width=6cm,
            height=2.3cm,
               enlarge x limits=0.1,
            symbolic x coords={6, 7, 8, 9, 10, 11, 12, 13, 14, 15},
            xtick=data,
            xticklabels=\empty,
              ytick={0,0.01,0.02},
                scaled y ticks={base 10:1},
            ymax=0.02,
            ymin=0,
            legend style={at={(0.7,1.15)},draw=none,fill=none,anchor=north,legend columns=1},
            legend cell align={left},
           % y tick label style={rotate=45,anchor=east},
            % legend style={at={(0.9,1)},
            %     anchor=north,legend columns=-1},
            % enlarge x limits={abs=0.65cm},
            % nodes near coords,
            % point meta=explicit symbolic
        ]
        \addplot+[fill=red!50,draw=black
            ] plot coordinates {
             % (1,39.37)
             % (2,30.27)
             % (3,20.27)
             % (4,20.27)
             % (5,20.27) 
             % (6,20.27)
             
             % (12,20.27)
             % (13,20.27) 
             % (14,20.27)
             % (15,20.27)
             % (16,20.27) 
             % (17,30.27)
             % (18,20.27)
           
             % (26,20.27)
             % (27,20.27)
             % (28,20.27) 
             % (29,20.27)
             % (30,20.27)
             % (1,99.99)

             % (2,99.99)
             % (3,99.99)
             % (4,99.99)
             % (5,99.99) 
             % (6,99.99)
             % (7,80.01)
             % (8,31.06)
             % (9,26.07)
             % (10,19.28)
             % (11,10.12)
             % (12,9.98)

             % (14,6.81)
             % (15,3.29)

             % (17,2.81)
             % (18,2.73)
             % (19,2.69)
             % (20,2.55)

             % (22,1.63)
             % (23,1.56)
             % (24,1.16)
             % (25,0.90)
             % (26,0.74)
             % (27,0.73)
             % (28,0.40) 
             % (29,0.39)
             % (30,0.22)
             (6, 0)%1
             (7, 0)%1
             (8,0)%1
             (9,0)%1
             (10,0.007)
             (11,0.005)
             (12,0.004)
             (13,0)%1
             (14,0)%1
             (15,0.003)
        };
%\addlegendentry{\texttt{FTCP}}

%\addlegendentry{\texttt{FairRF}}
        % \legend{Otherwise}
        \end{axis}

    \end{tikzpicture}
    % \caption{The runtime in all dataset on Causal Graph.}
    % \label{fig:runtime}
% \end{figure}
    \end{subfigure}
     ~\hfill
    \begin{subfigure}[t!]{0.45\columnwidth}
        \centering        
        % \begin{figure}
%     \centering
    \begin{tikzpicture}
        \begin{axis}[
            ybar,
            axis x line  = bottom,
            axis y line  = left  ,
             reverse legend,
         bar width=.3cm,
            width=6cm,
            height=2.3cm,
               enlarge x limits=0.1,
                  ylabel={Pr($\mathcal{C}$)},
            symbolic x coords={6, 7, 8, 9, 10, 11, 12, 13, 14, 15},
            xtick=data,
            xticklabels=\empty,
             ytick={0,0.02,0.04},
             scaled y ticks={base 10:1},
            ymax=0.04,
            ymin=0,
            legend style={at={(0.1,1.15)},draw=none,fill=none,anchor=north,legend columns=1},
            legend cell align={left},
           % y tick label style={rotate=45,anchor=east},
            % legend style={at={(0.9,1)},
            %     anchor=north,legend columns=-1},
            % enlarge x limits={abs=0.65cm},
            % nodes near coords,
            % point meta=explicit symbolic
        ]
        \addplot+[fill=teal!50,draw=black
            ] plot coordinates {
             % (1,39.37)
             % (2,30.27)
             % (3,20.27)
             % (4,20.27)
             % (5,20.27) 
             % (6,20.27)
             % (7,20.27)
             % (8,20.27) 
             % (9,39.37)
             % (10,30.27)
             % (11,20.27)
             % (12,20.27)
             % (13,20.27) 
             % (14,20.27)
             % (15,20.27)
             % (16,20.27) 
             % (17,30.27)
             % (18,20.27)
             % (19,20.27)
             % (20,20.27) 
             % (21,20.27)
             % (22,20.27)
             % (23,20.27) 
             % (24,20.27) 
             % (25,30.27)
             % (26,20.27)
             % (27,20.27)
             % (28,20.27) 
             % (29,20.27)
             % (30,20.27)
             % (6, 0.0002)%1
             % (7, 0.0006)%1
             % (8,0.0004)%1
             % (9,0.0005)%1
             % (10,0.0002)
             % (11,0.036)
             % (12,0.0002)
             % (13,0.0003)%1
             % (14,0.007)%1
             % (15,0.002)
             (6, 0.002)
             (7, 0.006)
             (8,0.004)
             (9,0.005)
             (10,0.002)
             (11,0.036)
             (12,0.002)
             (13,0.003)
             (14,0.007)
             (15,0.002)
             % (11,10.12)
             % (12,9.98)
             % (13,7.85)
             % (14,6.81)
             % (15,3.29)
             % (16,3.27) 
             % (17,2.81)
             % (18,2.73)
             % (19,2.69)
             % (20,2.55) 
             % (21,2.51)
             % (22,1.63)
             % (23,1.56)
             % (24,1.16)
             % (25,0.90)
             % (26,0.74)
             % (27,0.73)
             % (28,0.40) 
             % (29,0.39)
             % (30,0.22)
        };
%\addlegendentry{\texttt{FTCP}}

%\addlegendentry{\texttt{FairRF}}
        % \legend{CPBug-related}
        \end{axis}

\node at (3.6,0.6) {\textsc{Clang}}; 
               \begin{axis}[
            ybar,
            axis x line  = bottom,
            axis y line  = left  ,
             reverse legend,
           bar width=.3cm,
            width=6cm,
            height=2.3cm,
               enlarge x limits=0.1,
            symbolic x coords={6, 7, 8, 9, 10, 11, 12, 13, 14, 15},
            xtick=data,
            xticklabels=\empty,
                ytick={0,0.02,0.04},
                scaled y ticks={base 10:1},
            ymax=0.04,
            ymin=0,
            legend style={at={(0.7,1.15)},draw=none,fill=none,anchor=north,legend columns=1},
            legend cell align={left},
           % y tick label style={rotate=45,anchor=east},
            % legend style={at={(0.9,1)},
            %     anchor=north,legend columns=-1},
            % enlarge x limits={abs=0.65cm},
            % nodes near coords,
            % point meta=explicit symbolic
        ]
        \addplot+[fill=red!50,draw=black
            ] plot coordinates {
             % (1,39.37)
             % (2,30.27)
             % (3,20.27)
             % (4,20.27)
             % (5,20.27) 
             % (6,20.27)
             
             % (12,20.27)
             % (13,20.27) 
             % (14,20.27)
             % (15,20.27)
             % (16,20.27) 
             % (17,30.27)
             % (18,20.27)
           
             % (26,20.27)
             % (27,20.27)
             % (28,20.27) 
             % (29,20.27)
             % (30,20.27)
             % (1,99.99)

             % (2,99.99)
             % (3,99.99)
             % (4,99.99)
             % (5,99.99) 
             % (6,99.99)
             % (7,80.01)
             % (8,31.06)
             % (9,26.07)
             % (10,19.28)
             % (11,10.12)
             % (12,9.98)

             % (14,6.81)
             % (15,3.29)

             % (17,2.81)
             % (18,2.73)
             % (19,2.69)
             % (20,2.55)

             % (22,1.63)
             % (23,1.56)
             % (24,1.16)
             % (25,0.90)
             % (26,0.74)
             % (27,0.73)
             % (28,0.40) 
             % (29,0.39)
             % (30,0.22)
             (6, 0.002)
             (7, 0.006)
             (8,0.004)
             (9,0.005)
             (10,0.002)
             (11,0.036)
             (12,0.002)
             (13,0.003)
             (14,0.007)
             (15,0.002)
        };
%\addlegendentry{\texttt{FTCP}}

%\addlegendentry{\texttt{FairRF}}
        % \legend{Otherwise}
        \end{axis}

    \end{tikzpicture}
    % \caption{The runtime in all dataset on Causal Graph.}
    % \label{fig:runtime}
% \end{figure}
    \end{subfigure}
\vspace{-0.15cm}

     \begin{subfigure}[t!]{0.45\columnwidth}
      \centering  
        \caption{Prioritization in \approach}
    \end{subfigure}
     ~\hfill
    \begin{subfigure}[t!]{0.45\columnwidth}
        \centering        
         \caption{Random in \texttt{CPD} and \texttt{US}}
    \end{subfigure}
    \caption{\revision{Order of options to be tested for all systems (left to right). Pr($\mathcal{C}$) denotes the probability of being CPBug-related.}}
    \label{fig:rank}
      \vspace{-0.2cm}
\end{figure}

\subsection{Prioritizing Options}

\revision{To understand why prioritizing at the options level helps \approach~to significantly improve the testing efficiency, Figure~\ref{fig:rank} plots the first few options to be tested by all approaches. We see that, with \approach, the prioritized options generally have higher probabilities of being CPBug-related than those of the other two, within which up to 60\% options can trigger CPBugs (for \textsc{MySQL}) while only one option from \texttt{CPD} and \texttt{US} can do so (for \textsc{Gcc}). This considerably impacts the CPBug testing.}

%in \approach~contains 60\% CPBug-related options (6 out of 10), including most higher probability ones; while the randomly selected first 10 options to be tested in \texttt{CPD} involve no CPBug-related ones at all and are with very low probabilities. This would create a considerable impact on the CPBug testing.

\subsection{Prioritizing Search}

We also analyze why prioritizing the different bounds of search space when testing the numeric options can help. We found that the extreme search and midmost search are of great benefit therein. For example, \verb|innodb_buffer_pool_size| is a CPBug-related numeric option on \textsc{MySQL}. Yet, such a CPBug can only be discovered when we change it from a near-minimal value, i.e., 10MB, to one that is closer to its maximum value, i.e., 256GB. With \approach, such an option is categorized as the high-density region (\textbf{\textit{Characteristic 2}} and \textbf{\textit{4}}), thus \approach~would prioritize its bounds as extreme search first. This fits perfectly with its range of values that causes a CPBug, since with the extreme search, a configuration in a pair would be explored within 10\% close to its minimum value while the other would take a value close to 10\% of its maximum extreme. In contrast, \texttt{CPD} would fix one configuration to the option value of 1MB while increasing the other as 2MB, 4MB, and 8MB, etc, each pair of which needs to be tested. Unlike the others, \texttt{US} does not use a heuristic as it aims to sample randomly and uniformly. Therefore, for options like \verb|innodb_buffer_pool_size|, \approach~needs significantly less number of tests to reveal the CPBugs compared with the others, which might even fail to find the CPBugs due to exhaustion of budget.

\section{Threats to Validity}
\label{sec:tov}

\textbf{Threats to internal validity:} We set the parameters either adopting pragmatic values or following widely-used defaults, e.g., the inner budget of 100 tests for GA under a bound is a pragmatic setting, achieving a good balance between quality and cost. For confirming performance drop in the oracle, we set a minimum of 5\% change as prior work~\cite{DBLP:conf/sigsoft/ChenCWYHM23}. However, we agree that some settings might not be the best.

%In our approach, the actual performance tendencies of configuration options are derived from actual tests. We determine whether a performance tendency is present by calculating if the performance gap between configuration pairs reaches 5\%. This division may be too strict. For example, actual tests may be affected by other unknown system noises, resulting in performance gaps of 4\% and 6\% between different configuration pairs in a no-impact state. The former would be judged as having no performance tendency, while the latter would be judged as having a specific tendency. Even though the difference between them is small, they are classified into two different states. In future research, we will attempt more techniques for fuzzy boundary states.
 
\textbf{Threats to external validity:} For evaluating the estimation of CPBugs types, we use prior datasets of 12 systems~\cite{DBLP:conf/kbse/HeJLXYYWL20}. For testing the CPBugs, we use five systems with reproduced CPBugs. In both cases, the systems are of diverse languages, domains, and scales. \revision{We have also considered a wide range of workloads (2--10) and versions (4--31), which are the most commonly used ones from existing work~\cite{DBLP:conf/kbse/HeJLXYYWL20,tongkaw2016comparison,DBLP:conf/sigmod/CurinoJMB11,DBLP:journals/jss/LesoilABJ23}}. Indeed, more subjects might strengthen the conclusion. 

%{\color{black} Notably, default hyperparameters were used in this study, hence leaving their tuning for future work.}

%To evaluate our approach, we conducted experiments on five different large-scale software systems (i.e., MySQL, MariaDB, GCC, Clang, and httpd), which had already been evaluated in previous performance engineering studies. These systems target different fields: MySQL and MariaDB are database systems, GCC and Clang are compiler systems, and httpd is a web proxy system. They all use the same type of programming language, C/C++. Future research should involve software systems in multiple programming languages and more domains. In our experiments, we mainly used low, medium, and high workloads for testing. In reality, since real-world CPBugs are triggered using multiple benchmarks running multiple test cases, we cannot ensure that each test case runs under similar workloads. Future research should use other load test drivers and more complex workloads to evaluate our approach.

\textbf{Threats to construct validity:} We use several metrics, including precision, recall, and F1 score, together with the trajectory of finding CPBugs and the best efficiency of each approach. Yet, unintended programming errors or mis-considerations are always possible. 

%\revision{We also used the most commonly studied workloads for the systems in existing work ~\cite{DBLP:conf/kbse/HeJLXYYWL20,tongkaw2016comparison,DBLP:conf/sigmod/CurinoJMB11,DBLP:journals/jss/LesoilABJ23}. Indeed, we cannot ensure this is exhaustive, which might prevent the discovery of certain CPBugs.}

%In this experiment, although we tested dozens of metrics, most came from the type of test time, lacking discussion on multiple types of test metrics. Future research should use more metrics to evaluate our approach, combining various aspects of software systems (such as type, environment, and programming language) to select more suitable performance metrics.

\section{Related Work}
\label{sec:related}

\subsection{Implication of Configuration to Performance Issues}

A vast amount of early work has been conducted to understand the implications of configurations for performance issues. For example, Jin et al.~\cite{DBLP:conf/pldi/JinSSSL12} and Han et al.~\cite{DBLP:conf/esem/HanY16} reveal that 59\% of the performance problems can be traced back to configuration errors. Xiang et al.\cite{DBLP:conf/usenix/XiangHY0P20} further suggest that configuration option documentation is a significant resource for analyzing configuration-related performance expectations, which serve as a foundation for identifying CPBug oracle. Those studies provide insights into how configuration caused performance issues while \approach~automatically testing CPBugs.

\subsection{Performance Bug Testing}

There exist tools that detect general performance bugs using a fixed set of patterns, such as loops and memory access~\cite{DBLP:conf/icse/NistorSML13, DBLP:conf/icse/SongL17, DBLP:conf/icse/SuWYC019, DBLP:conf/icse/YangSLYC18, DBLP:conf/issta/YuP16,DBLP:conf/pldi/CoppaDF12}. To tackle unforeseen bottleneck patterns, Shen et al.~\cite{DBLP:conf/issta/ShenLPG15} propose a GA-based testing framework with contrast data mining. However, they are not related to configurations.

% to explore software inputs for finding performance bottlenecks.

%Similarly, Coppa et al.~\cite{DBLP:conf/pldi/CoppaDF12} propose a tool that detects certain patterns that cause performance bottlenecks.

Among configuration-related testing approaches, \texttt{ctest}~\cite{DBLP:conf/issta/ChengZMX21} is a tool that leverages existing regression testing code to prioritize the execution of test cases for misconfiguration-related performance issues. \texttt{DiagConfig}~\cite{DBLP:conf/sigsoft/ChenCWYHM23} leverages static code analysis and machine learning to detect performance bugs caused by misconfiguration. Yet, they aim for misconfiguration, which is user-induced performance issues while \approach~reveal CPBugs---the configuration performance issues that are unintentionally introduced by the developers of configurable systems. This work also advances \texttt{CPD}~\cite{DBLP:conf/kbse/HeJLXYYWL20}---a state-of-the-art CPBug testing tool---in several aspects:

%a CPBug testing tool that relies on rule mining and keyword search to estimate CPBug type and adopts an exponentially increased search strategy for testing the numeric options.

\begin{itemize}
    \item \revision{We additionally summarize \textbf{\textit{Characteristic 3}} (on the commonality of median range value for CPBug-related numeric options) and \textbf{\textit{Characteristic 4}} (on the more detailed categorization of CPBug-related numeric options), which have not been revealed by the work of \texttt{CPD}.}
    
    \item \revision{\texttt{CPD} predicts the oracle using rule mining and keyword search while \approach~does so via a RoBERTa, fine-tuned by configuration documentation. This, as shown in Section~\ref{sec:rq1}, has led to much superior accuracy.}
    
    \item \revision{\texttt{CPD} does not prioritize the options to be tested and a numeric option is tested by fixing it in one configuration as maximal/minimal value, while exponentially changing the value of the same option in the other configuration. In contrast, through exploiting the other RoBERTa fine-tuned by both documentation and code, \approach~designs dual-level prioritization that (1) prioritizes the options that are more likely to cause CPBugs to be tested first while (2) stochastically exploring the values of a numeric option in the pair using differently prioritized search bounds, according to the likelihood of the option being CPBugs-related and the observations from the \textbf{\textit{Characteristics 2--4}}. This has resulted in considerably improved efficiency.}
\end{itemize}

%the above work has not explicitly handled the testing order and hence is limited in the efficiency of CPBug testing. \approach, in contrast, significantly expedites the CPBug testing via prioritization at a dual level supported by neural language models.

\subsection{Configuration Performance Tuning}

Unlike testing for configuration-related performance bugs, configuration performance tuning aims to find the best configuration that reaches the optimal performance for deployment time benchmarking~\cite{DBLP:conf/icse/YeChen25,DBLP:journals/tse/Nair0MSA20,DBLP:conf/icse/0003XC021,DBLP:journals/pacmse/0001L24,DBLP:conf/sigsoft/0001L21,DBLP:journals/tse/ChenCL24} or runtime self-adaptation~\cite{DBLP:conf/icse/YeChen25,DBLP:conf/seams/Chen22,DBLP:conf/wcre/Chen22,DBLP:conf/icse/Kumar0BB20,DBLP:conf/icse/Chen19b,DBLP:journals/tosem/ChenLBY18,DBLP:journals/tsc/ChenB17}. Among others, \texttt{FLASH}~\cite{DBLP:journals/tse/Nair0MSA20} and \texttt{BOCA}~\cite{DBLP:conf/icse/0003XC021} are tuners based on Bayesian optimization to find optimal configuration. Chen and Li propose MMO~\cite{DBLP:journals/pacmse/0001L24,DBLP:conf/sigsoft/0001L21,DBLP:journals/tse/ChenCL24}---an alternative way to tune configuration via tuner-agnostic multi-objectivization.

Yet, configuration tuning differs from the testing that \approach~focus on in several aspects: the representation in configuration tuning is often a single configuration while for CPBug testing, we need to test a pair of configurations for revealing whether the actual performance matches the expectation. Further, CPBug testing examines each option in turn while configuration tuning changes several options simultaneously.

\section{Conclusion}
\label{sec:con}

This paper presents \approach, a general framework aiming to expedite CPBug testing via neural dual-level prioritization. \approach~builds two neural language models for estimating the CPBugs types oracle and inferring the probabilities of the options being CPBug-related, respectively. These models serve as the foundation for prioritization at two levels--- prioritizing the order of tested options and the order of search bounds for numeric options. Experiments on several real-world systems and against state-of-the-art tools/approaches reveal that:

\begin{itemize}
    \item \approach~estimates more reliable oracle of CPBug types;
    \item while significantly expedites the testing at both the options level and the search level for numeric options, with up to 1.73$\times$ and 88.88$\times$ speedup, respectively.
\end{itemize}

For future work, the static handling of workloads in \approach~can also be prioritized, placing the more vulnerable ones to be tested first. Extending \approach~to detect CPBugs by testing multiple options simultaneously is also fruitful.

\section*{Acknowledgement}
This work was supported by a NSFC Grant (62372084) and a UKRI Grant (10054084).

\balance

\bibliographystyle{IEEEtran}
\bibliography{mybibf}

\begin{thebibliography}{10}
\providecommand{\url}[1]{#1}
\csname url@samestyle\endcsname
\providecommand{\newblock}{\relax}
\providecommand{\bibinfo}[2]{#2}
\providecommand{\BIBentrySTDinterwordspacing}{\spaceskip=0pt\relax}
\providecommand{\BIBentryALTinterwordstretchfactor}{4}
\providecommand{\BIBentryALTinterwordspacing}{\spaceskip=\fontdimen2\font plus
\BIBentryALTinterwordstretchfactor\fontdimen3\font minus \fontdimen4\font\relax}
\providecommand{\BIBforeignlanguage}[2]{{%
\expandafter\ifx\csname l@#1\endcsname\relax
\typeout{** WARNING: IEEEtran.bst: No hyphenation pattern has been}%
\typeout{** loaded for the language `#1'. Using the pattern for}%
\typeout{** the default language instead.}%
\else
\language=\csname l@#1\endcsname
\fi
#2}}
\providecommand{\BIBdecl}{\relax}
\BIBdecl

\bibitem{10.1145/3702986}
\BIBentryALTinterwordspacing
J.~Gong and T.~Chen, ``Deep configuration performance learning: A systematic survey and taxonomy,'' \emph{ACM Trans. Softw. Eng. Methodol.}, vol.~34, no.~1, Dec. 2024. [Online]. Available: \url{https://doi.org/10.1145/3702986}
\BIBentrySTDinterwordspacing

\bibitem{DBLP:conf/kbse/HeJLXYYWL20}
\BIBentryALTinterwordspacing
H.~He, Z.~Jia, S.~Li, E.~Xu, T.~Yu, Y.~Yu, J.~Wang, and X.~Liao, ``Cp-detector: Using configuration-related performance properties to expose performance bugs,'' in \emph{35th {IEEE/ACM} International Conference on Automated Software Engineering, {ASE} 2020, Melbourne, Australia, September 21-25, 2020}.\hskip 1em plus 0.5em minus 0.4em\relax {IEEE}, 2020, pp. 623--634. [Online]. Available: \url{https://doi.org/10.1145/3324884.3416531}
\BIBentrySTDinterwordspacing

\bibitem{DBLP:journals/tse/ChenGongChen25}
P.~Chen, J.~Gong, and T.~Chen, ``Accuracy can lie: On the impact of surrogate model in configuration tuning,'' \emph{{IEEE} Trans. Software Eng.}, 2025.

\bibitem{DBLP:journals/tse/GongChen25}
J.~Gong, T.~Chen, and R.~Bahsoon, ``Dividable configuration performance learning,'' \emph{{IEEE} Trans. Software Eng.}, vol.~51, no.~1, pp. 106--134, 2025.

\bibitem{DBLP:conf/sigsoft/Gong023}
\BIBentryALTinterwordspacing
J.~Gong and T.~Chen, ``Predicting software performance with divide-and-learn,'' in \emph{Proceedings of the 31st {ACM} Joint European Software Engineering Conference and Symposium on the Foundations of Software Engineering, {ESEC/FSE} 2023, San Francisco, CA, USA, December 3-9, 2023}.\hskip 1em plus 0.5em minus 0.4em\relax {ACM}, 2023, pp. 858--870. [Online]. Available: \url{https://doi.org/10.1145/3611643.3616334}
\BIBentrySTDinterwordspacing

\bibitem{DBLP:journals/tosem/ChenL23a}
\BIBentryALTinterwordspacing
T.~Chen and M.~Li, ``Do performance aspirations matter for guiding software configuration tuning? an empirical investigation under dual performance objectives,'' \emph{{ACM} Trans. Softw. Eng. Methodol.}, vol.~32, no.~3, pp. 68:1--68:41, 2023. [Online]. Available: \url{https://doi.org/10.1145/3571853}
\BIBentrySTDinterwordspacing

\bibitem{DBLP:journals/pacmse/Gong024}
\BIBentryALTinterwordspacing
J.~Gong and T.~Chen, ``Predicting configuration performance in multiple environments with sequential meta-learning,'' \emph{Proc. {ACM} Softw. Eng.}, vol.~1, no. {FSE}, pp. 359--382, 2024. [Online]. Available: \url{https://doi.org/10.1145/3643743}
\BIBentrySTDinterwordspacing

\bibitem{DBLP:conf/issta/ChengZMX21}
\BIBentryALTinterwordspacing
R.~Cheng, L.~Zhang, D.~Marinov, and T.~Xu, ``Test-case prioritization for configuration testing,'' in \emph{{ISSTA} '21: 30th {ACM} {SIGSOFT} International Symposium on Software Testing and Analysis, Virtual Event, Denmark, July 11-17, 2021}, C.~Cadar and X.~Zhang, Eds.\hskip 1em plus 0.5em minus 0.4em\relax {ACM}, 2021, pp. 452--465. [Online]. Available: \url{https://doi.org/10.1145/3460319.3464810}
\BIBentrySTDinterwordspacing

\bibitem{DBLP:conf/icse/VelezJSAK22}
\BIBentryALTinterwordspacing
M.~Velez, P.~Jamshidi, N.~Siegmund, S.~Apel, and C.~K{\"{a}}stner, ``On debugging the performance of configurable software systems: Developer needs and tailored tool support,'' in \emph{44th {IEEE/ACM} 44th International Conference on Software Engineering, {ICSE} 2022, Pittsburgh, PA, USA, May 25-27, 2022}.\hskip 1em plus 0.5em minus 0.4em\relax {ACM}, 2022, pp. 1571--1583. [Online]. Available: \url{https://doi.org/10.1145/3510003.3510043}
\BIBentrySTDinterwordspacing

\bibitem{DBLP:conf/icse/WangJLZYXPL23}
\BIBentryALTinterwordspacing
T.~Wang, Z.~Jia, S.~Li, S.~Zheng, Y.~Yu, E.~Xu, S.~Peng, and X.~Liao, ``Understanding and detecting on-the-fly configuration bugs,'' in \emph{45th {IEEE/ACM} International Conference on Software Engineering, {ICSE} 2023, Melbourne, Australia, May 14-20, 2023}.\hskip 1em plus 0.5em minus 0.4em\relax {IEEE}, 2023, pp. 628--639. [Online]. Available: \url{https://doi.org/10.1109/ICSE48619.2023.00062}
\BIBentrySTDinterwordspacing

\bibitem{nygard2018release}
M.~Nygard, ``Release it!: design and deploy production-ready software,'' 2018.

\bibitem{DBLP:series/synthesis/2018Barroso}
\BIBentryALTinterwordspacing
L.~A. Barroso, U.~H{\"{o}}lzle, and P.~Ranganathan, \emph{The Datacenter as a Computer: Designing Warehouse-Scale Machines, Third Edition}, ser. Synthesis Lectures on Computer Architecture.\hskip 1em plus 0.5em minus 0.4em\relax Morgan {\&} Claypool Publishers, 2018. [Online]. Available: \url{https://doi.org/10.2200/S00874ED3V01Y201809CAC046}
\BIBentrySTDinterwordspacing

\bibitem{DBLP:conf/sosp/TangKVCWNDK15}
\BIBentryALTinterwordspacing
C.~Tang, T.~Kooburat, P.~Venkatachalam, A.~Chander, Z.~Wen, A.~Narayanan, P.~Dowell, and R.~Karl, ``Holistic configuration management at facebook,'' in \emph{Proceedings of the 25th Symposium on Operating Systems Principles, {SOSP} 2015, Monterey, CA, USA, October 4-7, 2015}, E.~L. Miller and S.~Hand, Eds.\hskip 1em plus 0.5em minus 0.4em\relax {ACM}, 2015, pp. 328--343. [Online]. Available: \url{https://doi.org/10.1145/2815400.2815401}
\BIBentrySTDinterwordspacing

\bibitem{DBLP:conf/sigsoft/XuJFZPT15}
\BIBentryALTinterwordspacing
T.~Xu, L.~Jin, X.~Fan, Y.~Zhou, S.~Pasupathy, and R.~Talwadker, ``Hey, you have given me too many knobs!: understanding and dealing with over-designed configuration in system software,'' in \emph{Proceedings of the 2015 10th Joint Meeting on Foundations of Software Engineering, {ESEC/FSE} 2015, Bergamo, Italy, August 30 - September 4, 2015}, E.~D. Nitto, M.~Harman, and P.~Heymans, Eds.\hskip 1em plus 0.5em minus 0.4em\relax {ACM}, 2015, pp. 307--319. [Online]. Available: \url{https://doi.org/10.1145/2786805.2786852}
\BIBentrySTDinterwordspacing

\bibitem{DBLP:journals/pacmse/0001L24}
\BIBentryALTinterwordspacing
T.~Chen and M.~Li, ``Adapting multi-objectivized software configuration tuning,'' \emph{Proc. {ACM} Softw. Eng.}, vol.~1, no. {FSE}, pp. 539--561, 2024. [Online]. Available: \url{https://doi.org/10.1145/3643751}
\BIBentrySTDinterwordspacing

\bibitem{DBLP:conf/icst/PlazarAPDC19}
\BIBentryALTinterwordspacing
Q.~Plazar, M.~Acher, G.~Perrouin, X.~Devroey, and M.~Cordy, ``Uniform sampling of {SAT} solutions for configurable systems: Are we there yet?'' in \emph{12th {IEEE} Conference on Software Testing, Validation and Verification, {ICST} 2019, Xi'an, China, April 22-27, 2019}.\hskip 1em plus 0.5em minus 0.4em\relax {IEEE}, 2019, pp. 240--251. [Online]. Available: \url{https://doi.org/10.1109/ICST.2019.00032}
\BIBentrySTDinterwordspacing

\bibitem{DBLP:conf/icse/NistorSML13}
\BIBentryALTinterwordspacing
A.~Nistor, L.~Song, D.~Marinov, and S.~Lu, ``Toddler: detecting performance problems via similar memory-access patterns,'' in \emph{35th International Conference on Software Engineering, {ICSE} '13, San Francisco, CA, USA, May 18-26, 2013}, D.~Notkin, B.~H.~C. Cheng, and K.~Pohl, Eds.\hskip 1em plus 0.5em minus 0.4em\relax {IEEE} Computer Society, 2013, pp. 562--571. [Online]. Available: \url{https://doi.org/10.1109/ICSE.2013.6606602}
\BIBentrySTDinterwordspacing

\bibitem{DBLP:journals/corr/abs-1907-11692}
\BIBentryALTinterwordspacing
Y.~Liu, M.~Ott, N.~Goyal, J.~Du, M.~Joshi, D.~Chen, O.~Levy, M.~Lewis, L.~Zettlemoyer, and V.~Stoyanov, ``Roberta: {A} robustly optimized {BERT} pretraining approach,'' \emph{CoRR}, vol. abs/1907.11692, 2019. [Online]. Available: \url{http://arxiv.org/abs/1907.11692}
\BIBentrySTDinterwordspacing

\bibitem{DBLP:conf/sigsoft/ChenCWYHM23}
\BIBentryALTinterwordspacing
Z.~Chen, P.~Chen, P.~Wang, G.~Yu, Z.~He, and G.~Mai, ``Diagconfig: Configuration diagnosis of performance violations in configurable software systems,'' in \emph{Proceedings of the 31st {ACM} Joint European Software Engineering Conference and Symposium on the Foundations of Software Engineering, {ESEC/FSE} 2023, San Francisco, CA, USA, December 3-9, 2023}, S.~Chandra, K.~Blincoe, and P.~Tonella, Eds.\hskip 1em plus 0.5em minus 0.4em\relax {ACM}, 2023, pp. 566--578. [Online]. Available: \url{https://doi.org/10.1145/3611643.3616300}
\BIBentrySTDinterwordspacing

\bibitem{DBLP:conf/icse/MuhlbauerSKDAS23}
\BIBentryALTinterwordspacing
S.~M{\"{u}}hlbauer, F.~Sattler, C.~Kaltenecker, J.~Dorn, S.~Apel, and N.~Siegmund, ``Analysing the impact of workloads on modeling the performance of configurable software systems,'' in \emph{45th {IEEE/ACM} International Conference on Software Engineering, {ICSE} 2023, Melbourne, Australia, May 14-20, 2023}.\hskip 1em plus 0.5em minus 0.4em\relax {IEEE}, 2023, pp. 2085--2097. [Online]. Available: \url{https://doi.org/10.1109/ICSE48619.2023.00176}
\BIBentrySTDinterwordspacing

\bibitem{DBLP:journals/nlpj/RoumeliotisTN24}
\BIBentryALTinterwordspacing
K.~I. Roumeliotis, N.~D. Tselikas, and D.~K. Nasiopoulos, ``Llms in e-commerce: {A} comparative analysis of {GPT} and llama models in product review evaluation,'' \emph{Nat. Lang. Process. J.}, vol.~6, p. 100056, 2024. [Online]. Available: \url{https://doi.org/10.1016/j.nlp.2024.100056}
\BIBentrySTDinterwordspacing

\bibitem{DBLP:conf/icsm/ZhangXTH0J20}
\BIBentryALTinterwordspacing
T.~Zhang, B.~Xu, F.~Thung, S.~A. Haryono, D.~Lo, and L.~Jiang, ``Sentiment analysis for software engineering: How far can pre-trained transformer models go?'' in \emph{{IEEE} International Conference on Software Maintenance and Evolution, {ICSME} 2020, Adelaide, Australia, September 28 - October 2, 2020}.\hskip 1em plus 0.5em minus 0.4em\relax {IEEE}, 2020, pp. 70--80. [Online]. Available: \url{https://doi.org/10.1109/ICSME46990.2020.00017}
\BIBentrySTDinterwordspacing

\bibitem{DBLP:conf/mascots/JamshidiC16}
\BIBentryALTinterwordspacing
P.~Jamshidi and G.~Casale, ``An uncertainty-aware approach to optimal configuration of stream processing systems,'' in \emph{24th {IEEE} International Symposium on Modeling, Analysis and Simulation of Computer and Telecommunication Systems, {MASCOTS} 2016, London, United Kingdom, September 19-21, 2016}.\hskip 1em plus 0.5em minus 0.4em\relax {IEEE} Computer Society, 2016, pp. 39--48. [Online]. Available: \url{https://doi.org/10.1109/MASCOTS.2016.17}
\BIBentrySTDinterwordspacing

\bibitem{llvm}
\url{https://clang.llvm.org/docs/LibASTMatchers.html}.

\bibitem{forrest1996genetic}
S.~Forrest, ``Genetic algorithms,'' \emph{ACM computing surveys (CSUR)}, vol.~28, no.~1, pp. 77--80, 1996.

\bibitem{DBLP:conf/icse/LiangChen25}
H.~Liang, Y.~Huang, and T.~Chen, ``The same only different: On information modality for configuration performance analysis,'' in \emph{47th {IEEE/ACM} International Conference on Software Engineering (ICSE)}.\hskip 1em plus 0.5em minus 0.4em\relax {IEEE}, 2025.

\bibitem{DBLP:journals/pvldb/LaoWLWZCCTW24}
\BIBentryALTinterwordspacing
J.~Lao, Y.~Wang, Y.~Li, J.~Wang, Y.~Zhang, Z.~Cheng, W.~Chen, M.~Tang, and J.~Wang, ``Gptuner: {A} manual-reading database tuning system via gpt-guided bayesian optimization,'' \emph{Proc. {VLDB} Endow.}, vol.~17, no.~8, pp. 1939--1952, 2024. [Online]. Available: \url{https://www.vldb.org/pvldb/vol17/p1939-tang.pdf}
\BIBentrySTDinterwordspacing

\bibitem{DBLP:conf/sigsoft/ChenWLLX20}
\BIBentryALTinterwordspacing
Q.~Chen, T.~Wang, O.~Legunsen, S.~Li, and T.~Xu, ``Understanding and discovering software configuration dependencies in cloud and datacenter systems,'' in \emph{{ESEC/FSE} '20: 28th {ACM} Joint European Software Engineering Conference and Symposium on the Foundations of Software Engineering, Virtual Event, USA, November 8-13, 2020}, P.~Devanbu, M.~B. Cohen, and T.~Zimmermann, Eds.\hskip 1em plus 0.5em minus 0.4em\relax {ACM}, 2020, pp. 362--374. [Online]. Available: \url{https://doi.org/10.1145/3368089.3409727}
\BIBentrySTDinterwordspacing

\bibitem{DBLP:conf/sigsoft/SiegmundGAK15}
\BIBentryALTinterwordspacing
N.~Siegmund, A.~Grebhahn, S.~Apel, and C.~K{\"{a}}stner, ``Performance-influence models for highly configurable systems,'' in \emph{Proceedings of the 2015 10th Joint Meeting on Foundations of Software Engineering, {ESEC/FSE} 2015, Bergamo, Italy, August 30 - September 4, 2015}, E.~D. Nitto, M.~Harman, and P.~Heymans, Eds.\hskip 1em plus 0.5em minus 0.4em\relax {ACM}, 2015, pp. 284--294. [Online]. Available: \url{https://doi.org/10.1145/2786805.2786845}
\BIBentrySTDinterwordspacing

\bibitem{DBLP:conf/icse/SiegmundKKABRS12}
\BIBentryALTinterwordspacing
N.~Siegmund, S.~S. Kolesnikov, C.~K{\"{a}}stner, S.~Apel, D.~S. Batory, M.~Rosenm{\"{u}}ller, and G.~Saake, ``Predicting performance via automated feature-interaction detection,'' in \emph{34th International Conference on Software Engineering, {ICSE} 2012, June 2-9, 2012, Zurich, Switzerland}, M.~Glinz, G.~C. Murphy, and M.~Pezz{\`{e}}, Eds.\hskip 1em plus 0.5em minus 0.4em\relax {IEEE} Computer Society, 2012, pp. 167--177. [Online]. Available: \url{https://doi.org/10.1109/ICSE.2012.6227196}
\BIBentrySTDinterwordspacing

\bibitem{DBLP:conf/icse/MedeirosKRGA16}
\BIBentryALTinterwordspacing
F.~Medeiros, C.~K{\"{a}}stner, M.~Ribeiro, R.~Gheyi, and S.~Apel, ``A comparison of 10 sampling algorithms for configurable systems,'' in \emph{Proceedings of the 38th International Conference on Software Engineering, {ICSE} 2016, Austin, TX, USA, May 14-22, 2016}, L.~K. Dillon, W.~Visser, and L.~A. Williams, Eds.\hskip 1em plus 0.5em minus 0.4em\relax {ACM}, 2016, pp. 643--654. [Online]. Available: \url{https://doi.org/10.1145/2884781.2884793}
\BIBentrySTDinterwordspacing

\bibitem{DBLP:journals/tse/ChenCL24}
\BIBentryALTinterwordspacing
P.~Chen, T.~Chen, and M.~Li, ``{MMO:} meta multi-objectivization for software configuration tuning,'' \emph{{IEEE} Trans. Software Eng.}, vol.~50, no.~6, pp. 1478--1504, 2024. [Online]. Available: \url{https://doi.org/10.1109/TSE.2024.3388910}
\BIBentrySTDinterwordspacing

\bibitem{DBLP:journals/tosem/ChenL23}
\BIBentryALTinterwordspacing
T.~Chen and M.~Li, ``The weights can be harmful: Pareto search versus weighted search in multi-objective search-based software engineering,'' \emph{{ACM} Trans. Softw. Eng. Methodol.}, vol.~32, no.~1, pp. 5:1--5:40, 2023. [Online]. Available: \url{https://doi.org/10.1145/3514233}
\BIBentrySTDinterwordspacing

\bibitem{tongkaw2016comparison}
S.~Tongkaw and A.~Tongkaw, ``A comparison of database performance of mariadb and mysql with oltp workload,'' in \emph{2016 IEEE conference on open systems (ICOS)}.\hskip 1em plus 0.5em minus 0.4em\relax IEEE, 2016, pp. 117--119.

\bibitem{DBLP:conf/sigmod/CurinoJMB11}
\BIBentryALTinterwordspacing
C.~Curino, E.~P.~C. Jones, S.~Madden, and H.~Balakrishnan, ``Workload-aware database monitoring and consolidation,'' in \emph{Proceedings of the {ACM} {SIGMOD} International Conference on Management of Data, {SIGMOD} 2011, Athens, Greece, June 12-16, 2011}, T.~K. Sellis, R.~J. Miller, A.~Kementsietsidis, and Y.~Velegrakis, Eds.\hskip 1em plus 0.5em minus 0.4em\relax {ACM}, 2011, pp. 313--324. [Online]. Available: \url{https://doi.org/10.1145/1989323.1989357}
\BIBentrySTDinterwordspacing

\bibitem{DBLP:journals/jss/LesoilABJ23}
\BIBentryALTinterwordspacing
L.~Lesoil, M.~Acher, A.~Blouin, and J.~J{\'{e}}z{\'{e}}quel, ``Input sensitivity on the performance of configurable systems an empirical study,'' \emph{J. Syst. Softw.}, vol. 201, p. 111671, 2023. [Online]. Available: \url{https://doi.org/10.1016/j.jss.2023.111671}
\BIBentrySTDinterwordspacing

\bibitem{DBLP:conf/pldi/JinSSSL12}
\BIBentryALTinterwordspacing
G.~Jin, L.~Song, X.~Shi, J.~Scherpelz, and S.~Lu, ``Understanding and detecting real-world performance bugs,'' in \emph{{ACM} {SIGPLAN} Conference on Programming Language Design and Implementation, {PLDI} '12, Beijing, China - June 11 - 16, 2012}, J.~Vitek, H.~Lin, and F.~Tip, Eds.\hskip 1em plus 0.5em minus 0.4em\relax {ACM}, 2012, pp. 77--88. [Online]. Available: \url{https://doi.org/10.1145/2254064.2254075}
\BIBentrySTDinterwordspacing

\bibitem{DBLP:conf/esem/HanY16}
\BIBentryALTinterwordspacing
X.~Han and T.~Yu, ``An empirical study on performance bugs for highly configurable software systems,'' in \emph{Proceedings of the 10th {ACM/IEEE} International Symposium on Empirical Software Engineering and Measurement, {ESEM} 2016, Ciudad Real, Spain, September 8-9, 2016}.\hskip 1em plus 0.5em minus 0.4em\relax {ACM}, 2016, pp. 23:1--23:10. [Online]. Available: \url{https://doi.org/10.1145/2961111.2962602}
\BIBentrySTDinterwordspacing

\bibitem{DBLP:conf/usenix/XiangHY0P20}
\BIBentryALTinterwordspacing
C.~Xiang, H.~Huang, A.~Yoo, Y.~Zhou, and S.~Pasupathy, ``Pracextractor: Extracting configuration good practices from manuals to detect server misconfigurations,'' in \emph{Proceedings of the 2020 {USENIX} Annual Technical Conference, {USENIX} {ATC} 2020, July 15-17, 2020}, A.~Gavrilovska and E.~Zadok, Eds.\hskip 1em plus 0.5em minus 0.4em\relax {USENIX} Association, 2020, pp. 265--280. [Online]. Available: \url{https://www.usenix.org/conference/atc20/presentation/xiang}
\BIBentrySTDinterwordspacing

\bibitem{DBLP:conf/icse/SongL17}
\BIBentryALTinterwordspacing
L.~Song and S.~Lu, ``Performance diagnosis for inefficient loops,'' in \emph{Proceedings of the 39th International Conference on Software Engineering, {ICSE} 2017, Buenos Aires, Argentina, May 20-28, 2017}, S.~Uchitel, A.~Orso, and M.~P. Robillard, Eds.\hskip 1em plus 0.5em minus 0.4em\relax {IEEE} / {ACM}, 2017, pp. 370--380. [Online]. Available: \url{https://doi.org/10.1109/ICSE.2017.41}
\BIBentrySTDinterwordspacing

\bibitem{DBLP:conf/icse/SuWYC019}
\BIBentryALTinterwordspacing
P.~Su, S.~Wen, H.~Yang, M.~Chabbi, and X.~Liu, ``Redundant loads: a software inefficiency indicator,'' in \emph{Proceedings of the 41st International Conference on Software Engineering, {ICSE} 2019, Montreal, QC, Canada, May 25-31, 2019}, J.~M. Atlee, T.~Bultan, and J.~Whittle, Eds.\hskip 1em plus 0.5em minus 0.4em\relax {IEEE} / {ACM}, 2019, pp. 982--993. [Online]. Available: \url{https://doi.org/10.1109/ICSE.2019.00103}
\BIBentrySTDinterwordspacing

\bibitem{DBLP:conf/icse/YangSLYC18}
\BIBentryALTinterwordspacing
J.~Yang, P.~Subramaniam, S.~Lu, C.~Yan, and A.~Cheung, ``How \emph{not} to structure your database-backed web applications: a study of performance bugs in the wild,'' in \emph{Proceedings of the 40th International Conference on Software Engineering, {ICSE} 2018, Gothenburg, Sweden, May 27 - June 03, 2018}, M.~Chaudron, I.~Crnkovic, M.~Chechik, and M.~Harman, Eds.\hskip 1em plus 0.5em minus 0.4em\relax {ACM}, 2018, pp. 800--810. [Online]. Available: \url{https://doi.org/10.1145/3180155.3180194}
\BIBentrySTDinterwordspacing

\bibitem{DBLP:conf/issta/YuP16}
\BIBentryALTinterwordspacing
T.~Yu and M.~Pradel, ``Syncprof: detecting, localizing, and optimizing synchronization bottlenecks,'' in \emph{Proceedings of the 25th International Symposium on Software Testing and Analysis, {ISSTA} 2016, Saarbr{\"{u}}cken, Germany, July 18-20, 2016}, A.~Zeller and A.~Roychoudhury, Eds.\hskip 1em plus 0.5em minus 0.4em\relax {ACM}, 2016, pp. 389--400. [Online]. Available: \url{https://doi.org/10.1145/2931037.2931070}
\BIBentrySTDinterwordspacing

\bibitem{DBLP:conf/pldi/CoppaDF12}
\BIBentryALTinterwordspacing
E.~Coppa, C.~Demetrescu, and I.~Finocchi, ``Input-sensitive profiling,'' in \emph{{ACM} {SIGPLAN} Conference on Programming Language Design and Implementation, {PLDI} '12, Beijing, China - June 11 - 16, 2012}, J.~Vitek, H.~Lin, and F.~Tip, Eds.\hskip 1em plus 0.5em minus 0.4em\relax {ACM}, 2012, pp. 89--98. [Online]. Available: \url{https://doi.org/10.1145/2254064.2254076}
\BIBentrySTDinterwordspacing

\bibitem{DBLP:conf/issta/ShenLPG15}
\BIBentryALTinterwordspacing
D.~Shen, Q.~Luo, D.~Poshyvanyk, and M.~Grechanik, ``Automating performance bottleneck detection using search-based application profiling,'' in \emph{Proceedings of the 2015 International Symposium on Software Testing and Analysis, {ISSTA} 2015, Baltimore, MD, USA, July 12-17, 2015}, M.~Young and T.~Xie, Eds.\hskip 1em plus 0.5em minus 0.4em\relax {ACM}, 2015, pp. 270--281. [Online]. Available: \url{https://doi.org/10.1145/2771783.2771816}
\BIBentrySTDinterwordspacing

\bibitem{DBLP:conf/icse/YeChen25}
Y.~Ye, T.~Chen, and M.~Li, ``Distilled lifelong self-adaptation for configurable systems,'' in \emph{47th {IEEE/ACM} International Conference on Software Engineering (ICSE)}.\hskip 1em plus 0.5em minus 0.4em\relax {IEEE}, 2025.

\bibitem{DBLP:journals/tse/Nair0MSA20}
\BIBentryALTinterwordspacing
V.~Nair, Z.~Yu, T.~Menzies, N.~Siegmund, and S.~Apel, ``Finding faster configurations using {FLASH},'' \emph{{IEEE} Trans. Software Eng.}, vol.~46, no.~7, pp. 794--811, 2020. [Online]. Available: \url{https://doi.org/10.1109/TSE.2018.2870895}
\BIBentrySTDinterwordspacing

\bibitem{DBLP:conf/icse/0003XC021}
\BIBentryALTinterwordspacing
J.~Chen, N.~Xu, P.~Chen, and H.~Zhang, ``Efficient compiler autotuning via bayesian optimization,'' in \emph{43rd {IEEE/ACM} International Conference on Software Engineering, {ICSE} 2021, Madrid, Spain, 22-30 May 2021}.\hskip 1em plus 0.5em minus 0.4em\relax {IEEE}, 2021, pp. 1198--1209. [Online]. Available: \url{https://doi.org/10.1109/ICSE43902.2021.00110}
\BIBentrySTDinterwordspacing

\bibitem{DBLP:conf/sigsoft/0001L21}
\BIBentryALTinterwordspacing
T.~Chen and M.~Li, ``Multi-objectivizing software configuration tuning,'' in \emph{{ESEC/FSE} '21: 29th {ACM} Joint European Software Engineering Conference and Symposium on the Foundations of Software Engineering, Athens, Greece, August 23-28, 2021}, D.~Spinellis, G.~Gousios, M.~Chechik, and M.~D. Penta, Eds.\hskip 1em plus 0.5em minus 0.4em\relax {ACM}, 2021, pp. 453--465. [Online]. Available: \url{https://doi.org/10.1145/3468264.3468555}
\BIBentrySTDinterwordspacing

\bibitem{DBLP:conf/seams/Chen22}
\BIBentryALTinterwordspacing
T.~Chen, ``Planning landscape analysis for self-adaptive systems,'' in \emph{International Symposium on Software Engineering for Adaptive and Self-Managing Systems, {SEAMS} 2022, Pittsburgh, PA, USA, May 22-24, 2022}, B.~R. Schmerl, M.~Maggio, and J.~C{\'{a}}mara, Eds.\hskip 1em plus 0.5em minus 0.4em\relax {ACM/IEEE}, 2022, pp. 84--90. [Online]. Available: \url{https://doi.org/10.1145/3524844.3528060}
\BIBentrySTDinterwordspacing

\bibitem{DBLP:conf/wcre/Chen22}
\BIBentryALTinterwordspacing
{Tao Chen}, ``Lifelong dynamic optimization for self-adaptive systems: Fact or fiction?'' in \emph{{IEEE} International Conference on Software Analysis, Evolution and Reengineering, {SANER} 2022, Honolulu, HI, USA, March 15-18, 2022}.\hskip 1em plus 0.5em minus 0.4em\relax {IEEE}, 2022, pp. 78--89. [Online]. Available: \url{https://doi.org/10.1109/SANER53432.2022.00022}
\BIBentrySTDinterwordspacing

\bibitem{DBLP:conf/icse/Kumar0BB20}
\BIBentryALTinterwordspacing
S.~Kumar, T.~Chen, R.~Bahsoon, and R.~Buyya, ``{DATESSO:} self-adapting service composition with debt-aware two levels constraint reasoning,'' in \emph{{SEAMS} '20: {IEEE/ACM} 15th International Symposium on Software Engineering for Adaptive and Self-Managing Systems, Seoul, Republic of Korea, 29 June - 3 July, 2020}, S.~Honiden, E.~D. Nitto, and R.~Calinescu, Eds.\hskip 1em plus 0.5em minus 0.4em\relax {ACM}, 2020, pp. 96--107. [Online]. Available: \url{https://doi.org/10.1145/3387939.3391604}
\BIBentrySTDinterwordspacing

\bibitem{DBLP:conf/icse/Chen19b}
\BIBentryALTinterwordspacing
T.~Chen, ``All versus one: an empirical comparison on retrained and incremental machine learning for modeling performance of adaptable software,'' in \emph{Proceedings of the 14th International Symposium on Software Engineering for Adaptive and Self-Managing Systems, SEAMS@ICSE 2019, Montreal, QC, Canada, May 25-31, 2019}, M.~Litoiu, S.~Clarke, and K.~Tei, Eds.\hskip 1em plus 0.5em minus 0.4em\relax {ACM}, 2019, pp. 157--168. [Online]. Available: \url{https://doi.org/10.1109/SEAMS.2019.00029}
\BIBentrySTDinterwordspacing

\bibitem{DBLP:journals/tosem/ChenLBY18}
\BIBentryALTinterwordspacing
T.~Chen, K.~Li, R.~Bahsoon, and X.~Yao, ``{FEMOSAA:} feature-guided and knee-driven multi-objective optimization for self-adaptive software,'' \emph{{ACM} Trans. Softw. Eng. Methodol.}, vol.~27, no.~2, pp. 5:1--5:50, 2018. [Online]. Available: \url{https://doi.org/10.1145/3204459}
\BIBentrySTDinterwordspacing

\bibitem{DBLP:journals/tsc/ChenB17}
\BIBentryALTinterwordspacing
T.~Chen and R.~Bahsoon, ``Self-adaptive trade-off decision making for autoscaling cloud-based services,'' \emph{{IEEE} Trans. Serv. Comput.}, vol.~10, no.~4, pp. 618--632, 2017. [Online]. Available: \url{https://doi.org/10.1109/TSC.2015.2499770}
\BIBentrySTDinterwordspacing

\end{thebibliography}
% \bibliography{IEEEabrv,mybibf}

\end{document}